\documentclass[11pt,a4paper]{article}
\pdfoutput=1

\usepackage{jheppub-nosort,amsthm}

\PassOptionsToPackage{svgnames}{xcolor}
\usepackage{xcolor}
\usepackage{amsmath,amssymb,amsthm,amscd}
\usepackage{physics}
\usepackage{tikz}
\usetikzlibrary{decorations.pathmorphing}
\tikzset{node distance=2cm, auto}
\tikzset{snake it/.style={decorate, decoration=snake}}
\usepackage{epsfig}
\usepackage{psfrag,comment}
\usepackage{graphicx}
\usepackage{caption}
\usepackage{subcaption}
\usepackage{mathrsfs}
\usepackage[normalem]{ulem}
\usepackage{bm}
\usepackage{lscape}
\usepackage{diagbox}
\usepackage{upgreek}

\usepackage{url}
\newcommand\doilink[1]{\href{http://dx.doi.org/#1}{#1}}
\newcommand\arxivlink[1]{\href{http://arxiv.org/abs/#1}{#1}}

\newcommand{\CA}{{\mathcal A}}

\newcommand{\CC}{{\mathcal C}}

\newcommand{\CE}{{\mathcal E}}
\newcommand{\CF}{{\mathcal F}}

\newcommand{\CI}{{\mathcal I}}

\newcommand{\CN}{{\mathcal N}}

\newcommand{\CS}{{\mathcal S}}

\newcommand{\CX}{{\mathcal X}}
\newcommand{\CY}{{\mathcal Y}}
\newcommand{\CZ}{{\mathcal Z}}

\newcommand{\NCC}{{\mathscr C}}

\newcommand{\NCF}{{\mathscr F}}

\newcommand{\NCO}{{\mathscr O}}

\newcommand{\NCZ}{{\mathscr Z}}

\def\BN{{\mathbb N}}
\def\BZ{{\mathbb Z}}
\def\BR{{\mathbb R}}
\def\BC{{\mathbb C}}
\def\BP{{\mathbb P}}
\def\BT{{\mathbb T}}
\def\BS{{\mathbb S}}
\def\BH{{\mathbb H}}

\newcommand{\be}{\begin{equation}}
\newcommand{\ee}{\end{equation}}
\newcommand{\ba}{\begin{aligned}}
\newcommand{\ea}{\end{aligned}}
\newcommand{\bea}{\begin{eqnarray}}
\newcommand{\eea}{\end{eqnarray}}
\newcommand{\bean}{\begin{eqnarray*}}
\newcommand{\eean}{\end{eqnarray*}}

\def\r{\right\rangle}

\def\1{\mathbf{1}}
\def\0{|\1\r}

\newcommand{\rme}{{\mathrm{e}}}
\newcommand{\rmi}{{\mathrm{i}}}
\newcommand{\rmd}{{\mathrm{d}}}

\DeclareMathOperator{\arccosh}{arccosh}

\def\XXint#1#2#3{{\setbox0=\hbox{$#1{#2#3}{\int}$}
     \vcenter{\hbox{$#2#3$}}\kern-.5\wd0}}

\makeatletter
\newsavebox\myboxA
\newsavebox\myboxB
\newlength\mylenA

\newcommand*\widebar[2][0.75]{%
    \sbox{\myboxA}{$\m@th#2$}%
    \setbox\myboxB\null
    \ht\myboxB=\ht\myboxA%
    \dp\myboxB=\dp\myboxA%
    \wd\myboxB=#1\wd\myboxA
    \sbox\myboxB{$\m@th\overline{\copy\myboxB}$}
    \setlength\mylenA{\the\wd\myboxA}
    \addtolength\mylenA{-\the\wd\myboxB}%
    \ifdim\wd\myboxB<\wd\myboxA%
       \rlap{\hskip 0.8\mylenA\usebox\myboxB}{\usebox\myboxA}%
    \else
        \hskip -0.5\mylenA\rlap{\usebox\myboxA}{\hskip 0.5\mylenA\usebox\myboxB}%
    \fi}
\makeatother

\newdimen\tableauside\tableauside=1.0ex
\newdimen\tableaurule\tableaurule=0.4pt
\newdimen\tableaustep
\def\phantomhrule#1{\hbox{\vbox to0pt{\hrule height\tableaurule width#1\vss}}}
\def\phantomvrule#1{\vbox{\hbox to0pt{\vrule width\tableaurule height#1\hss}}}
\def\sqr{\vbox{%
  \phantomhrule\tableaustep
  \hbox{\phantomvrule\tableaustep\kern\tableaustep\phantomvrule\tableaustep}%
  \hbox{\vbox{\phantomhrule\tableauside}\kern-\tableaurule}}}
\def\squares#1{\hbox{\count0=#1\noindent\loop\sqr
  \advance\count0 by-1 \ifnum\count0>0\repeat}}
\def\tableau#1{\vcenter{\offinterlineskip
  \tableaustep=\tableauside\advance\tableaustep by-\tableaurule
  \kern\normallineskip\hbox
    {\kern\normallineskip\vbox
      {\gettableau#1 0 }%
     \kern\normallineskip\kern\tableaurule}%
  \kern\normallineskip\kern\tableaurule}}
\def\gettableau#1{\ifnum#1=0\let\next=\null\else
\squares{#1}\let\next=\gettableau\fi\next}
\definecolor{cottoncandy}{rgb}{1.0, 0.74, 0.85}
\definecolor{cornellred}{rgb}{0.7, 0.11, 0.11}
\definecolor{darktangerine}{rgb}{1.0, 0.66, 0.07}

\newcommand{\picscale}{0.8}
\definecolor{iceberg}{rgb}{0.44, 0.65, 0.82}
\newcommand{\DiscFZZT}[1]{\mathord{
\begin{tikzpicture}[baseline={([yshift=-0.8ex]current bounding box.center)}, line width=1, scale=\picscale]
\draw[line width=2pt] (0,0) circle (0.8cm);
\fill[iceberg] (0,0) circle (0.8cm);
\node at (1,0.6) {\footnotesize $#1$};
\end{tikzpicture}}}
\newcommand{\DiscZZ}[1]{\mathord{
\begin{tikzpicture}[baseline={([yshift=-0.8ex]current bounding box.center)}, line width=1, scale=\picscale]
\draw[line width=2pt] (0,0) circle (0.8cm);
\fill[iceberg] (0,0) circle (0.8cm);
\node at (1.35,0.6) {\footnotesize $#1$};
\end{tikzpicture}}}
\newcommand{\AnnulusFZZT}[2]{\mathord{
\begin{tikzpicture}[baseline={([yshift=-0.8ex]current bounding box.center)}, line width=1, scale=\picscale]
\draw[line width=2pt] (0,0) circle (0.6cm);
\draw[line width=2pt] (0,0) circle (1.2cm);
\fill[iceberg, even odd rule] (0,0) circle[radius=1.2cm] circle[radius=0.6cm];
\node at (0.2,0.15) {\footnotesize $#1$};
\node at (1.4,0.95) {\footnotesize $#2$};
\end{tikzpicture}}}

\title{All the D-Branes of Resurgence}

\author[a,b]{Ricardo~Schiappa,}
\affiliation[a]{Isaac Newton Institute for Mathematical Sciences,\\ University of Cambridge, Cambridge CB3 0EH, United Kingdom\\}
\affiliation[b]{CAMGSD, Departamento de Matem\'atica, Instituto Superior T\'ecnico,\\ Universidade de Lisboa, 1049-001 Lisboa, Portugal\\}
\emailAdd{ricardo.schiappa@}

\author[a,b]{Maximilian~Schwick,}
\emailAdd{maximilian.schwick@}

\author[a,b]{Noam~Tamarin\,}
\emailAdd{noam.tamarin@tecnico.ulisboa.pt}

\abstract{
It was recently shown how to account for all instantons of hermitian matrix models via (anti-) eigenvalue-tunneling---including both exponentially-suppressed and exponentially-enhanced transseries-transmonomials which are predicted by resurgence. Matrix-model eigen-\ value-tunneling corresponds to ZZ-branes. The present work shows how matrix-model \textit{anti}-eigenvalues correspond to \textit{negative-tension} ZZ-branes; and how to compute generic nonperturbative sectors---with both ZZ and negative-tension-ZZ branes---in the minimal-string free-energy. Negative-tension D-branes are herein a \textit{requirement} of resurgence. This results in the construction of minimal-string free-energy transseries and the analytic computation of their resurgent Stokes data. Calculations are presented via Liouville boundary conformal field theory and via (matching) matrix model analysis. Minimal-string results are extended to Jackiw--Teitelboim gravity. Building on the matrix model analysis, one extension towards topological string theory is obtained via the remodeling-conjecture---which allows for addressing one-cut, toric Calabi--Yau geometries. Building on the Liouville theory calculation, one other extension towards critical string theory is obtained via the $\BH_3^+$--Liouville correspondence---which allows for addressing negative-tension D-instantons in AdS spacetime. Throughout, checks of the construction and formulae are made in several examples, against both Borel resurgent analysis and string-equation transseries data.
}


\arxivnumber{2301.05214}

\begin{document}

\maketitle

\vfill

\eject

\allowdisplaybreaks

\section{Introduction and Summary}\label{sec:intro}

D-branes stand as one of the great discoveries in string theory \cite{p94, p95}. They open a remarkable nonperturbative window with wide applications ranging from black-hole microstate counting \cite{sv96} to the celebrated AdS/CFT large $N$ duality \cite{m97}. Playing such prominent roles, it is clear that finding all the D-branes in a given closed-string (curved) background is of paramount importance in the full description of the theory. This, however, is not always an easy task \cite{mms01}. The literature abounds with very interesting examples, of which an extremely partial list includes, \textit{e.g.}, D-branes in minimal string theory \cite{fzz00, t00, zz01}, D-branes in WZW models \cite{ks96, as98, fffs99, s99b, mms01}, D-branes in AdS spacetimes \cite{s99a, bp00, gks01, lop01, pst01, r05, gkv21}, D-branes in Calabi--Yau geometries \cite{ooy96, a04}, and so on and on. If in addition we recall the myriad of different D-branes one may encounter (\textit{e.g.}, different RR charges \cite{p95}, S-branes \cite{gs02}, ghost D-branes \cite{ot06}, and the like), this immediately begs for the question: in specific problems, how can one be sure if we indeed found them all?

On a different line of work, and in fact much predating the discovery of D-branes, the results in \cite{bw69, bw73, bpv78a, bpv78b} sparked a plethora of research into the large-order growth of perturbation theory in one-dimensional quantum-mechanics---therein regarded as another open-window into the nonperturbative content of these many interesting quantum problems \cite{l77, bl-gz-j77, z81}. Work across the years has since shown how this is a rather generic feature, and how digging deeper into asymptotic growths allows for unveilings of the full instanton or renormalon content of many distinct quantum and field theories. Translating this line of thought into string theory---historically defined in an inherently perturbative fashion with \textit{asymptotic} string-theoretic perturbative-expansion \cite{gp88}---one is immediately led to recall how the nature of its perturbative asymptotic growth was in fact a precursor to the discovery of D-branes \cite{s90}. Might this then also be a general strategy towards unveiling the \textit{full} D-brane content of \textit{any} string theory?

One context where to better frame this question is that of closed string theories associated to double-scaled multicritical hermitian matrix-models \cite{gm90a, ds90, bk90, d90, gm90b}; models which were at the time dubbed ``solvable'' due to recursion-relations---obtained out of their corresponding string equations---iteratively yielding their full perturbative data (see, \textit{e.g.}, \cite{g91, gm93, dgz93} for reviews). Associated to these perturbative data, a fair amount of nonperturbative data alongside their semiclassical interpretation as matrix-model eigenvalue tunneling, was also obtained at the same time \cite{d91, d92}---and these nonperturbative data were further validated against the leading large-order growth of the corresponding perturbative sectors \cite{gz90b, gz91, ez93}. A second wave of interest in matrix-model descriptions of string theories refocused attention upon minimal\footnote{Where the worldsheet matter-content is given by the minimal models of \cite{bpz84}, and where the KdV times \cite{gd75} are now retuned from the multicritical to the conformal background (see, \textit{e.g.}, \cite{mss91, ss03, gs21}).} string theories \cite{ss03, ss04b}, with similar properties to the aforementioned multicritical models (see, \textit{e.g.}, \cite{n04, ss04a} for reviews). One important novelty in the minimal string story is the clear identification of matrix model and string theoretic quantities; in particular, the identification of D-branes: eigenvalue-tunneling corresponds to ZZ-branes \cite{zz01}, and the matrix-model holomorphic effective-potential describes FZZT-branes \cite{fzz00, t00}. For example, many properties of ZZ-brane instanton contributions follow quite straightforwardly from matrix model calculations \cite{akk03, ss03, hhikkmt04, st04, iy05, iky05}. Matching of matrix model and boundary conformal field theory calculations in these models has further seen recent progress in \cite{emms22a, emms22b}. But this is not the end of the story.

That there is more to this story than described above was sparked by \cite{m06}, setting-up large-order analyses for (off critical) matrix models and topological strings, later carried through in \cite{msw07, msw08} beyond leading order. Resurgence \cite{e81} enters the game in \cite{m08}, where resurgent transseries neatly package together the full perturbative and nonperturbative multi-instanton content of matrix models, unveiled in the aforementioned references (see, \textit{e.g.}, \cite{m12, abs18} for reviews). This line of research finally led to \cite{gikm10}, which pioneered analyzing the asymptotics of \textit{multi-instanton} sectors (in the specific Painlev\'e~I example), and where it was made clear that these transseries are \textit{resonant}, \textit{i.e.}, instanton actions always arise in \textit{symmetric} pairs. This was extended to (off critical) matrix models and further examples in \cite{ps09, asv11, sv13, as13, gs21} (resonance being \textit{generic} for multicritical and minimal string models \cite{gs21}). Yet, the lingering question concerned the matrix-model semiclassical interpretation of resonance. It was recently shown in \cite{mss22} that whereas eigenvalue tunneling in the physical sheet of the matrix-model large-$N$ spectral-curve describes the $\sim \exp \left( - 1/g_{\text{s}} \right)$ multi-instanton sectors of a resonant resurgent-transseries, it is eigenvalue tunneling into the \textit{non-physical sheet} which describes the $\sim \exp \left( + 1/g_{\text{s}} \right)$ resonant pairs (with $g_{\text{s}}$ the string coupling). Now, the former correspond to minimal string ZZ-branes. One question we address in this work is how the latter correspond to minimal-string\footnote{The reader should bear in mind that whereas minimal string theory naturally includes negative-tension D-branes, these are \textit{not} the ones which we are referring to. Rather, our claim is that each positive-tension D-brane will now have a negative-tension counterpart, and vice-versa. In other words, just like for the instanton actions, also positive and negative-tension D-branes will always arise in \textit{symmetric} pairs.} \textit{negative-tension} ZZ-branes. Such ``ghost'' or ``negative'' branes\footnote{Mostly they have been studied in the superstring context, with both NS and R sectors. As a quick reminder, if we denote a D-brane boundary state by $\ket{\text{D}} = \ket{\text{NS}} + \ket{\text{R}}$ then its anti-D-brane counterpart will be $\ket{\text{anti-D}} = \ket{\text{NS}} - \ket{\text{R}}$, but its negative-tension pair is instead $\ket{\text{negative-D}} = - \ket{\text{D}} = - \ket{\text{NS}} - \ket{\text{R}}$ with an overall rather than a relative minus sign. In this paper we only have overall minus signs, which will turn out to be quite non-trivial.} have previously appeared in the literature \cite{v01, ot06, v14, dhjv16}. Herein, \textit{negative-tension D-branes turn out to be a requirement of resurgence}.

One important aspect to the resurgent transseries constructions deals with their non-linear Stokes data (precisely quantifying Stokes-phenomenon jumps of the asymptotic expansions). This is what allows us to extend transseries from asymptotic series to full-fledged functions; but Stokes data is generically very hard to compute analytically. In the aforementioned Painlev\'e~I example, one Stokes coefficient was originally known---computed as the one-loop coefficient around the one-instanton sector associated to eigenvalue tunneling \cite{d92, msw07} (extension for all multicritical and minimal string models appeared in, \textit{e.g.}, \cite{gs21}). But resonance predicts \textit{infinitely} many other resurgent Stokes coefficients \cite{gikm10, asv11, sv13, as13, abs18}. Recently, this infinite set of (transcendental) numbers was found in analytical closed-form via analysis of resonant resurgence \cite{bssv22}, and then received a direct matrix-model calculation via (anti) eigenvalue tunneling in \cite{mss22}. Our present work gives yet another take on these non-linear resurgent Stokes data, with a direct boundary conformal field theory (B-CFT) calculation via negative-tension D-branes.

One last comment concerning Stokes phenomenon connects to the results in \cite{mmss04}. Let us recall their important main point: in the semiclassical approximation observables live on branched, possibly multi-sheeted, geometries (\textit{e.g.}, the large $N$ spectral curve)---but exact quantum observables should be entire functions; which is only possible when starting from semiclassics if Stokes phenomena intervenes in order to achieve full analyticity \cite{mmss04}. A somewhat similar phenomenon was also observed in \cite{mz16}. It is now clear that full analyticity entails knowledge of all possible Stokes jumps, everywhere on the complex plane---which at the transseries level translates to knowledge of the complete set of transmonomials at play (be them different eigenvalue types, different D-branes, or whatever else). Our present construction yields the \textit{complete} semiclassical interpretation of the resurgent structure of minimal strings; fully allowing implementation of \textit{all possible} Stokes phenomena at play and finally construct exact, entire quantum observables.

\paragraph{The Contents of the Paper:}

We begin in section~\ref{sec:ZZminimalstring} with a complete BCFT analysis of $(p,q)$ minimal string theory, including both standard and negative-tension D-branes. This is essentially the BCFT version of the matrix model calculations in \cite{mss22} and very much constructs upon \cite{m03, kopss04} for the computation of disk and (regularized) annulus amplitudes (in both ZZ and FZZT cases). In fact, we immediately address the need for such a regularization in subsection~\ref{subsec:regularization-one-instanton}, which we set-up (building upon \cite{kopss04}) via careful analytic considerations---and further compare to the recent string field theory regularization in \cite{emms22a}. Using such properly regularized amplitudes immediately supports for both tension types of ZZ-branes, as discussed in the ensuing subsection~\ref{subsec:bcft-resonant-pairs}. Calculations addressing different possible types of ZZ- and negative-tension ZZ-brane contributions, across different possible sheets of the FZZT moduli space in Liouville theory, are discussed in subsection~\ref{subsec:generic-ZZ-BCFT}---including explicit nonperturbative \textit{mixed} sectors of the minimal-string free-energy transseries. A generalization to \textit{pure} (negative-tension) multi-ZZ-instantons is then computed in subsection~\ref{subsec:ZZ+negativeZZ}. Our ZZ-brane results are straightforwardly extended to FZZT-branes in subsection~\ref{subsec:FZZT-BCFT}. It is important to point out that subsection~\ref{subsec:generic-ZZ-BCFT} further makes it rather explicit how an \textit{a priori} seemingly-harmless overall minus sign, associated to a negative-tension D-brane, can become rather non-trivial. As we shall see, these negative-signs appear in exponents in the integrands of Liouville amplitudes, thus leading to the appearance of integrand-poles where one would otherwise naively expect to find integrand-zeroes---and hence to rather different final integrals for the Liouville BCFT amplitudes wherever they are featured. The avid reader may take a sneak preview of this phenomenon in formulae \eqref{eq:20-contribution-minimal-strings} and \eqref{eq:11-contribution-minimal-strings}.

In section~\ref{sec:NPminimalstring} we turn to the matrix model analysis of minimal strings, now focusing upon $(p,q)=(2,2k-1)$ as we mainly address the one-matrix model case. This is essentially the (degenerate hyperelliptic) multi-pinched extension of the analysis in \cite{mss22}---which can become quite intricate at times. The calculation is generically discussed in subsection~\ref{subsec:double-scaled-geometry}, in the context of the double-scaling limit; and then is explicitly applied to the $(2,2k-1)$ minimal string theory in subsection~\ref{subsec:22k-1minstring}. These matrix model results fully match the earlier BCFT results in section~\ref{sec:ZZminimalstring}, as will be shown in subsection~\ref{subsec:22k-1minstring}. On top of this, they are further matched against results arising from string equations (as described in appendices~\ref{app:minimal-string-equation-setup} and~\ref{app:minimal-string-25}), so that our proposal is in fact \textit{triple} checked---as will be shown in subsection~\ref{subsec:BCFT-MM-stringeqs}. Further supporting evidence is gathered with resurgent large-order tests for Stokes data and Borel--Pad\'e analysis for Borel singularities, also in subsection~\ref{subsec:BCFT-MM-stringeqs}. This supports the overall consistency of all proposal and results. The minimal string results are then extended to Jackiw--Teitelboim (JT) gravity \cite{t83, j85, ap14, ms16, msy16, sw17} in subsection~\ref{subsec:JTgravity}, by simple application of our generic double-scaled formulae to the spectral curve of its matrix model \cite{sss19}. The resonant nature of JT gravity, which has already been addressed in \cite{gs21}, is further supported by direct Borel--Pad\'e analysis of its Borel singularities, and by the explicit large $k$ limit of the earlier $(2,2k-1)$ results. A very short discussion on the extension of our matrix model calculations to the case of the two-matrix model is included in subsection~\ref{subsec:2matrixmodel}. Throughout section~\ref{sec:NPminimalstring} we include many explicit calculations of non-trivial, multiple ZZ- and multiple negative-tension ZZ-instanton contributions to the minimal-string free-energy transseries, alongside analytic computations of their non-linear Stokes data. Matrix model calculations many times involve rather intricate integrals, and we included one generic such integral in appendix~\ref{app:matrix-formula} for the convenience of the reader who wants to reproduce our calculations.

Having gathered strong and compelling evidence for the resurgence requirement of negative-tension D-branes within minimal string theory, one is then led to wonder if the same may hold true across generic string theories. This is addressed in section~\ref{sec:topcritstrings}, where we discuss possible extensions towards topological string theory in toric backgrounds and towards critical string theory in AdS spacetimes. The topological string extension is very much based on the fact that the results in \cite{mss22} alongside our results in section~\ref{sec:NPminimalstring} are model independent, in the sense that they only depend on spectral-curve data---and may hence be applicable in broader contexts. For example, for topological string theory in toric Calabi--Yau geometries it is well known that applying the topological recursion \cite{eo07a} to the B-model \textit{mirror curve} remarkably yields the topological string free-energy expansion \cite{bkmp07}. This procedure was later extended to topological-string nonperturbative instanton-calculus in \cite{msw07}, addressing the one-instanton sector of the local-curve toric geometry. Subsection~\ref{subsec:top-strings} revisits the local curve at nonperturbative level, showing how an analysis of Borel singularities reveals the resonant nature of its nonperturbative free energy. Formulae in \cite{mss22} and our section~\ref{sec:NPminimalstring} are then applied to this problem following \cite{m06, bkmp07, msw07}, yielding many predictions for the resonant nonperturbative sectors of this model (which are further checked against their double-scaling limit towards the Painlev\'e~I transseries). A very brief discussion concerning topological strings on local $\BP^2$ and local $\BP^1 \times \BP^1$ is also included. Turning to critical string theory, with focus upon AdS spacetimes, at first seems harder. But in subsection~\ref{subsec:ads-strings} we build upon the remarkable $\BH_3^+$--Liouville correspondence \cite{rt05, hr06, hs07}, allowing for use of our Liouville results from section~\ref{sec:ZZminimalstring} in addressing D-branes in euclidean $\text{AdS}_3$. In this light, Liouville FZZT branes correspond to $\text{AdS}_2$ branes in $\text{AdS}_3$, and the Liouville ZZ branes have D-instanton analogues in $\text{AdS}_3$ \cite{r05}. Building on our discussion of resonant negative-tension D-branes in Liouville theory earlier on, we can thus infer on the existence of negative-tension D-instantons in $\text{AdS}_3$. The evidence is circumstantial as we mainly rely on the $\BH_3^+$--Liouville map---implying that due to the natural added computational intricacies, the results herein are not as exhaustive as in the minimal or topological string examples. We nonetheless believe they amount to clear and supporting evidence, and further work along these directions will be reported in the near future.

\paragraph{A Resurgence Requirement:}

Let us stress how our overall results are a \textit{direct requirement of resurgence}, simply given the nature of the large-order growth of the string perturbative-expansion. With hindsight, in fact, one could even claim that our results are obviously expected. Consider the perturbative expansions for either the string-theoretic free energy or the multi-resolvent\footnote{These are the Laplace transforms of multiple macroscopic-loop operator insertions \cite{bdss90, mss91, eggls23}.} correlation functions,
\begin{align}
\label{eq:F-generic}
F &\simeq \sum_{g=0}^{+\infty} F_g (t)\, g_{\text{s}}^{2g-2}, \\
\label{eq:W-generic}
W_{h} \left( x_1, \ldots, x_h \right) &\simeq \sum_{g=0}^{+\infty} W_{g,h} \left( x_1, \ldots, x_h; t \right) g_{\text{s}}^{2g-2+h}
\end{align}
\noindent
(with $t$ some background-geometry modulus). These expansions are asymptotic, with leading large-order growths $F_g (t) \sim \left( 2g \right)!$ and $W_{g,h} (t) \sim \left( 2g+h \right)!$. While it is very well-known how the naive expectation where such growths would lead to nonperturbative exponential corrections of the type $\sim \exp \left( -1/g_{\text{s}}^2 \right)$ is incorrect---rather leading to the extremely well-known D-brane-type nonperturbative exponential corrections $\sim \exp \left( -1/g_{\text{s}} \right)$---it seems that the need for the resonant siblings $\sim \exp \left( +1/g_{\text{s}} \right)$ has somehow escaped detailed scrutiny; but see the discussions in \cite{asv11, abs18}. In fact these resonant pairs are required simply so as to keep the above asymptotic perturbative expansions in \textit{even} powers of the string coupling. This is a straightforward statement on the Borel plane (see, \textit{e.g.}, \cite{abs18}). Taking a Borel transform $\mathcal{B} : \mathbb{C}[[g_{\text{s}}]] \to \mathbb{C}\lbrace s\rbrace$ of the above asymptotic series \eqref{eq:F-generic}-\eqref{eq:W-generic} immediately yields parity-fixed functions\footnote{The resonant nature of multi-resolvent correlation functions is addressed in \cite{eggls23}.},
\begin{equation}
\mathcal{B} \left[F\right] (s) = - \mathcal{B} \left[F\right] (-s), \qquad \mathcal{B} \left[W_h\right] (s) = (-1)^{h-1} \mathcal{B} \left[W_h\right](-s).
\end{equation}
\noindent
This simple statement immediately implies that Borel singularities are symmetrically distributed upon the complex Borel plane $s \in \BC$, \textit{i.e.}, it immediately implies the aforementioned resonant behavior of their associated nonperturbative exponential contributions. As already mentioned, this then leads to the minus signs which end up swapping zeros and poles inside integrands over D-brane moduli space (as will be later be made explicit in, \textit{e.g.}, formulae \eqref{eq:20-contribution-minimal-strings} and \eqref{eq:11-contribution-minimal-strings}).

\paragraph{Moving Forward in Future Work:}

At the root of the string-theoretic results reported in this paper lies the study of the adequate analytic regularization of ZZ- and FZZT-brane amplitudes in Liouville BCFT\footnote{Analytic continuation of Liouville CFT was studied in \cite{hmw11}.}. Given the centrality of Liouville theory in the whole string theory construct \cite{p81a, p81b}, it would certainly not come as a big surprise if the present results would be transverse to larger classes of backgrounds. We give evidence along this direction for strings in toric Calabi--Yau and Anti-de~Sitter backgrounds. Nonetheless, further evidence and further backgrounds are certainly required in order to make any definite statements. For example, in upcoming work we plan to report on the matrix-model \textit{multi-cut} analysis, building on \cite{bde00, e08, msw08, em08}. Other models and examples we plan to address in the near future include Chern--Simons matrix models, following upon \cite{m02, akmv02, hy03}; and more intricate toric Calabi--Yau geometries such as local $\BP^2$ or local $\BP^1 \times \BP^1$, following upon \cite{m06, bkmp07}. Hopefully these will stand as strong stepping-stones on the road to address ABJM gauge theory \cite{abjm08} along the lines of \cite{kwy09, dt09, mp09, dmp10, dmp11}; or the more intricate topological-string matrix-models associated to generic toric Calabi--Yau threefolds constructed in \cite{ghm14, mz15, kmz15, z18}. This seems to us as a solid line of research also in light of the large amount of work which exists in the literature tackling resurgence and transseries within topological string theory; \textit{e.g.}, \cite{m06, msw07, m08, msw08, ps09, kmr10, dmp11, cesv13, gmz14, cesv14, ars14, c15, csv16, cms16, gm21, gm22a, gm22b}. Another rather standard line of research going forward would be to work out the supersymmetric extension of our present bosonic results, which should be straightforward albeit possibly more technically involved. As already mentioned in \cite{mss22}, it would furthermore be very interesting to automatize our whole resonant-resurgence construction within the (nonperturbative) topological recursion framework \cite{eo07a, eggls23}. Finally, we would like to understand which---if any---would be the consequences of our results on what concerns generic nonperturbative corrections to the full gauge-theoretic large $N$ expansion \cite{th74} and their eventual role within the AdS/CFT correspondence \cite{m97}.

\section{On All ZZ-Brane Amplitudes of Minimal Strings}\label{sec:ZZminimalstring}

Let us begin by addressing the full resurgence content of D-branes in minimal string theory, building upon \cite{m03, ss03, kopss04, mmss04}. This first requires swiftly recalling some elementary data concerning these models (see, \textit{e.g.}, \cite{n04, ss04a, g10} for reviews and a list of references). The closed-string world-sheet CFT is composed of a $(p,q)$ minimal model \cite{bpz84} describing the background; of Liouville theory describing the conformal mode
\be
\label{eq:Liouville-bulk-action}
\CS_{\text{L}} [\varphi] = \frac{1}{4\pi} \int \rmd^2 \sigma \left( \left( \partial_{a} \varphi \right)^2 + 4\pi \mu\, \rme^{2b\varphi} \right);
\ee
\noindent
and of the usual reparametrization ghosts
\be
\CS_{\text{gh}} [\mathfrak{b},\mathfrak{c}] = \frac{1}{2\pi} \int \rmd^2 \sigma\, \mathfrak{b} \bar{\partial} \mathfrak{c}.
\ee
\noindent
Herein, $\mu$ is the bulk cosmological constant and $b>0$ is the Liouville coupling constant.  The matter, Liouville, and ghost central charges are
\be
c_{p,q} = 1 - 6\, \frac{\left( p-q \right)^2}{pq} < 1, \qquad c_{\text{L}} = 1 + 6 \left( b + \frac{1}{b}\right)^2 > 25, \qquad c_{\text{gh}} = -26,
\ee
\noindent
and the on-shell requirement of vanishing total central charge yields
\be
b^2 = \frac{p}{q}.
\ee

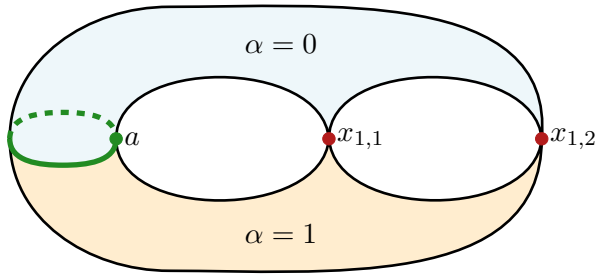
\begin{figure}
\centering
	\begin{tikzpicture}
	\begin{scope}[scale=0.7,  shift={({-5.8},{0})}]
	\draw[fill=LightBlue,fill opacity=0.2, line width=1pt] (0,0)   to [out=90,in=95] (4,0)
	to [out=90,in=95] (8,0)
	to [out=85,in=0] (1,2.5)
    to [out=180,in=90] (-2,0)
    to [out=270, in=180] (-1, -0.5)
    to [out=0, in=270] cycle;
    \draw[fill=darktangerine,fill opacity=0.2, line width=1pt] (-2,0)
    to [out=270,in=180] (1,-2.5)
    to [out=0,in=270] (8,0)
    to [out=265,in=275] (4,0)
    to [out=265,in=270] (0,0)
    to [out=270, in=0] (-1, -0.5)
    to [out=180, in=270] cycle;
    \draw[color=ForestGreen, line width=2pt] (-2,0) to [out=270, in=180] (-1, -0.5)
    to [out=0, in=270] (0,0);
    \draw[dashed, color=ForestGreen, line width=2pt] (-2,0) to [out=90, in=180] (-1, 0.5)
    to [out=0, in=90] (0,0);
\draw[ForestGreen, fill=ForestGreen] (0,0) circle (.7ex);
\draw[cornellred, fill=cornellred] (4,0) circle (.7ex);
\draw[cornellred, fill=cornellred] (8,0) circle (.7ex); 
\node at (0.3, 0) {$a$};
\node at (4.6, 0) {$x_{1, 1}$}; 
\node at (8.6, 0) {$x_{1, 2}$};
\node at (3.1,1.8) {$\alpha=0$}; 
\node at (3.1,-1.8) {$\alpha=1$}; 
\end{scope}
	\end{tikzpicture}
\caption{Example of the $(2,5)$ minimal-string FZZT Riemann surface. Its two sheets are separated by the cut (in {\color{ForestGreen}green}) starting at $a$ and ending at $\infty$ (from the matrix model point-of-view to be discussed later, this is a double-scaled spectral curve). Furthermore, we have labeled the two sheets with $\alpha=0$ (physical sheet in {\color{LightBlue}blue}) and with $\alpha=1$ ({\color{darktangerine}orange}); following \eqref{eq:sheet-labels}. Throughout this section we shall return to similar figures to illustrate our discussion.}
\label{fig:minimalstringspectralcurve-start}
\end{figure}

Moving towards minimal-string D-branes, these also factorize into D-branes of minimal models and D-branes of Liouville theory. The latter then split into ZZ-branes \cite{zz01} and FZZT-branes \cite{fzz00, t00}. The FZZT-branes are labeled by a continuous parameter, which is sometimes taken to be $\mu_{\text{B}}$, the boundary cosmological constant of the boundary term in the Liouville action
\be
\label{eq:Liouville-boundary-action}
\CS_{\text{L,B}} [\varphi] = \mu_{\text{B}} \oint \rme^{b\varphi}.
\ee
\noindent
More conveniently, we shall use the uniformization variable
\be
\label{eq:uniformization}
\upzeta = \cosh \frac{1}{p} \arccosh \frac{\mu_{\text{B}}}{\sqrt{\mu}},
\ee
\noindent
and denote FZZT-branes by $\ket{\upzeta}_{\text{FZZT}}$ with $\upzeta \in \BC$. It is also convenient to introduce variables $x$ and $y$ (which will play more prominent roles in section~\ref{sec:NPminimalstring} as the spectral curve $y=y(x)$) as \cite{ss03}
\bea
\label{eq:uniformization-minimal-string-x}
x &\equiv& \frac{\mu_{\text{B}}}{\sqrt{\mu}} \equiv T_p (\upzeta), \\
\label{eq:uniformization-minimal-string-y}
y &\equiv& \mu^{-\frac{1}{2b^2}}\, \frac{\partial \mathsf{A}_{\text{D}}}{\partial \mu_{\text{B}}} \equiv T_q (\upzeta).
\eea
\noindent
Herein $T_{p} \left( \cos \theta \right) = \cos p \theta$ are Chebyshev polynomials of first kind, the variables $x$ and $y$ satisfy $T_q (x) = T_p (y)$ essentially by definition, and $\mathsf{A}_{\text{D}}$ is the FZZT disk amplitude. We have visualized this Riemann surface in figure~\ref{fig:minimalstringspectralcurve-start} for $(p,q)=(2,5)$. One may equally write (with pictorial notation which will be handy as we move on) \cite{ss03}
\begin{equation}
\label{eq:FZZTdisc}
\mathsf{A}_{\text{D}} \left(\upzeta\right) = \DiscFZZT{\upzeta} = \mu^{\frac{p+q}{2p}}\,\int^{x(\upzeta)} \text{d}x\, y(x).
\end{equation}
\noindent
This is the full D-brane tension, including a factor of $g_{\text{s}}$ which we may then identify as $g_{\text{s}} = \mu^{-\frac{p+q}{2p}}$. From now on, we follow \cite{ss03, kopss04} and rescale $\mu$ by $\pi \gamma(b^2)$ with $\gamma (x) = \Gamma(x)/\Gamma(1-x)$.

The algebraic relation $T_q (x) = T_p (y)$ describes a Riemann surface $\Sigma_{p,q}$ which is a $p$-sheeted covering of the $x$ complex-plane, and we can label these different sheets by simply considering the inverse-map to $T_p (\upzeta)$, which is the set \cite{fim06}
\begin{equation}
\label{eq:sheet-labels}
T_p^{-1} = \lbrace \upzeta_0,\ldots,\upzeta_{p-1} \rbrace. 
\end{equation}
\noindent
Herein $\upzeta_{\alpha}$ with $\alpha=0,\ldots,p-1$ live in the different sheets, in the uniformization-cover language (see figure~\ref{fig:minimalstringspectralcurve-start} again). Further introducing \cite{kopss04}
\begin{equation}
\label{eq:upzeta-to-sigma-map}
\upzeta = \cosh \left( \frac{\pi b \sigma}{p} \right),
\end{equation}
\noindent
the relation between sheets may be explicitly expressed as
\begin{align}
\label{eq:upzeta_alpha}
\upzeta_{\alpha} = \cosh \left(\frac{\pi b \sigma}{p} + \frac{2\pi\rmi\alpha}{p} \right) = \cosh \left( \frac{\pi b}{p} \left(\sigma + \frac{2\rmi\alpha}{b} \right) \right).
\end{align}
\noindent
In particular, jumping in-between sheets may be understood via the shift $\sigma \mapsto \sigma_{\alpha} = \sigma + \frac{2\rmi\alpha}{b}$. For the upcoming discussions it will also prove useful to have the expressions for the FZZT \textit{disk} and \textit{annulus} amplitudes explicitly given in terms of the uniformization variable \cite{m03, ss03, kopss04},
\begin{align}
\label{eq:FZZT-disk}
\mathsf{A}_{\text{D}} (\upzeta) &= \DiscFZZT{\upzeta} = \mu^{\frac{p+q}{2p}}\, \frac{p}{2} \left( \frac{T_{q+p}(\upzeta)}{q+p} - \frac{T_{q-p}(\upzeta)}{q-p} \right), \\
\label{eq:StandardFZZTAnnulus}
\mathsf{A}_{\text{A}} (\upzeta,\tilde{\upzeta}) &= \AnnulusFZZT{\upzeta}{\tilde{\upzeta}} = \log \left(\frac{\upzeta-\tilde{\upzeta}}{x(\upzeta)-x(\tilde{\upzeta})}\right).
\end{align}

Having described FZZT-branes, let us move on to ZZ-branes. These are differences of FZZT-branes \cite{zz01, m03},
\be
\label{eq:ZZ-from-FZZTs}
\ket{m,n}_{\text{ZZ}} = \ket{\cos \pi \left( \frac{m}{p} - \frac{n}{q} \right)}_{\text{FZZT}} - \ket{\cos \pi \left( \frac{m}{p} + \frac{n}{q} \right)}_{\text{FZZT}},
\ee
\noindent
where both FZZT boundary-states share the same value of $\mu_{\text{B}} = (-1)^{m} \sqrt{\mu} \cos \pi b^2 n$ \cite{m03}. As such, ZZ-branes are labeled by two discrete (integer) parameters $m$ and $n$, and we shall denote them by\footnote{There is a connection between the labeling of sheets $\alpha$ for FZZT branes and the labeling of pinches $(m,n)$ for ZZ branes which will soon be made clear in subsection~\ref{subsec:regularization-one-instanton}.}
\be
\ket{m,n}_{\text{ZZ}} \qquad \text{with} \qquad 1 \le m \le p-1, \quad 1 \le n \le q-1, \quad q m - p n > 0,
\ee
\noindent
and where reflexivity further implies $\ket{m,n}_{\text{ZZ}} = \ket{p-m,q-n}_{\text{ZZ}}$. In addition the ZZ-branes are directly related to the singularities of the FZZT Riemann surface $\Sigma_{p,q}$ underlying the algebraic relation $T_q (x) = T_p (y)$ (this is a genus-zero Riemann surface with $\frac{1}{2} \left( p-1 \right) \left( q-1 \right)$ singularities) which are located at \cite{ss03}
\begin{equation}
\label{eq:xmn,ymn,upzetamn}
\left( x_{mn}, y_{mn} \right) = \left( (-1)^m \cos\frac{\pi n p}{q}, (-1)^n \cos\frac{\pi m q}{p} \right), \qquad \upzeta_{mn}^{\pm} = \cos\frac{\pi \left( m q \pm n p \right)}{p q}.
\end{equation}
\noindent
Note that it is sometimes more convenient to use the parameter $\sigma$ in \eqref{eq:upzeta-to-sigma-map} rather than $\upzeta$, for which the above singularities correspond to
\begin{equation}
\sigma \left( m, \pm n \right) = \rmi \left( \frac{m}{b} \pm n b \right).
\end{equation}
\noindent
The ZZ disk amplitudes are then given as period-integrals
\begin{equation}
\label{eq:ZZdisc}
\mathsf{A}_{\text{D}}(m,n) = \DiscZZ{(n,m)} = \mu^{\frac{p+q}{2p}}\, \oint_{B_{m n}} \text{d}x\, y(x),
\end{equation}
\noindent
where $B_{mn}$ is a $B$-cycle between the cut of the Riemann surface and the singular point $(x_{mn}, y_{mn})$ (\textit{e.g.}, one such example will later be illustrated in figure~\ref{fig:minimalstringspectralcurve-resonant-cycles}).

\subsection{Boundary CFT and Analytic Regularization of D-Instantons}\label{subsec:regularization-one-instanton}

The combinatorics of multiple, disconnected worldsheet Dirichlet-boundaries famously exponentiate \cite{p94}, yielding the D-brane nonperturbative one-instanton contribution\footnote{We denote with ``new caligraphic'' notation quantities computed via BCFT on the string worldsheet.} to the string free energy (a ratio of partition functions)
\begin{equation}
\label{eq:1instZZcontribution}
\NCF_{\text{nonpert}}^{(1)} = \frac{\NCZ_{\text{nonpert}}^{(1)}}{\NCZ_{\text{pert}}} \simeq \exp \Bigg( \mathsf{A}_{\text{D}} (m,n) + \frac{1}{2}\mathsf{A}_{\text{A}} (m,n;m,n) + \cdots \Bigg).
\end{equation}
\noindent
In this section, we want to understand the single ZZ-instanton contribution to the minimal-string partition-function or free-energy in a purely conformal field theoretic fashion. However, the BCFT annulus amplitude $\mathsf{A}_{\text{A}}(m,n;m,n)$ of two identical ZZ-branes is known to diverge \cite{m03, kopss04}. Explicitly, denoting the divergent ZZ annulus with a tilde, the original BCFT calculations predict \cite{kopss04}
\begin{equation}
\label{eq:divergentZZAnnulus}
\widetilde{\mathsf{A}}_{\text{A}} (m,n;m^{\prime},n^{\prime}) = \log \frac{\left( \upzeta^{+}_{mn} - \upzeta^{+}_{m^{\prime}n^{\prime}} \right) \left( \upzeta^{-}_{mn} - \upzeta^{-}_{m^{\prime}n^{\prime}} \right)}{\left( \upzeta^{+}_{mn} - \upzeta^{-}_{m^{\prime}n^{\prime}} \right) \left( \upzeta^{-}_{mn} - \upzeta^{+}_{m^{\prime}n^{\prime}} \right)},
\end{equation}
\noindent
which diverges as one sets $(m^{\prime},n^{\prime})=(m,n)$.

This problem has been addressed rather recently, in the string field theory framework \cite{emms22a, emms22b}---where this divergence was resolved, and a finite result for the annulus amplitude from the worldsheet perspective was obtained which perfectly matches the corresponding matrix model calculations in, \textit{e.g.}, \cite{d92, akk03, hhikkmt04, st04, iy05, iky05, msw07, msw08}. It reads \cite{emms22b} 
\begin{equation}
\label{eq:convergentZZAnnulus}
\mathsf{A}_{\text{A}} (m,n;m,n) = 2\, \log \sqrt{\frac{g_{\text{s}}}{8\pi \mathsf{A}_{\text{D}} (m,n)}\, \frac{\cot^{2} \left( \frac{\pi n}{q} \right) - \cot^{2} \left( \frac{\pi m}{p} \right)}{q^2-p^2}}.
\end{equation}
\noindent
This regularized result was obtained by reconsidering the integration over the annulus modulus, $t$, of the ZZ annulus. This (or rather its exponentiation) was given in \cite{emms22b} as 
\begin{equation}
\label{eq:string-field-theory-regularization}
\mathsf{A}_{\text{A}}(m,n;m,n) = \int_{0}^{+\infty} \frac{\text{d}t}{t}\, \mathsf{Z}_{p,q}\, \mathsf{Z}_{\text{gh}}\, \mathsf{Z}_{\text{L}} = \int_{0}^{+\infty} \frac{\text{d}t}{t}\, \sum_{\ell=1}^{n}\sum_{k=1}^{m} F_{2k-1,2\ell-1}(t),
\end{equation}
\noindent
where the authors identified the $(\ell,k)=(1,1)$ contribution as problematic: expanding $F_{1,1}(t)$ as an infinite sum it was observed that the zero-mode in the matter character produces a term $\sim \rme^{2\pi t}-2$ which is leading and diverges as $t \to +\infty$. In string field theory language, this breakdown of the integration is understood as an invalid use of Siegel gauge for the zero-modes. Rewriting those terms using the full path-integral and undoing  Siegel gauge then removes this divergence. We refer the interested reader to \cite{emms22a, emms22b} for complete details on this calculation.

In the following we will introduce\footnote{A very similar calculation to the one that follows was obtained from purely string field theoretic considerations in \cite{fim06}, and which further followed upon earlier work in \cite{fy96}.} a different---purely BCFT---approach to obtaining the correct \textit{finite} answer to the annulus amplitude, in a setting which will turn out to be very natural given the connection to matrix models in \cite{mss22} and in section~\ref{sec:NPminimalstring}. This way of performing the calculation arises due to the relation between ZZ and FZZT boundary states \cite{m03}, and thus we first take a short detour to recast the above ZZ discussion in terms of FZZT amplitudes.

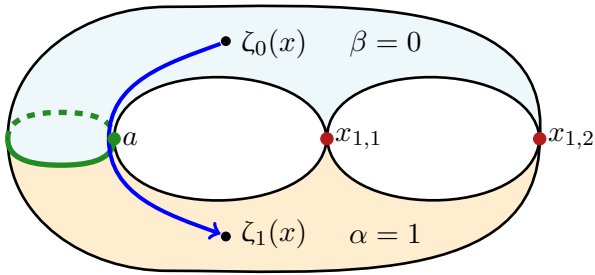
\begin{figure}
\centering
	\begin{tikzpicture}
	\begin{scope}[scale=0.7,  shift={({-5.8},{0})}]
	\draw[fill=LightBlue,fill opacity=0.2, line width=1pt] (0,0)   to [out=90,in=95] (4,0)
	to [out=90,in=95] (8,0)
	to [out=85,in=0] (1,2.5)
    to [out=180,in=90] (-2,0)
    to [out=270, in=180] (-1, -0.5)
    to [out=0, in=270] cycle;
    \draw[fill=darktangerine,fill opacity=0.2, line width=1pt] (-2,0)
    to [out=270,in=180] (1,-2.5)
    to [out=0,in=270] (8,0)
    to [out=265,in=275] (4,0)
    to [out=265,in=270] (0,0)
    to [out=270, in=0] (-1, -0.5)
    to [out=180, in=270] cycle;
    \draw[color=ForestGreen, line width=2pt] (-2,0) to [out=270, in=180] (-1, -0.5)
    to [out=0, in=270] (0,0);
    \draw[dashed, color=ForestGreen, line width=2pt] (-2,0) to [out=90, in=180] (-1, 0.5)
    to [out=0, in=90] (0,0);
    \draw[blue, line width=1.5pt] (-0.1,0) to [out=90, in=200] (2, 1.8);
    \draw[blue, line width=1.5pt, ->] (-0.1,0) to [out=270, in=160] (2, -1.8);
\draw[fill=black] (2.1,1.85) circle (.5ex);
\draw[fill=black] (2.1,-1.85) circle (.5ex);
\node at (3, 1.8) {$\upzeta_{0}(x)$};
\node at (3, -1.8) {$\upzeta_{1}(x)$};
\draw[ForestGreen, fill=ForestGreen] (0,0) circle (.7ex);
\draw[cornellred, fill=cornellred] (4,0) circle (.7ex);
\draw[cornellred, fill=cornellred] (8,0) circle (.7ex); 
\node at (0.3, 0) {$a$};
\node at (4.6, 0) {$x_{1, 1}$}; 
\node at (8.6, 0) {$x_{1, 2}$};
\node at (5.1,1.8) {$\beta=0$}; 
\node at (5.1,-1.8) {$\alpha=1$}; 
\end{scope}
	\end{tikzpicture}
\caption{Visualization of the disk contribution $\CA_{\alpha \beta}^{[-1]}(x)$ in \eqref{eq:1instFZZTcontribution} as a {\color{blue}contour integral} over the FZZT Riemann surface. Here we illustrate again the $(2,5)$ example as in the previous figure~\ref{fig:minimalstringspectralcurve-start}.}
\label{fig:minimalstringspectralcurve-difference-disks}
\end{figure}

Recall \eqref{eq:ZZ-from-FZZTs}. If we further view ZZ annulus amplitudes as differences of FZZT annulus amplitudes on the pinches of the Riemann surface $\Sigma_{p,q}$, then we may obtain the former via saddle-point integrations over this FZZT moduli space around a pinch $x_{mn}$. Note how this space is parameterized by $x=\frac{\mu_{\text{B}}}{\sqrt{\mu}}$, but at the same time the choice of sheet changes the nature of the FZZT-brane under consideration---in fact, as we shall extensively explore below. In other words, the analogue of \eqref{eq:1instZZcontribution} but in the FZZT language is\footnote{The ratio-of-partition-functions notation is now in full analogy with the nonperturbative contributions  calculated in \cite{msw07, mss22}, where similar ratios were computed from string-equation transseries and from matrix integrals. All these matched each-other, as they will now further match against \eqref{eq:1instFZZTcontribution} in the present paper.} 
\begin{align}
\label{eq:1instFZZTcontribution}
\frac{\NCZ_{\alpha\beta}}{\NCZ_{\text{pert}}} \simeq \frac{1}{2\pi} \int_{\CC_{mn}} \text{d}x\, \exp \Bigg( \CA_{\alpha \beta}^{[-1]}(x) + \CA^{[0]}_{\alpha\beta}(x) + \cdots \Bigg).
\end{align}
\noindent
Herein $\alpha$ and $\beta$ label the different sheets, as in \eqref{eq:sheet-labels}, and the superscripts in the disk and annulus amplitudes label their corresponding $g_{\text{s}}$ powers. In addition, the contour $\CC_{mn}$ is the steepest-descent contour associated to the saddle $x_{mn}$. These amplitudes have matrix model analogues which were discussed in \cite{msw07, mss22}. For $\alpha \neq \beta$ we define the difference of FZZT disk amplitudes as
\begin{equation}
\label{eq:differences-FZZT-disks}
\CA_{\alpha \beta}^{[-1]}(x) \equiv \mathsf{A}_{\text{D}} \left(\upzeta_{\alpha}(x)\right) -\mathsf{A}_{\text{D}} \left(\upzeta_{\beta}(x)\right) = \DiscFZZT{\upzeta_{\alpha}} -\, \DiscFZZT{\upzeta_{\beta}}.
\end{equation}
\noindent
Using formula \eqref{eq:FZZTdisc} we can give a nice interpretation to the above expression as an integration over the FZZT Riemann surface starting at $\upzeta_{\beta}$ and ending at $\upzeta_{\alpha}$. This is visualized in figure~\ref{fig:minimalstringspectralcurve-difference-disks}. In addition we need the linear combination of FZZT annulus amplitudes (with all possible FZZT annuli, this is the particular combination whose saddles will reproduce the ZZ annulus \eqref{eq:divergentZZAnnulus} on the pinches of the spectral curve\footnote{In other words, via \eqref{eq:differences-FZZT-disks} and \eqref{eq:ZZ-from-FZZTs}, one very schematically obtains
\bea
{}_{\text{ZZ}}\bra{\alpha,\beta} \ket{\alpha,\beta}_{\text{ZZ}} &=& \left( {}_{\text{FZZT}} \bra{\alpha} - {}_{\text{FZZT}} \bra{\beta} \right) \left( \ket{\alpha}_{\text{FZZT}} - \ket{\beta}_{\text{FZZT}} \right) = \\
&=& {}_{\text{FZZT}} \bra{\alpha} \ket{\alpha}_{\text{FZZT}} - {}_{\text{FZZT}} \bra{\alpha} \ket{\beta}_{\text{FZZT}} - {}_{\text{FZZT}} \bra{\beta} \ket{\alpha}_{\text{FZZT}} + {}_{\text{FZZT}} \bra{\beta} \ket{\beta}_{\text{FZZT}}. \nonumber
\eea
})
\begin{equation}
\label{eq:FZZT-Annulus-Combination-1-inst}
\CA_{\alpha \beta}^{[0]}(x) = \frac{1}{2} \left( \AnnulusFZZT{\upzeta_{\alpha}}{\upzeta_{\alpha}} -\, \AnnulusFZZT{\upzeta_{\alpha}}{\upzeta_{\beta}} -\, \AnnulusFZZT{\upzeta_{\beta}}{\upzeta_{\alpha}} +\, \AnnulusFZZT{\upzeta_{\beta}}{\upzeta_{\beta}} \right).
\end{equation}
\noindent
The above global factor of $1/2$ comes from the cumulant expansion. In addition, the $1/{2\pi}$ prefactor in \eqref{eq:1instFZZTcontribution} originates from the $\text{U}(1)$ symmetry of the FZZT boundaries (see as well \cite{msy21}). There are essentially two distinct types of amplitudes appearing in the above expression:
\begin{enumerate}
\item The first of these has both boundary parameters living on the same sheet, $\alpha$, is regular, and is given via \eqref{eq:StandardFZZTAnnulus} by \cite{kopss04}: 
\begin{equation}
\label{eq:FZZT-annulus-same-sheet}
\mathsf{A}_{\text{A}} (\upzeta_{\alpha}, \upzeta_{\alpha}) = \AnnulusFZZT{\upzeta_{\alpha}}{\upzeta_{\alpha}} = \left. \log \left( \frac{\upzeta_{\alpha}-\tilde{\upzeta}_{\alpha}}{x (\upzeta_{\alpha})-x(\tilde{\upzeta}_{\alpha})}\right) \right|_{\tilde{\upzeta}_{\alpha}\to \upzeta_{\alpha}} = - \log \left( p\, U_{p-1}(\upzeta_{\alpha}) \right).
\end{equation}
\noindent
Herein $U_{p-1} \left( \cos \theta \right) \sin \theta = \sin p \theta$ are Chebyshev polynomials of the second kind.
\item The second type of annulus amplitudes sees its boundaries living on different sheets, labeled by $\upzeta_{\alpha}$ and $\upzeta_{\beta}$, but projecting to the same physical location $x$ via the uniformization map \eqref{eq:uniformization-minimal-string-x}-\eqref{eq:uniformization-minimal-string-y}. Such amplitude is divergent \cite{kopss04}. This is the divergence already alluded to---given by \eqref{eq:string-field-theory-regularization} prior to treating the zero-mode divergence---and which was addressed for ZZ-branes in \cite{emms22a}. This can be seen as follows: recall that the divergence of the ZZ annulus in \eqref{eq:string-field-theory-regularization} originates in $F_{1,1}(t)$ behaving like $\sim \rme^{2\pi t} - 2$; which exactly maps to the FZZT divergence discussed herein when spelling out $\CA_{\alpha\beta}^{[0]}$ before the $t$ integration,
\begin{equation}
\CA_{\alpha \beta}^{[0]} = \int_{0}^{+\infty}\frac{\text{d}t}{2t}\, \mathsf{Z}_{p,q}\, \mathsf{Z}_{\text{gh}}\, \Big\{ \mathsf{Z}_{\text{L}}(\upzeta_{\alpha}, \upzeta_{\alpha}) - \mathsf{Z}_{\text{L}}(\upzeta_{\alpha}, \upzeta_{\beta}) - \mathsf{Z}_{\text{L}}(\upzeta_{\beta}, \upzeta_{\alpha}) + \mathsf{Z}_{\text{L}}(\upzeta_{\beta}, \upzeta_{\beta}) \Big\}.
\end{equation}
\noindent
Here $\mathsf{Z}_{\text{L}}$ is the annulus Liouville partition function, $\mathsf{Z}_{p,q}$ the minimal-matter contribution, and $\mathsf{Z}_{\text{gh}}$ captures the ghosts. Around the ZZ singular points it is well known how to evaluate the above and indeed we find back the leading expression $\sim \rme^{2\pi t} - 2$, regularized in \cite{emms22a}. 

We propose that, on the FZZT-brane side, the required regularization comes down to:
\begin{equation}
\label{eq:RegularizedFZZTAnnulus}
\mathsf{A}_{\text{A}} (\upzeta_{\alpha}, \upzeta_{\beta}) = \AnnulusFZZT{\upzeta_{\alpha}}{\upzeta_{\beta}} = \log \left( \upzeta_{\alpha}-\upzeta_{\beta} \right).
\end{equation}
\noindent
This expression for the FZZT annulus has its history: it is discussed at length in the original computation of the annulus \cite{kopss04}. An additional exposition motivated by matrix model calculations is provided in \cite{mmss04}, and a detailed string field theory analysis in \cite{fim06}.
\end{enumerate}

Having established a (alternative) regularization procedure let us check it by testing if our purely BCFT calculation matches the string field theory considerations in \cite{emms22a, emms22b}. Combining all of the results above, we find for the contribution of the annuli
\begin{equation}
\label{eq:1-inst-FZZT-annulus-contribution}
\CA^{[0]}_{\alpha \beta}(x) = -\frac{1}{2} \log \left( -p^2\,U_{p-1}(\upzeta_{\alpha})\, U_{p-1}(\upzeta_{\beta}) \left(\upzeta_{\alpha}-\upzeta_{\beta}\right)^2 \right).
\end{equation}
\noindent
Further recall that the FZZT disk amplitudes contain a factor of $\mu^{\frac{p+q}{2p}}$, which we associate to $g_{\text{s}}^{-1}$. Next we have to evaluate the integral \eqref{eq:1instFZZTcontribution}, which we do using saddle-point integration. Let us perform this computation explicitly as it will clarify the relation between the sheets labeled by $\alpha, \beta$ and the saddle-point labeled by $m,n$. The saddles of this integral are given by the zeros of $\partial_x \CA^{[-1]}_{\alpha\beta}$, which from \eqref{eq:FZZTdisc} may be expressed as
\begin{align}
y(\upzeta_{\alpha})-y(\upzeta_{\beta})=0.
\end{align}
\noindent
In terms of the uniformization variable $\upzeta$ this is solved by
\begin{equation}
\label{eq:saddle-zeta}
\upzeta^{\star} = \cos\left(\pi\left(-\frac{\alpha+\beta}{p}+\frac{n}{q}\right)\right), \qquad n\in\lbrace1,\ldots,q-1\rbrace,
\end{equation}
\noindent
or, explicitly evaluated on the different sheets we have \cite{fim06} (recall \eqref{eq:xmn,ymn,upzetamn}),
\begin{align}
\label{eq:saddle-zeta-alpha}
\upzeta_{\alpha}^{\star} &= \cos\left(\pi\left(\frac{\beta-\alpha}{p}-\frac{n}{q}\right)\right), \\
\label{eq:saddle-zeta-beta}
\upzeta_{\beta}^{\star} &= \cos\left(\pi\left(\frac{\beta-\alpha}{p}+\frac{n}{q}\right)\right).
\end{align}
\noindent
In particular the identification with \eqref{eq:xmn,ymn,upzetamn} comes from noting that the saddle-labeling (which does \textit{not} distinguish sheets) is given by $m=|\beta-\alpha|$. In this way, one can identify $\zeta^{\star}_{\alpha,\beta}$ with $\zeta^{\pm}_{m n}$ depending on the ordering\footnote{For example, for $\beta > \alpha$ these coincide with $\zeta^{-}_{mn}$, $\zeta^{+}_{mn}$, while for $\beta < \alpha$ they coincide with $\zeta^{+}_{mn}$, $\zeta^{-}_{mn}$.} of $\alpha,\beta$. Moreover, upon swapping $\alpha,\beta$ one effectively swaps $\zeta^{\pm}_{m n}$ with  $\zeta^{\mp}_{m n}$, as would be expected. Notice how the sheets $\alpha,\beta$ interplay with the label $m$ but not directly with the label $n$ of the \textit{a priori} chosen saddle-point $x_{mn}$ in this identification. This calculation again underlines how we may view the ZZ disk amplitudes (labeled by $(m,n)$) as disk ``FZZT difference'' amplitudes (labeled by $(\alpha,\beta)$) on the saddles of the integral \eqref{eq:1instFZZTcontribution}. The information about the sheets that touch at $x_{mn}=x(\upzeta^{\pm}_{mn})=x(\upzeta^{\star})$ is completely encoded in $\mathcal{A}_{\alpha\beta}^{[-1]}(x)$ (see figures~\ref{fig:minimalstringspectralcurve-start}, \ref{fig:minimalstringspectralcurve-difference-disks} and~\ref{fig:minimalstringspectralcurve-resonant-cycles}). Recalling that $\CA^{[-1]}_{\alpha \beta}(x)$ contains a factor of $1/g_{\text{s}}$, we may finally perform the saddle-point integration to find
\begin{align}
\frac{\NCZ_{\alpha \beta}}{\NCZ_{\text{pert}}} &\simeq \frac{1}{2\pi} \int_{\CC_{mn}} \text{d}x\, \exp \Bigg( \CA^{[-1]}_{\alpha \beta}(x) + \CA^{[0]}_{\alpha \beta}(x) + \cdots \Bigg) = \nonumber \\
&\simeq \frac{1}{\sqrt{-2\pi\, \partial_x^2\CA^{[-1]}_{\alpha \beta}(x_{mn})}}\, \rme^{\CA^{[-1]}_{\alpha \beta}(x_{mn})}\, \exp\left( \CA^{[0]}_{\alpha \beta}(x_{mn})\right) + \cdots.
\label{eq:ZabZpertratio}
\end{align}
\noindent
Making explicit use of the result in \eqref{eq:1-inst-FZZT-annulus-contribution} for the differences of FZZT annuli, and of the ZZ disk amplitude \eqref{eq:ZZdisc}, we find 
\begin{equation}
\frac{\NCZ_{\alpha \beta}}{\NCZ_{\text{pert}}} \simeq \frac{1}{\sqrt{-2\pi\, \partial_x^2\CA^{[-1]}_{\alpha \beta}(x_{mn})}}\, \exp \left\{ -\frac{1}{2} \log \left( -p^2\,U_{p-1}(\upzeta^{\star}_{\alpha})\, U_{p-1}(\upzeta^{\star}_{\beta}) \left(\upzeta^{\star}_{\alpha}-\upzeta^{\star}_{\beta}\right)^2 \right) \right\} \rme^{\mathsf{A}_{\text{D}}(m,n)} + \cdots.
\end{equation}
\noindent
To make the comparison against \eqref{eq:convergentZZAnnulus} it remains to express the second derivative of the disk amplitude in terms of the disk itself, which is evaluated on the saddle to be 
\noindent
\begin{equation}
\partial_x^2\CA^{[-1]}_{\alpha \beta}(x_{mn}) = \frac{p^2-q^2}{p^2 \sin^2 \left(\frac{n\pi p}{q} \right)}\, \CA^{[-1]}_{\alpha \beta}(x_{mn}).
\end{equation} 
\noindent
Substitution into the above result and evaluating everything on the pinches $\upzeta_{\alpha}, \upzeta_{\beta}$ we obtain 
\begin{equation}
\label{eq:1-instanton-result}
\frac{\NCZ_{\alpha \beta}}{\NCZ_{\text{pert}}} \simeq \frac{1}{\sqrt{-\mathsf{A}_{\text{D}}(m,n)}}\, \sqrt{\frac{\cos \left(\frac{2\pi n}{q}\right) - \cos \left(\frac{2\pi m}{p}\right)}{16\pi \left(p^2-q^2\right) \sin^2 \left(\frac{\pi m}{p}\right) \sin^2 \left(\frac{\pi n}{q}\right)}}\, \rme^{\mathsf{A}_{\text{D}}(m,n)} + \cdots,
\end{equation}
\noindent
precisely matching the result in \eqref{eq:convergentZZAnnulus}. To see this in an concrete example, let us specialize the above to the $(p,q)=(2,2k-1)$ case and be fully explicit (this will also be useful later, when doing comparisons in section~\ref{sec:NPminimalstring}). Noting that for these cases $m=1$, we first explicitly evaluate the appropriate differences of \eqref{eq:FZZT-disk} on the pinches to find the known
\begin{equation}
\label{eq:22km1-bcft-Instanton-Action}
\mathsf{A}_{\text{D}}(1,n) = \frac{2}{g_{\text{s}}}\, (-1)^{k+n} \left(\frac{1}{2k+1}+\frac{1}{2k-3}\right) \sin\frac{2\pi n}{2k-1}.
\end{equation}
\noindent
Substituting this into \eqref{eq:1-instanton-result}, it then becomes 
\begin{equation}
\label{eq:22km1-1-instanton-result}
\left. \frac{\NCZ_{\alpha \beta}}{\NCZ_{\text{pert}}}\right|_{(2,2k-1)} \simeq \sqrt{g_{\text{s}}\, \frac{(-1)^{k+n} \cot \left(\frac{\pi n}{2k-1}\right)}{64\pi \left(2k-1\right) \sin ^2\left(\frac{\pi n}{2k-1}\right)}}\, \rme^{\mathsf{A}_{\text{D}}(1,n)} + \cdots.
\end{equation}

To conclude this subsection let us emphasize again that the above calculation was a purely BCFT calculation. Moreover, the FZZT annulus regularization \eqref{eq:RegularizedFZZTAnnulus}, which had already been discussed in the literature, is now further supported by the demonstration that it is in fact required so as to match to the corresponding string field theory ZZ regularization in \eqref{eq:convergentZZAnnulus}.

\subsection{Resonant Resurgence from the Worldsheet}\label{subsec:bcft-resonant-pairs}

\begin{figure}
\centering
	\begin{tikzpicture}
	\begin{scope}[scale=0.7,  shift={({-5.8},{0})}]
	\draw[fill=LightBlue,fill opacity=0.2, line width=1pt] (0,0)   to [out=90,in=95] (4,0)
	to [out=90,in=95] (8,0)
	to [out=85,in=0] (1,2.5)
    to [out=180,in=90] (-2,0)
    to [out=270, in=180] (-1, -0.5)
    to [out=0, in=270] cycle;
    \draw[fill=darktangerine,fill opacity=0.2, line width=1pt] (-2,0)
    to [out=270,in=180] (1,-2.5)
    to [out=0,in=270] (8,0)
    to [out=265,in=275] (4,0)
    to [out=265,in=270] (0,0)
    to [out=270, in=0] (-1, -0.5)
    to [out=180, in=270] cycle;
    \draw[color=ForestGreen, line width=2pt] (-2,0) to [out=270, in=180] (-1, -0.5)
    to [out=0, in=270] (0,0);
    \draw[dashed, color=ForestGreen, line width=2pt] (-2,0) to [out=90, in=180] (-1, 0.5)
    to [out=0, in=90] (0,0);
    \draw[blue, line width=1.5pt] (0,0) to [out=90, in=180] (1.8, 1.6);
    \draw[blue, line width=1.5pt] (1.8, 1.6) to [out=0, in=90] (4, 0);
    \draw[blue, line width=1.5pt, ->] (0,0) to [out=270, in=180] (1.8, -1.6);
    \draw[blue, line width=1.5pt] (1.8, -1.6) to [out=0, in=270] (4, 0);
\draw[ForestGreen, fill=ForestGreen] (0,0) circle (.7ex);
\draw[cornellred, fill=cornellred] (4,0) circle (.7ex);
\draw[cornellred, fill=cornellred] (8,0) circle (.7ex); 
\node at (0.3, 0) {$a$};
\node at (4.6, 0) {$x_{1, 1}$}; 
\node at (8.6, 0) {$x_{1, 2}$};
\node at (3.5, 1.8) {$\beta=0$}; 
\node at (3.5, -1.8) {$\alpha=1$};  
\end{scope}
\begin{scope}[scale=0.7,  shift={({5.8},{0})}]
	\draw[fill=LightBlue,fill opacity=0.2, line width=1pt] (0,0)   to [out=90,in=95] (4,0)
	to [out=90,in=95] (8,0)
	to [out=85,in=0] (1,2.5)
    to [out=180,in=90] (-2,0)
    to [out=270, in=180] (-1, -0.5)
    to [out=0, in=270] cycle;
    \draw[fill=darktangerine,fill opacity=0.2, line width=1pt] (-2,0)
    to [out=270,in=180] (1,-2.5)
    to [out=0,in=270] (8,0)
    to [out=265,in=275] (4,0)
    to [out=265,in=270] (0,0)
    to [out=270, in=0] (-1, -0.5)
    to [out=180, in=270] cycle;
    \draw[color=ForestGreen, line width=2pt] (-2,0) to [out=270, in=180] (-1, -0.5)
    to [out=0, in=270] (0,0);
    \draw[dashed, color=ForestGreen, line width=2pt] (-2,0) to [out=90, in=180] (-1, 0.5)
    to [out=0, in=90] (0,0);
    \draw[orange, line width=1.5pt,->] (0,0) to [out=90, in=180] (1.8, 1.6);
    \draw[orange, line width=1.5pt] (1.8, 1.6) to [out=0, in=90] (4, 0);
    \draw[orange, line width=1.5pt] (0,0) to [out=270, in=180] (1.8, -1.6);
    \draw[orange, line width=1.5pt] (1.8, -1.6) to [out=0, in=270] (4, 0);
\draw[ForestGreen, fill=ForestGreen] (0,0) circle (.7ex);
\draw[cornellred, fill=cornellred] (4,0) circle (.7ex);
\draw[cornellred, fill=cornellred] (8,0) circle (.7ex); 
\node at (0.3, 0) {$a$};
\node at (4.6, 0) {$x_{1, 1}$}; 
\node at (8.6, 0) {$x_{1, 2}$}; 
\node at (3.5, -1.8) {$\alpha=1$}; 
\node at (3.5, 1.8) {$\beta=0$}; 
\end{scope}
	\end{tikzpicture}
\caption{The $(2,5)$ FZZT surface revisited. On the left-plot we show the {\color{blue}integration cycle} associated to the configuration $\NCZ_{10}$ (this is \eqref{eq:1instFZZTcontribution} with $\alpha=1$, $\beta=0$, which culminates in \eqref{eq:1-instanton-result}). On the right-plot we show the exactly {\color{orange}switched configuration} $\NCZ_{01}$ (again \eqref{eq:1instFZZTcontribution}, this time around with $\alpha=0$, $\beta=1$, or, equivalently, as visualized above, \eqref{eq:negative-1-instanton-result} with $\alpha=1$, $\beta=0$).}
\label{fig:minimalstringspectralcurve-resonant-cycles}
\end{figure}
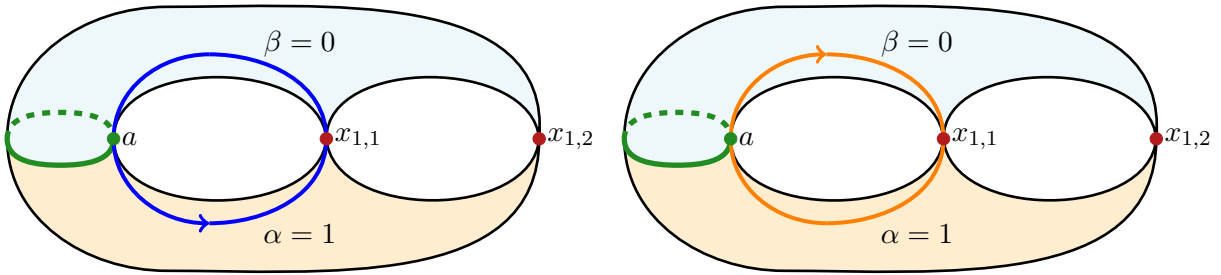

Having established the one-instanton calculation purely from BCFT, let us investigate it more closely in light of the resonant (anti) eigenvalue pairs recently proposed in \cite{mss22}. Indeed considering the general expression \eqref{eq:1instFZZTcontribution} and swapping sheets (meaning exchanging $\alpha$ and $\beta$) will produce a minus sign in the disk contribution---immediately evident in formula\footnote{Note that exchanging $\alpha$ and $\beta$ basically exchanges $\upzeta^{+}_{mn}$ with $\upzeta^{-}_{mn}$; see formulae \eqref{eq:saddle-zeta-alpha}-\eqref{eq:saddle-zeta-beta}.} \eqref{eq:differences-FZZT-disks}. This is consistent with reversing the integration direction of the $B_{mn}$ $B$-cycles which correspond with each saddle-point $x_{nm}$ of the minimal-string FZZT Riemann surface \eqref{eq:uniformization-minimal-string-x}-\eqref{eq:uniformization-minimal-string-y} (again visualized in figure~\ref{fig:minimalstringspectralcurve-resonant-cycles} in our usual example). Therefore, the above immediately translates to the result of the saddle-point approximation \eqref{eq:1-instanton-result} as $\mathsf{A}_{\text{D}}(m,n) \to - \mathsf{A}_{\text{D}}(m,n)$. This is shown in figure \ref{fig:minimalstringspectralcurve-resonant-cycles}. We thus find 
\begin{equation}
\label{eq:negative-1-instanton-result}
\frac{\NCZ_{\beta \alpha}}{\NCZ_{\text{pert}}} \simeq \frac{1}{\sqrt{\mathsf{A}_{\text{D}}(m,n)}}\, \sqrt{\frac{\cos \left(\frac{2\pi n}{q}\right) - \cos \left(\frac{2\pi m}{p}\right)}{16\pi \left(p^2-q^2\right) \sin^2 \left(\frac{\pi m}{p}\right) \sin^2 \left(\frac{\pi n}{q}\right)}}\, \rme^{-\mathsf{A}_{\text{D}}(m,n)} + \cdots,
\end{equation}
\noindent
which indeed yields the resonant sibling of \eqref{eq:1-instanton-result}. This is consistent with the results of \cite{mss22} and we can hence be explicit about the resonant pair of D-instanton actions associated to a saddle $x_{mn}$, given by
\begin{equation}
\label{eq:Adand-Ad}
\mathsf{A}_{\text{D}}(m,n) \quad \text{ and } \quad -\mathsf{A}_{\text{D}}(m,n).
\end{equation}
\noindent
This establishes the parallel that while eigenvalues correspond to ZZ-branes, anti-eigenvalues now correspond to (pairwise) negative-tension ZZ-branes fully explicitly\footnote{\label{footnote:ExplainingMinusSign1}Let us stress an important point on conventions. As mentioned in appendix~B of \cite{kopss04}, there is a relative minus sign between the ZZ disk-amplitudes on the BCFT side, and the holomorphic effective-potential on the matrix-model side. This difference implies a reversal of the direction of integration between the definition of ZZ branes on the BCFT side, and the definition of eigenvalue tunneling on the matrix model side. In other words, there will be a reversal of directions between the cycles drawn throughout our present section~\ref{sec:ZZminimalstring}, describing the usual $\exp(\mathsf{A}_{\text{D}}(m,n) )$ ZZ branes in BCFT language and starting at the physical sheet and ending at the unphysical sheet; and the cycles implicit throughout the upcoming section~\ref{sec:NPminimalstring} (explicitly shown in \cite{mss22}), describing the usual instanton contributions as eigenvalue tunneling from the cut to the physical sheet on the matrix model side.}. Let us explore this next. 

Just to finish this discussion we specialize to the case $(2,2k-1)$ and insert \eqref{eq:22km1-bcft-Instanton-Action} to find
\begin{equation}
\label{eq:22km1-negative-1-instanton-result}
\left.\frac{\NCZ_{\beta \alpha}}{\NCZ_{\text{pert}}} \right|_{(2,2k-1)} \simeq \sqrt{g_{\text{s}}\, \frac{(-1)^{k+n+1} \cot \left(\frac{\pi n}{2k-1}\right)}{64\pi \left(2 k-1\right) \sin^2\left(\frac{\pi n}{2k-1}\right)}}\, \rme^{-\mathsf{A}_{\text{D}}(1,n)} + \cdots.
\end{equation}

\subsection{On Boundary CFT for Generic ZZ-Instanton Contributions}\label{subsec:generic-ZZ-BCFT}

Having established a BCFT formulation of the one D-instanton contribution in our previous subsection~\ref{subsec:regularization-one-instanton}, we can now ask what a general two D-instanton calculation might look like. Interestingly, we can capture the general two D-instanton amplitude in a single double-integral, which, depending on the choice of sheets for the disks and annuli and the saddle-points we integrate around, will produce different nonperturbative contributions. As we shall later see in section~\ref{sec:NPminimalstring} these contributions \textit{precisely match} the matrix model expectations. In the following we will view the $\alpha\beta$ sheet-labels as belonging to the FZZT labeled by $x$ (integrated around the saddle $x_{mn}$) whereas $\gamma\delta$ label the $\tilde{x}$-FZZT (integrated around the saddle $x_{\tilde{m}\tilde{n}}$). This is visualized throughout figures~\ref{fig:minimalstringspectralcurve-difference-disks}, \ref{fig:minimalstringspectralcurve-resonant-cycles}, \ref{fig:minimalstringspectralcurve-20-contours} and~\ref{fig:minimalstringspectralcurve-11-contours}. The integral to consider reads
\begin{equation}
\label{eq:general-double-instanton-configuration}
\frac{\NCZ_{\alpha \beta, \gamma \delta}}{\NCZ_{\text{pert}}} \simeq \int_{\CC_{mn}}\frac{\text{d}x}{2\pi} \int_{\CC_{\tilde{m}\tilde{n}}}\frac{\text{d}\tilde{x}}{2\pi}\, \exp \Bigg( \CA^{[-1]}_{\alpha \beta}(x) + \CA^{[-1]}_{\gamma \delta}(\tilde{x}) + \CA^{[0]}_{\alpha \beta}(x) + \CA^{[0]}_{\gamma \delta}(\tilde{x}) + \CA^{[0]}_{\alpha \beta, \gamma \delta}(x,\tilde{x}) + \cdots \Bigg),
\end{equation}
\noindent
where, in addition to the definitions in subsection~\ref{subsec:regularization-one-instanton}, we have introduced the following combination of FZZT annulus amplitudes (compare to formula \eqref{eq:FZZT-Annulus-Combination-1-inst}---again, with all possible FZZT annuli, this is the particular combination whose saddles will reproduce the ZZ annulus \eqref{eq:divergentZZAnnulus} on the pinches of the spectral curve)
\begin{equation}
\label{eq:FZZTAnnulusDifferences}
\CA^{[0]}_{\alpha \beta, \gamma \delta}(x, \tilde{x}) = \AnnulusFZZT{\upzeta_{\alpha}}{\tilde{\upzeta}_{\gamma}} -\, \AnnulusFZZT{\upzeta_{\alpha}}{\tilde{\upzeta}_{\delta}} -\, \AnnulusFZZT{\upzeta_{\beta}}{\tilde{\upzeta}_{\gamma}} +\, \AnnulusFZZT{\upzeta_{\beta}}{\tilde{\upzeta}_{\delta}}.
\end{equation}
\noindent
As mentioned above, depending on the choice of sheets in the form of $\alpha, \beta, \gamma, \delta$ and the saddle-points $x_{mn}$ and $x_{\tilde{m}\tilde{n}}$, we will produce different nonperturbative contributions. We may split them into the following three cases:
\begin{enumerate}
\item \underline{$(\tilde{m},\tilde{n})=(m,n)$, $\gamma=\alpha, \delta=\beta, \alpha \neq \beta$:} This corresponds to a $(2,0)$ configuration for the nonperturbative saddle $x_{mn}$, \textit{i.e.}, both FZZT branes localize around the same saddle (with the same sheet configuration) and one ends-up with two ZZ branes and no negative-tension ZZ branes---which also explains the $(2,0)$ notation. This is visualized in figure~\ref{fig:minimalstringspectralcurve-20-contours}. Let us evaluate the integral \eqref{eq:general-double-instanton-configuration} under the above assumptions. It is useful to first rewrite \eqref{eq:FZZTAnnulusDifferences} for $\CA^{[0]}_{\alpha \beta,\alpha \beta}(x, \tilde{x})$ as
\begin{align}
\label{eq:split-regular-term-vandermonde}
\CA^{[0]}_{\alpha \beta,\alpha \beta} (x, \tilde{x}) = \widehat{\CA}^{[0]}_{\alpha \beta,\alpha \beta}(x, \tilde{x}) + \log \left\{ \left( x(\upzeta_{\alpha})-x(\tilde{\upzeta}_{\alpha})\right) \left( x(\upzeta_{\beta})-x(\tilde{\upzeta}_{\beta}) \right) \right\},
\end{align}
\noindent
where we have introduced the ``regular'' difference of FZZT annulus amplitudes which, using 
\begin{equation}
\frac{x(\upzeta_{\alpha})-x(\tilde{\upzeta}_{\alpha})}{\upzeta_{\alpha}-\tilde{\upzeta}_{\alpha}} = p\, U_{p-1}(\tilde{\upzeta}_{\alpha}) + \frac{1}{2} x^{\prime\prime}(\tilde{\upzeta}_{\alpha}) \left(\upzeta_{\alpha}-\tilde{\upzeta}_{\alpha}\right) + \cdots
\end{equation}
\noindent
evaluated around $\upzeta=\tilde{\upzeta}$, reads:
\begin{align}
\widehat{\CA}^{[0]}_{\alpha \beta,\alpha \beta} (x, \tilde{x}) = -\log \left( -p^2\,U_{p-1}(\tilde{\upzeta}_{\alpha})\, U_{p-1}(\tilde{\upzeta}_{\beta}) \left(\tilde{\upzeta}_{\alpha}-\tilde{\upzeta}_{\beta}\right)^2 \right) + o (\upzeta-\tilde{\upzeta}).
\end{align}
\noindent
Further, note how the logarithmic term in \eqref{eq:split-regular-term-vandermonde} is effectively producing a matrix-model-like Vandermonde determinant in the integrand of \eqref{eq:general-double-instanton-configuration}. In fact, the integral for the $(2,0)$ configuration may now be written as\footnote{Since we wish to integrate over the moduli space of FZZT-branes exactly once, and given that both FZZT-branes are indistinguishable, we have to insert an additional global prefactor of $\frac{1}{2}$.}
\begin{align}
\label{eq:20-contribution-minimal-strings}
\frac{\NCZ_{\alpha \beta, \alpha \beta}}{\NCZ_{\text{pert}}} &\simeq \frac{1}{2} \int_{\CC_{mn}} \frac{\text{d}x}{2\pi} \int_{\CC_{mn}} \frac{\text{d}\tilde{x}}{2\pi} \left(x-\tilde{x}\right)^2 \times \\
&\times \exp \Bigg( \CA^{[-1]}_{\alpha \beta}(x) + \CA^{[-1]}_{\alpha \beta}(\tilde{x}) + \CA^{[0]}_{\alpha \beta}(x) + \CA^{[0]}_{\alpha \beta}(\tilde{x}) + \widehat{\CA}^{[0]}_{\alpha \beta, \alpha \beta}(x, \tilde{x}) + o(g_{\text{s}}) \Bigg). \nonumber
\end{align}
\noindent
Performing the saddle-point evaluation in analogy to the computation in subsection~\ref{subsec:regularization-one-instanton} we immediately arrive at\footnote{Note how the integration over $x,\tilde{x}$, together with the Vandermonde term in its integrand, produces a square in the denominator when compared with \eqref{eq:ZabZpertratio}.}
\begin{align}
\frac{\NCZ_{\alpha \beta, \alpha \beta}}{\NCZ_{\text{pert}}} \simeq \frac{1}{2\pi \left(-\partial^2_x\CA^{[-1]}_{\alpha \beta}(x_{mn}) \right)^2}\, \exp \left( 2\CA^{[0]}_{\alpha \beta}(x_{mn}) + \widehat{\CA}^{[0]}_{\alpha \beta, \alpha \beta}(x_{mn})\right) \rme^{2\mathsf{A}_{\text{D}}(m,n)} + \cdots.
\end{align}
\noindent
Using our explicit expressions for $\widehat{\CA}^{[0]}_{\alpha \beta, \alpha \beta}$, $\CA^{[0]}_{\alpha \beta}$ and $\partial^2_x \CA^{[-1]}_{\alpha \beta}$ we finally obtain 
\begin{equation}
\frac{\NCZ_{\alpha \beta, \alpha \beta}}{\NCZ_{\text{pert}}} \simeq \left( - \frac{\cot^2 \left(\frac{\pi m}{p}\right) - \cot^2 \left(\frac{\pi n}{q}\right)}{4\sqrt{2\pi}\, \mathsf{A}_{\text{D}}(m,n) \left(p^2-q^2\right)}\right)^2 \rme^{2\mathsf{A}_{\text{D}}(m,n)} + \cdots.
\end{equation}
\noindent
This is in precise agreement with results in the literature; \textit{e.g.}, \cite{akk03, hhikkmt04, st04, iky05, msw08, emms22b}. Further specializing this formula to the $(2,2k-1)$ case, we find
\begin{equation}
\label{eq:22km1-cft-20-sector}
\left.\frac{\NCZ_{\alpha \beta, \alpha \beta}}{\NCZ_{\text{pert}}}\right|_{(2,2k-1)} \simeq g^2_{\text{s}}\, \frac{\cot^2\left(\frac{\pi n}{2k-1}\right)}{2048 \pi \left(2k-1\right)^2 \sin^4 \left(\frac{\pi n}{2k-1}\right)}\, \rme^{2\mathsf{A}_{\text{D}}(1,n)} + \cdots.
\end{equation}

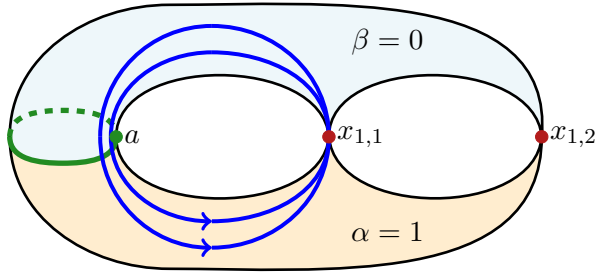
\begin{figure}
\centering
	\begin{tikzpicture}
	\begin{scope}[scale=0.7,  shift={({-5.8},{0})}]
	\draw[fill=LightBlue,fill opacity=0.2, line width=1pt] (0,0)   to [out=90,in=95] (4,0)
	to [out=90,in=95] (8,0)
	to [out=85,in=0] (1,2.5)
    to [out=180,in=90] (-2,0)
    to [out=270, in=180] (-1, -0.5)
    to [out=0, in=270] cycle;
    \draw[fill=darktangerine,fill opacity=0.2, line width=1pt] (-2,0)
    to [out=270,in=180] (1,-2.5)
    to [out=0,in=270] (8,0)
    to [out=265,in=275] (4,0)
    to [out=265,in=270] (0,0)
    to [out=270, in=0] (-1, -0.5)
    to [out=180, in=270] cycle;
    \draw[color=ForestGreen, line width=2pt] (-2,0) to [out=270, in=180] (-1, -0.5)
    to [out=0, in=270] (0,0);
    \draw[dashed, color=ForestGreen, line width=2pt] (-2,0) to [out=90, in=180] (-1, 0.5)
    to [out=0, in=90] (0,0);
    \draw[blue, line width=1.5pt] (-0.1,0) to [out=90, in=180] (1.8, 1.6);
    \draw[blue, line width=1.5pt] (1.8, 1.6) to [out=0, in=90] (4, 0);
    \draw[blue, line width=1.5pt, ->] (-0.1,0) to [out=270, in=180] (1.8, -1.6);
    \draw[blue, line width=1.5pt] (1.8, -1.6) to [out=0, in=270] (4, 0);
    \draw[blue, line width=1.5pt] (-0.3,0) to [out=90, in=180] (1.8, 2.1);
    \draw[blue, line width=1.5pt] (1.8, 2.1) to [out=0, in=90] (4, 0);
    \draw[blue, line width=1.5pt, ->] (-0.3,0) to [out=270, in=180] (1.8, -2.1);
    \draw[blue, line width=1.5pt] (1.8, -2.1) to [out=0, in=270] (4, 0);
\draw[ForestGreen, fill=ForestGreen] (0,0) circle (.7ex);
\draw[cornellred, fill=cornellred] (4,0) circle (.7ex);
\draw[cornellred, fill=cornellred] (8,0) circle (.7ex); 
\node at (0.3, 0) {$a$};
\node at (4.6, 0) {$x_{1, 1}$}; 
\node at (8.6, 0) {$x_{1, 2}$};
\node at (5.1,1.8) {$\beta=0$}; 
\node at (5.1,-1.8) {$\alpha=1$}; 
\end{scope}
	\end{tikzpicture}
\caption{ Example of the $\NCZ_{\alpha\beta,\alpha\beta}$ configuration with $\beta=0$ and $\alpha=1$, for the $x_{1,1}$ saddle of the $(2,5)$ minimal string. The figure illustrates how each pair $\alpha\beta$ denotes one full {\color{blue}$B$-cycle} associated to the saddle-point $x_{1,1}$, where the cycle starts on sheet $\beta=0$ and ends on sheet $\alpha=1$.}
\label{fig:minimalstringspectralcurve-20-contours}
\end{figure}

\item \underline{$(\tilde{m},\tilde{n})=(m,n)$, $\gamma=\beta, \delta=\alpha, \alpha \neq \beta$:} Having swapped from $\gamma\delta=\alpha\beta$ to $\gamma\delta=\beta\alpha$ this now corresponds to a $(1,1)$ configuration, where we still have one ZZ-brane but now with one  \textit{negative-tension} ZZ-brane associated to the nonperturbative saddle $x_{mn}$. Upon saddle-point integration we again find two times the same $B$-cycle but now one of them has its integration-direction reversed. See figure~\ref{fig:minimalstringspectralcurve-11-contours} for a visualization of the sheets and corresponding contours. This case has recently been covered from the matrix model side in \cite{mss22}, but it is a novelty as far as the BCFT calculations in the literature are concerned. Using precisely the same methods as above, and again splitting the annulus contributions into a ``regular'' and a ``logarithmic'' part as in \eqref{eq:split-regular-term-vandermonde}, we now obtain
\begin{align}
\label{eq:regular-split-11}
\CA^{[0]}_{\alpha \beta, \beta \alpha}(x, \tilde{x}) = \widehat{\CA}^{[0]}_{\alpha \beta, \beta \alpha}(x, \tilde{x}) - \log \left\{ \left( x(\upzeta_{\alpha})-x(\tilde{\upzeta}_{\beta}) \right) \left( x(\upzeta_{\beta})-x(\tilde{\upzeta}_{\alpha}) \right) \right\}.
\end{align}
\noindent
In particular we have
\begin{equation}
\widehat{\CA}^{[0]}_{\alpha \beta, \beta \alpha}(x, \tilde{x}) = - \widehat{\CA}^{[0]}_{\alpha \beta, \alpha \beta}(x, \tilde{x}).
\end{equation}
\noindent
In fact we further find the overall property, exactly consistent with known matrix model results \cite{mss22}, of
\begin{equation}
\CA^{[0]}_{\alpha \beta, \alpha \beta}(x, \tilde{x}) = - \CA^{[0]}_{\alpha \beta, \beta \alpha}(x, \tilde{x}).
\end{equation}
\noindent
The novelty now is to note how the logarithmic term in \eqref{eq:regular-split-11} (which has the \textit{opposite} sign from the corresponding logarithmic term in \eqref{eq:split-regular-term-vandermonde}) will effectively produce an \textit{inverse} Vandermonde-like contribution in the integrand of \eqref{eq:general-double-instanton-configuration}; \textit{i.e.}, a double pole contribution. The integral for the $(1,1)$ configuration is written as
\begin{align}
\label{eq:11-contribution-minimal-strings}
\frac{\NCZ_{\alpha \beta, \beta \alpha}}{\NCZ_{\text{pert}}} &\simeq \int \frac{\text{d}x}{2\pi} \int \frac{\text{d}\tilde{x}}{2\pi}\, \frac{1}{\left(x-\tilde{x}\right)^2} \times \\
&\times \exp \Bigg( \CA^{[-1]}_{\alpha \beta}(x) + \CA^{[-1]}_{\beta \alpha}(\tilde{x}) +  \CA^{[0]}_{\alpha \beta}(x) + \CA^{[0]}_{\beta \alpha}(\tilde{x}) + \widehat{\CA}^{[0]}_{\alpha \beta, \beta \alpha}(x, \tilde{x}) + o(g_{\text{s}}) \Bigg). \nonumber
\end{align}
\noindent
Comparing \eqref{eq:11-contribution-minimal-strings} above with our earlier \eqref{eq:20-contribution-minimal-strings} we finally see how the seemingly harmless negative-tension minus-sign has grown to prominence: it has traded an \textit{integrand zero} in \eqref{eq:20-contribution-minimal-strings} with an \textit{integrand pole} in \eqref{eq:11-contribution-minimal-strings}, hence leading to a rather \textit{different end result} for the amplitude. In order to compute this integral properly we need to reconsider our admissible integration contours in light of this pole in the integrand. As explained in \cite{mss22} these contours are sorted by their behavior at infinity  and by which saddles they cross---we no longer just have to consider steepest-descent contours but also closed contours that pick up residue contributions. As it turns out, to lowest order in $g_{\text{s}}$ we will not encounter any saddle-point integrals but simply find a residue contribution arising from the poles in the integrand\footnote{This is rather a subtle calculation, which was addressed with very explicit details in \cite{mss22}.}. Skipping the details\footnote{Let us nonetheless stress this point slightly more so as to avoid any possible confusion. To leading order, the contributions from this integral are \textit{not} arising from saddle contours $\mathcal{C}_{m n}$ (like those arising for example in \eqref{eq:ZabZpertratio}), but rather leadings orders arise from \textit{residue} contours. That these contributions even exist is a highly nontrivial fact that we shall not dwell upon herein. Indeed, this requires a detailed analysis of the contour deformations associated to an integral with poles in its integrand, as was thoroughly demonstrated in \cite{mss22}.} in \cite{mss22}, the bottom line is that the integration contour for $x$ is a closed contour encircling the endpoint $a$ of the cut of the FZZT surface and our saddle-point $x_{mn}$, with $\tilde{x}$ the steepest-ascent contour. Then after performing the residue integration one first finds\footnote{With a slight abuse of notation: although  $x_{m,\pm n}$ designate the same point, these should be understood as also signaling the sheet through which the contour reaches the saddle.}
\begin{align}
\frac{\NCZ_{\alpha \beta, \beta \alpha}}{\NCZ_{\text{pert}}} \simeq \frac{\rmi}{2\pi} \int^{x_{m,n}}_{x_{m,-n}} \text{d}\tilde{x} \left. \left( \partial_{x}\CA^{[-1]}_{\beta \alpha}(x) + \partial_{x}\CA^{[0]}_{\beta \alpha}(x) + \partial_{x}\widehat{\CA}^{[0]}_{\alpha \beta, \beta \alpha}(x, \tilde{x}) + o(g_{\text{s}}) \right) \right|_{x=\tilde{x}},
\end{align}
\noindent
where we note that herein we take into account contributions up to first-order in the power-series expansion of $\widehat{\CA}^{[0]}_{\alpha \beta, \beta \alpha}(x, \tilde{x})$. Interestingly, the non-differentiated annulus amplitudes appearing in the above exponential have exactly canceled at this lowest order in $g_{\text{s}}$. Moreover, their derivatives satisfy the relation
\begin{equation}
\left. \partial_{x} \widehat{\CA}^{[0]}_{\alpha \beta, \beta \alpha}(x, \tilde{x}) \right|_{x=\tilde{x}} = - \partial_{\tilde{x}} \CA^{[0]}_{\beta \alpha}(\tilde{x}).
\end{equation}
\noindent
Thus, the only term that actually survives at this lowest order is the ZZ disk amplitude; in which case we finally find
\begin{align}
\label{eq:cft-result-11}
\frac{\NCZ_{\alpha \beta, \beta \alpha}}{\NCZ_{\text{pert}}} \simeq -\frac{\rmi}{2\pi}\,\mathsf{A}_{\text{D}}(m,n)  + 0 + o(g_{\text{s}}).
\end{align}
\noindent
Illustrating this result for the $(2,2k-1)$ case, one has
\begin{align}
\label{eq:22km1-cft-result-11}
\left.\frac{\NCZ_{\alpha \beta, \beta \alpha}}{\NCZ_{\text{pert}}}\right|_{(2,2k-1)} \simeq -\frac{\rmi}{\pi g_{\text{s}}}\, (-1)^{k+n} \left(\frac{1}{2 k+1}+\frac{1}{2 k-3}\right) \sin \left(\frac{2 \pi  n}{2 k-1}\right) + 0 + o(g_{\text{s}}).
\end{align}
\noindent
Observe that we have obtained the two lowest orders in $g_{\text{s}}$ but to compute the next order the bi-annulus amplitude would be required. This caveat may be circumvented by resorting to the matrix model. Indeed, as we shall see in section~\ref{sec:NPminimalstring}, this coincides with the result obtained from the double-scaling limit of the matrix model result in \cite{mss22}. As another independent check, in section~\ref{sec:NPminimalstring} we will further verify this result by comparing against direct calculations from the string equations for the arbitrary $(2,2k-1)$ minimal string. 

\begin{figure}
\centering
	\begin{tikzpicture}
	\begin{scope}[scale=0.7,  shift={({-5.8},{0})}]
	\draw[fill=LightBlue,fill opacity=0.2, line width=1pt] (0,0)   to [out=90,in=95] (4,0)
	to [out=90,in=95] (8,0)
	to [out=85,in=0] (1,2.5)
    to [out=180,in=90] (-2,0)
    to [out=270, in=180] (-1, -0.5)
    to [out=0, in=270] cycle;
    \draw[fill=darktangerine,fill opacity=0.2, line width=1pt] (-2,0)
    to [out=270,in=180] (1,-2.5)
    to [out=0,in=270] (8,0)
    to [out=265,in=275] (4,0)
    to [out=265,in=270] (0,0)
    to [out=270, in=0] (-1, -0.5)
    to [out=180, in=270] cycle;
    \draw[color=ForestGreen, line width=2pt] (-2,0) to [out=270, in=180] (-1, -0.5)
    to [out=0, in=270] (0,0);
    \draw[dashed, color=ForestGreen, line width=2pt] (-2,0) to [out=90, in=180] (-1, 0.5)
    to [out=0, in=90] (0,0);
    \draw[orange, line width=1.5pt,->] (-0.3,0) to [out=90, in=180] (1.8, 2.1);
    \draw[orange, line width=1.5pt] (1.8, 2.1) to [out=0, in=90] (4, 0);
    \draw[orange, line width=1.5pt] (-0.3,0) to [out=270, in=180] (1.8, -2.1);
    \draw[orange, line width=1.5pt] (1.8, -2.1) to [out=0, in=270] (4, 0);
    \draw[blue, line width=1.5pt] (-0.1,0) to [out=90, in=180] (1.8, 1.6);
    \draw[blue, line width=1.5pt] (1.8, 1.6) to [out=0, in=90] (4, 0);
    \draw[blue, line width=1.5pt,->] (-0.1,0) to [out=270, in=180] (1.8, -1.6);
    \draw[blue, line width=1.5pt] (1.8, -1.6) to [out=0, in=270] (4, 0);
\draw[ForestGreen, fill=ForestGreen] (0,0) circle (.7ex);
\draw[cornellred, fill=cornellred] (4,0) circle (.7ex);
\draw[cornellred, fill=cornellred] (8,0) circle (.7ex); 
\node at (0.3, 0) {$a$};
\node at (4.6, 0) {$x_{1, 1}$}; 
\node at (8.6, 0) {$x_{1, 2}$};
\node at (5.1,1.8) {$\beta=0$}; 
\node at (5.1,-1.8) {$\alpha=1$}; 
\end{scope}
	\end{tikzpicture}
\caption{Example of the $\NCZ_{\alpha\beta,\beta\alpha}$ configuration with $\alpha=1$ and $\beta=0$, for the $x_{1,1}$ saddle of the $(2,5)$ minimal string. The figure illustrates how each pair $\alpha,\beta$ denotes one full $B$-cycle associated to the saddle point $x_{1,1}$. In comparison with figure~\ref{fig:minimalstringspectralcurve-20-contours}, we now see how for the first pair $\alpha,\beta$, the integration cycle ({\color{blue}blue}) starts on sheet $\beta=0$ and ends on sheet $\alpha=1$; whereas for the second pair the integration cycle ({\color{orange}orange}) reverses direction and starts on sheet $\alpha$ and ends on sheet $\beta$.}
\label{fig:minimalstringspectralcurve-11-contours}
\end{figure}
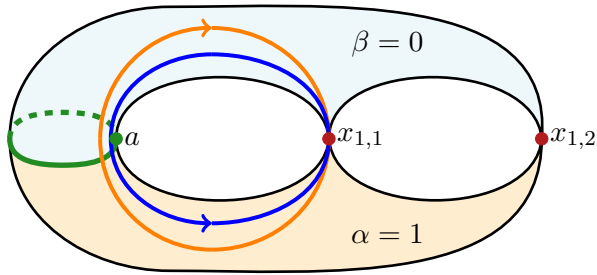

\item \underline{$(\tilde{m},\tilde{n})\neq(m,n)$:} When considering distinct saddles, there is now no \textit{a priori} restriction on the sheets. For example, this can correspond to configurations of the type $(1,0)(1,0)$ or any permutations thereof---where two (negative-tension) ZZ-branes occupy two distinct saddles. Two such possible configurations are illustrated in figure~\ref{fig:minimalstringspectralcurve-double-saddle-contours}. For standard ZZ-branes this case has been studied \cite{emms22b}, but which now needs extension to the new missing configurations. $\CA^{[0]}_{\alpha \beta, \gamma \delta}$ needs no further treatment and we may proceed directly to the evaluation of the integral. In this case one can simply write \eqref{eq:FZZTAnnulusDifferences} directly and obtain
\begin{equation}
\CA^{[0]}_{\alpha \beta, \gamma \delta}(x, \tilde{x}) = \log \frac{\left(\upzeta_{\alpha}-\tilde{\upzeta}_{\gamma}\right) \left(\upzeta_{\beta}-\tilde{\upzeta}_{\delta}\right)}{\left(\upzeta_{\alpha}-\tilde{\upzeta}_{\delta}\right) \left(\upzeta_{\beta}-\tilde{\upzeta}_{\gamma}\right)}.
\end{equation}
\noindent
The integration contours are similar to the first case, only with two distinct saddles. Such integral is simply evaluated via standard saddle-point techniques, and, to lowest order in $g_{\text{s}}$, it is written as
\begin{align}
\hspace{-27pt}
\frac{\NCZ_{\alpha \beta, \gamma \delta}}{\NCZ_{\text{pert}}} \simeq \int_{\CC_{mn}}\frac{\text{d}x}{2\pi} \int_{\CC_{mn}}\frac{\text{d}\tilde{x}}{2\pi}\, \exp \Bigg( \CA^{[-1]}_{\alpha \beta}(x) + \CA^{[-1]}_{\gamma \delta}(\tilde{x}) + \CA^{[0]}_{\alpha \beta}(x) + \CA^{[0]}_{\gamma \delta}(\tilde{x}) + \CA^{[0]}_{\alpha \beta, \gamma \delta}(x, \tilde{x}) + o(g_{\text{s}}) \Bigg).
\end{align}
\noindent
The final result follows, and is most neatly written in terms of one-instanton contributions of both saddles as
\begin{align}
\frac{\NCZ_{\alpha \beta, \gamma \delta}}{\NCZ_{\text{pert}}} \simeq \frac{\NCZ_{\alpha \beta}}{\NCZ_{\text{pert}}} \times \frac{\NCZ_{\gamma \delta}}{\NCZ_{\text{pert}}} \times \frac{\left(\upzeta^{\star}_{\alpha}-\upzeta^{\star}_{\gamma}\right) \left(\upzeta^{\star}_{\beta}-\upzeta^{\star}_{\delta}\right)}{\left(\upzeta^{\star}_{\alpha}-\upzeta^{\star}_{\delta}\right) \left(\upzeta^{\star}_{\beta}-\upzeta^{\star}_{\gamma}\right)} + o(g_{\text{s}}).
\end{align}
\noindent
Once again, we further specialize to the $(2,2k-1)$ example. We find
\begin{align}
\label{eq:22km1mixedinst}
\left.\frac{\NCZ_{\alpha \beta, \gamma \delta}}{\NCZ_{\text{pert}}}\right|_{(2,2k-1)} &\simeq g_{\text{s}}\, \frac{\rmi^{n+\tilde{n}} \left( \csc \left(\frac{\pi n}{2k-1}\right) - \csc \left(\frac{\pi \tilde{n}}{2k-1}\right) \right)^2}{128\pi \left(2k-1\right) \left( \sin \left(\frac{\pi n}{2k-1}\right) + \sin \left(\frac{\pi \tilde{n}}{2k-1}\right) \right)^2} \times \\
&
\times \sqrt{\abs{\sin \left(\frac{2\pi n}{2k-1}\right) \sin \left(\frac{2\pi \tilde{n}}{2k-1}\right)}}\, \rme^{\mathsf{A}_{\text{D}}(1,n)+\mathsf{A}_{\text{D}}(1,\tilde{n})}+\cdots. \nonumber
\end{align}

\begin{figure}
\centering
	\begin{tikzpicture}
	\begin{scope}[scale=0.7,  shift={({-5.8},{0})}]
	\draw[fill=LightBlue,fill opacity=0.2, line width=1pt] (0,0)   to [out=90,in=95] (4,0)
	to [out=90,in=95] (8,0)
	to [out=85,in=0] (1,2.5)
    to [out=180,in=90] (-2,0)
    to [out=270, in=180] (-1, -0.5)
    to [out=0, in=270] cycle;
    \draw[fill=darktangerine,fill opacity=0.2, line width=1pt] (-2,0)
    to [out=270,in=180] (1,-2.5)
    to [out=0,in=270] (8,0)
    to [out=265,in=275] (4,0)
    to [out=265,in=270] (0,0)
    to [out=270, in=0] (-1, -0.5)
    to [out=180, in=270] cycle;
    \draw[color=ForestGreen, line width=2pt] (-2,0) to [out=270, in=180] (-1, -0.5)
    to [out=0, in=270] (0,0);
    \draw[dashed, color=ForestGreen, line width=2pt] (-2,0) to [out=90, in=180] (-1, 0.5)
    to [out=0, in=90] (0,0);
    
    \draw[blue, line width=1.5pt] (-0.1,0) to [out=90, in=180] (1.8, 1.6);
    \draw[blue, line width=1.5pt] (1.8, 1.6) to [out=0, in=90] (4, 0);
    \draw[blue, line width=1.5pt,->] (-0.1,0) to [out=270, in=180] (1.8, -1.6);
    \draw[blue, line width=1.5pt] (1.8, -1.6) to [out=0, in=270] (4, 0);

    \draw[blue, line width=1.5pt] (-0.2,0) to [out=90, in=180] (1.5, 1.8);
    \draw[blue, line width=1.5pt] (1.5, 1.8) to [out=0, in=90] (8, 0);
    \draw[blue, line width=1.5pt,->] (-0.2,0) to [out=270, in=180] (1.5, -1.8);
    \draw[blue, line width=1.5pt] (1.5, -1.8) to [out=0, in=270] (8, 0);
\draw[ForestGreen, fill=ForestGreen] (0,0) circle (.7ex);
\draw[cornellred, fill=cornellred] (4,0) circle (.7ex);
\draw[cornellred, fill=cornellred] (8,0) circle (.7ex); 
\node at (0.3, 0) {$a$};
\node at (4.6, 0) {$x_{1, 1}$}; 
\node at (8.6, 0) {$x_{1, 2}$};
\node at (0.4,2.0) {$\beta=0$}; 
\node at (0.4,-2.0) {$\alpha=1$}; 
\node at (5.4,1.5) {$\delta=0$}; 
\node at (5.4,-1.5) {$\gamma=1$}; 
\end{scope}
\begin{scope}[scale=0.7,  shift={({5.8},{0})}]
	\draw[fill=LightBlue,fill opacity=0.2, line width=1pt] (0,0)   to [out=90,in=95] (4,0)
	to [out=90,in=95] (8,0)
	to [out=85,in=0] (1,2.5)
    to [out=180,in=90] (-2,0)
    to [out=270, in=180] (-1, -0.5)
    to [out=0, in=270] cycle;
    \draw[fill=darktangerine,fill opacity=0.2, line width=1pt] (-2,0)
    to [out=270,in=180] (1,-2.5)
    to [out=0,in=270] (8,0)
    to [out=265,in=275] (4,0)
    to [out=265,in=270] (0,0)
    to [out=270, in=0] (-1, -0.5)
    to [out=180, in=270] cycle;
    \draw[color=ForestGreen, line width=2pt] (-2,0) to [out=270, in=180] (-1, -0.5)
    to [out=0, in=270] (0,0);
    \draw[dashed, color=ForestGreen, line width=2pt] (-2,0) to [out=90, in=180] (-1, 0.5)
    to [out=0, in=90] (0,0);
    
    \draw[blue, line width=1.5pt] (-0.1,0) to [out=90, in=180] (1.8, 1.6);
    \draw[blue, line width=1.5pt] (1.8, 1.6) to [out=0, in=90] (4, 0);
    \draw[blue, line width=1.5pt,->] (-0.1,0) to [out=270, in=180] (1.8, -1.6);
    \draw[blue, line width=1.5pt] (1.8, -1.6) to [out=0, in=270] (4, 0);
  \draw[orange, line width=1.5pt,->] (-0.2,0) to [out=90, in=180] (1.5, 1.8);
    \draw[orange, line width=1.5pt] (1.5, 1.8) to [out=0, in=90] (8, 0);
    \draw[orange, line width=1.5pt] (-0.2,0) to [out=270, in=180] (1.5, -1.8);
    \draw[orange, line width=1.5pt] (1.5, -1.8) to [out=0, in=270] (8, 0);
\draw[ForestGreen, fill=ForestGreen] (0,0) circle (.7ex);
\draw[cornellred, fill=cornellred] (4,0) circle (.7ex);
\draw[cornellred, fill=cornellred] (8,0) circle (.7ex); 
\node at (0.3, 0) {$a$};
\node at (4.6, 0) {$x_{1, 1}$}; 
\node at (8.6, 0) {$x_{1, 2}$};
\node at (0.4,2.0) {$\beta=0$}; 
\node at (0.4,-2.0) {$\alpha=1$}; 
\node at (5.4,1.5) {$\gamma=0$}; 
\node at (5.4,-1.5) {$\delta=1$}; 
\end{scope}
	\end{tikzpicture}
\caption{Our usual example. On the left-plot we illustrate the $\NCZ_{\alpha\beta,\gamma\delta}$ configuration for $\alpha=1,\beta=0$ at the $x_{1,1}$ saddle, and $\gamma=1,\delta=0$ at the $x_{1,2}$ saddle. Similarly to previous figures, the pair $\alpha\beta$ denotes one full {\color{blue}$B$-cycle} associated to the saddle-point $x_{1,1}$, while the pair $\gamma\delta$ denotes one full {\color{blue}$B$-cycle} now associated to the saddle-point $x_{1,2}$. On the right-plot we illustrate the $\NCZ_{\alpha\beta,\gamma\delta}$ configuration for $\alpha=1,\beta=0$ at the $x_{1,1}$ saddle, and $\gamma=0,\delta=1$ at the $x_{1,2}$ saddle. The pair $\alpha\beta$ still denotes one full {\color{blue}$B$-cycle} associated to the saddle-point $x_{1,1}$, while now the pair $\gamma\delta$ denotes one full {\color{orange}$B$-cycle} with reversed direction of integration associated to the saddle-point $x_{1,2}$.}
\label{fig:minimalstringspectralcurve-double-saddle-contours}
\end{figure}
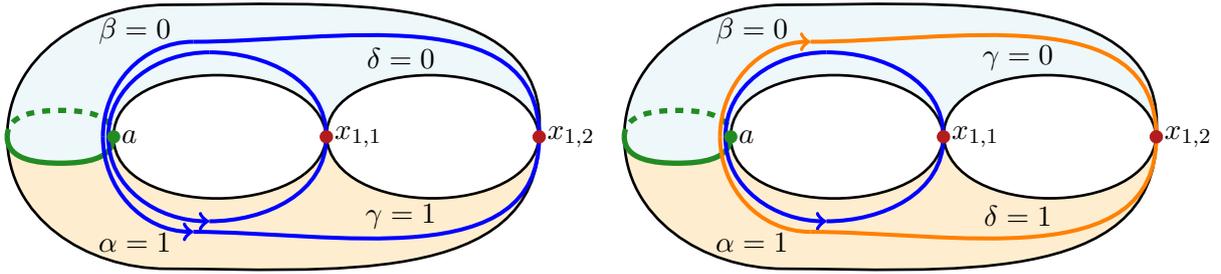

\end{enumerate}

We conclude the discussion emphasizing our main message once again. In the above calculations, we have found a highly non-trivial distinction between case~1, a familiar result previously explored in the literature, and case~2, a more recent result explored in the matrix model context in \cite{mss22}. It is striking that these rather distinct features (\textit{e.g.}, poles versus zeros in the integrand) have been obtained by allowing a seemingly minuscule change in our considerations: the flipping of signs of the (FZZT differences) disk and annulus amplitudes, $\CA^{[-1]}_{\alpha \beta}$ and $\CA^{[0]}_{\alpha \beta, \beta \alpha}$. Such seemingly small distinction at the level of FZZT amplitudes, or indeed ZZ amplitudes, might naively and \textit{a priori} not be expected to yield such different results. However, as we have seen, this minus sign has the effect of turning zeros into poles in the integrand and vice-versa, which ends up having highly non-trivial effects when it comes to performing the integration. It would be very interesting to explore the possibility of obtaining the very same results as in case 2 without the use of saddle-point integration of FZZT amplitudes, but rather by directly considering ZZ amplitudes (\textit{e.g.}, as was done in \cite{emms22a, emms22b}). If indeed possible, the repercussions of the minus sign will have to manifest themselves in some different, perhaps string field theoretic way.

\paragraph{Multiple Mixed Configurations:}

Moving towards multiple \textit{mixed} (negative-tension) D-instanton configurations leads to increasingly harder integrals. Having explained our main BCFT message in the (above) context of double-integrals, let us next outline the calculations for the $(2,1)$ and the $(1,1)(1,0)$ \textit{triple}-integral configurations. The procedure is now always the same, albeit increasingly more technically involved. We find:
\begin{enumerate}
\item \underline{The $(2,1)$ Sector:} First consider the nonperturbative contributions to the free energy arising from the $(2,1)$ sector. These have recently been computed using matrix integrals in \cite{mss22}, where they were of particular interest also in demonstrating the universality of the $\gamma_{\text{E}}$ Euler--Mascheroni constant in the Stokes data of hermitian matrix models. Herein we shall obtain the same conclusion, now in BCFT. To set-up this calculation one needs to consider an integral over D-brane moduli space which yields two ZZ-branes on a pinch, $x_{mn}$, alongside one negative-tension ZZ-brane on the same pinch. Based on our earlier experience, we consider the integral
\begin{align}
\label{eq:21-integral-minimmal-string}
\hspace{-35pt}
\frac{\NCZ_{\alpha\beta, \alpha\beta, \beta\alpha}}{\NCZ_{\text{pert}}} &\simeq \frac{1}{2} \int_{\CC_{mn}} \frac{\text{d}x_1}{2\pi} \int_{\CC_{mn}} \frac{\text{d}x_2}{2\pi}  \int_{\bar{\CC}_{mn}} \frac{\text{d}\tilde{x}}{2\pi}\, \frac{\left(x_1-x_2\right)^2}{\left(x_1-\tilde{x}\right)^2\left(x_2-\tilde{x}\right)^2}\, \exp \Bigg( \CA^{[-1]}_{\alpha \beta}(x_1) + \CA^{[-1]}_{\alpha \beta}(x_2) + \CA^{[-1]}_{\beta \alpha}(\tilde{x}) + \nonumber \\
&
+ \CA^{[0]}_{\alpha \beta}(x_1) + \CA^{[0]}_{\alpha \beta}(x_2) + \CA^{[0]}_{\beta \alpha}(\tilde{x}) + \widehat{\CA}^{[0]}_{\alpha \beta, \alpha \beta}(x_1, x_2) + \widehat{\CA}^{[0]}_{\alpha \beta, \beta \alpha}(x_1, \tilde{x}) +\widehat{\CA}^{[0]}_{\alpha \beta, \beta \alpha}(x_2, \tilde{x}) + o(g_{\text{s}}) \Bigg),
\end{align}
\noindent
where the integration contours $\CC_{mn}$ are steepest-descent contours associated to the saddle $x_{mn}$, together with the corresponding residue contributions \cite{mss22}, and $\bar{\CC}_{mn}$ is the steepest-ascent contour associated to $x_{mn}$. Note how the above combination of (regular) annuli is chosen so that they precisely match the six possible ZZ interactions that we expect---and each of these is then rewritten in terms of the appropriate FZZT differences which will produce the corresponding ZZ interaction on the saddles of the above triple-integral. The evaluation of the integral \eqref{eq:21-integral-minimmal-string} now follows in the spirit of \cite{mss22}---indeed, once the BCFT analysis leads to the relevant diagrammatics as an integral over D-brane moduli space, the evaluation of this integral is then completely analogous to its matrix model counterpart. Using the expressions for the annulus established above, and subtracting-out product contributions (\textit{i.e.}, focusing on the true novel contributions associated to the $(2,1)$ sector; see \cite{mss22}) we then find
\begin{align}
\label{eq:res-21-sector-minimal-string}
\frac{\NCZ_{\alpha \beta, \beta \alpha, \alpha\beta}}{\NCZ_{\text{pert}}} - \frac{\NCZ_{\alpha \beta, \beta \alpha}}{\NCZ_{\text{pert}}}\, \frac{\NCZ_{\alpha\beta}}{\NCZ_{\text{pert}}} &\simeq \frac{\rmi}{8 \left(2\pi\right)^2}\, \sqrt{\frac{2\pi}{-\partial_x^2\CA^{[-1]}_{\alpha \beta}(x_{mn}) \left(x_{mn}+1\right)^{2}}}\, \rme^{\mathsf{A}_{\text{D}}(m,n)} \times \\
&
\hspace{-20pt}
\times \left\{ 2 \gamma_{\text{E}} + \log \left( 2^8 \left(\partial_x^2\CA^{[-1]}_{\alpha \beta}(x_{mn})\right)^2 \left(x_{mn}+1\right)^{4} \right) \right\} + o(g_{\text{s}}^{3/2}). \nonumber
\end{align}
\noindent
Let us specialize to the friendly $(2,2k-1)$ case. The above formula becomes:
\begin{align}
\label{eq:BCFT21final}
\frac{\NCZ_{\alpha \beta, \beta \alpha, \alpha\beta}}{\NCZ_{\text{pert}}} - \frac{\NCZ_{\alpha \beta, \beta \alpha}}{\NCZ_{\text{pert}}}\, \frac{\NCZ_{\alpha\beta}}{\NCZ_{\text{pert}}} &\simeq \frac{\rmi}{32\pi^2}\, \sqrt{g_{\text{s}}\, \frac{(-1)^{k+n} \pi \cot \left(\frac{\pi  n}{2k-1}\right)}{\left(2k-1\right) \sin^2 \left(\frac{\pi n}{2k-1}\right)}}\, \rme^{\mathsf{A}_{\text{D}}(1,n)} \times \\
&
\hspace{-60pt}
\times \left\{ 2 \gamma_{\text{E}} + \log \left( \frac{2^{10}}{g_{\text{s}}^2}\, \left(2k-1\right)^2 \sin^4 \left(\frac{\pi n}{2k-1}\right) \tan^2 \left(\frac{\pi n}{2k-1}\right) \right) \right\} + o(g_{\text{s}}^{3/2}). \nonumber
\end{align}
\item \underline{The $(1,1)(1,0)$ Sector:} Finally, let us address the $(1,1)(1,0)$ nonperturbative contribution: where we consider a ZZ-brane of type $(n_1,m_1)$, alongside its negative-tension counterpart; and an additional ZZ-brane of different type $(n_2, m_2) \neq (n_1,m_1)$. This leads to an integral of the form
\begin{align}
\hspace{-30pt}
\frac{\NCZ_{\alpha \beta, \beta \alpha, \gamma\delta}}{\NCZ_{\text{pert}}} &\simeq \frac{1}{2} \int_{\CC_{m_1 n_1}} \frac{\text{d}x}{2\pi} \int_{\bar{\CC}_{m_1 n_1}} \frac{\text{d}\tilde{x}}{2\pi} \int_{\CC_{m_2 n_2}} \frac{\text{d}y}{2\pi}\, \frac{\left(x-y\right)^2}{\left(x-\tilde{x}\right)^2\left(y-\tilde{x}\right)^2}\, \exp\Bigg( \CA^{[-1]}_{\alpha \beta}(x) + \CA^{[-1]}_{\beta \alpha}(\tilde{x}) + \nonumber \\
&
\hspace{-30pt}
+ \CA^{[-1]}_{\gamma\delta}(y) + \CA^{[0]}_{\alpha \beta}(x) + \CA^{[0]}_{\beta \alpha}(\tilde{x}) + \CA^{[0]}_{\gamma\delta}(y) + \widehat{\CA}^{[0]}_{\alpha \beta, \gamma\delta}(x, y) + \widehat{\CA}^{[0]}_{\alpha \beta, \beta \alpha}(x, \tilde{x}) + \widehat{\CA}^{[0]}_{\gamma\delta, \beta \alpha}(y, \tilde{x}) + o(g_{\text{s}}) \Bigg),
\end{align}
\noindent
where $\CC_{m_1 n_1}$ is the steepest-descent contour (together with a residue contribution exactly as in \eqref{eq:11-contribution-minimal-strings}) associated to the $x_{m_1n_1}$ saddle; whereas $\bar{\CC}_{m_1 n_1}$ is the corresponding steepest-ascent contour (and the same for $\CC_{m_2 n_2}$). The $\alpha\beta$ labels relate to the saddle-point $x_{m_1 n_1}$ whereas the $\gamma\delta$ ones describe the multi-sheeted structure around $x_{m_2 n_2}$. Following the methods developed in \cite{mss22}, we first notice that---in analogy with the above $(2,1)$ sector---it makes sense to factor-out specific product contributions which add nothing new to the calculation, and rather focus on the true novel contributions associated to this sector. Evaluating the integral at lowest $g_{\text{s}}$ order yields
\begin{align}
\frac{\NCZ_{\alpha \beta, \beta \alpha, \gamma\delta}}{\NCZ_{\text{pert}}} - \frac{\NCZ_{\alpha \beta, \beta \alpha}}{\NCZ_{\text{pert}}}\, \frac{\NCZ_{\gamma\delta}}{\NCZ_{\text{pert}}} &\simeq \frac{\NCZ_{\gamma\delta}}{\NCZ_{\text{pert}}} \times \\
&
\hspace{-150pt}
\times \frac{\rmi}{2\pi} \int_{\bar{\CC}_{m_1 n_1}} \text{d}\tilde{x} \left(\frac{2}{\left(\tilde{x}-x_{m_2 n_2}\right)} + \left[ \partial_x \widehat{\CA}^{[0]}_{\alpha\beta}(x) + \partial_x\widehat{\CA}^{[0]}_{\alpha\beta,\gamma\delta}(x, x_{m_2 n_2}) + \partial_x\widehat{\CA}^{[0]}_{\alpha\beta,\beta\alpha}(x, \tilde{x}) \right]_{x=\tilde{x}}\right) + \cdots. \nonumber
\end{align}
\noindent
Notice that the second and fourth terms in the integrand above in fact cancel. Then computing the remaining integral leads to (displaying the different structures at play)
\begin{align}
\frac{\NCZ_{\alpha \beta, \beta \alpha, \gamma\delta}}{\NCZ_{\text{pert}}} - \frac{\NCZ_{\alpha \beta, \beta \alpha}}{\NCZ_{\text{pert}}}\, \frac{\NCZ_{\gamma\delta}}{\NCZ_{\text{pert}}} &\simeq \\
&
\hspace{-55pt}
\simeq \frac{\NCZ_{\gamma\delta}}{\NCZ_{\text{pert}}}\, \frac{\rmi}{2\pi}  \left( 2 \log \left( x_{m_1n_1}-x_{m_2n_2} \right) + \widehat{\CA}^{[0]}_{\alpha\beta,\gamma\delta} (x_{m_1n_1}, x_{m_2 n_2})\right) + \cdots = \nonumber \\
&
\hspace{-55pt}
= \frac{\NCZ_{\gamma\delta}}{\NCZ_{\text{pert}}}\, \frac{\rmi}{2\pi}\, \CA^{[0]}_{\alpha\beta,\gamma\delta} (x_{m_1n_1}, x_{m_2 n_2}) + \cdots \nonumber \\
&
\hspace{-55pt}
= \frac{\NCZ_{\gamma\delta}}{\NCZ_{\text{pert}}}\, \frac{\rmi}{2\pi}\, \log \left( \frac{\left( \cos \left(\frac{\pi m_1}{p}-\frac{\pi n_1}{q}\right) - \cos \left(\frac{\pi m_2}{p}-\frac{\pi n_2}{q}\right) \right)}{\left( \cos \left(\frac{\pi m_1}{p}+\frac{\pi n_1}{q}\right) - \cos \left(\frac{\pi m_2}{p}-\frac{\pi n_2}{q}\right) \right)} \right. \times \nonumber \\
&
\hspace{+55pt}
\times \left. \frac{\left( \cos \left(\frac{\pi m_1}{p}+\frac{\pi n_1}{q}\right) - \cos \left(\frac{\pi m_2}{p}+\frac{\pi n_2}{q}\right) \right)}{\left( \cos \left(\frac{\pi m_1}{p}-\frac{\pi n_1}{q}\right) - \cos \left(\frac{\pi m_2}{p}+\frac{\pi n_2}{q} \right) \right)} \right) + \cdots. \nonumber
\end{align}
\noindent
As usual, let us specialize to the case $(2,2k-1)$. We then find
\begin{align}
\label{eq:BCFT1110final}
\frac{\NCZ_{\alpha \beta, \beta \alpha, \gamma\delta}}{\NCZ_{\text{pert}}} - \frac{\NCZ_{\alpha \beta, \beta \alpha}}{\NCZ_{\text{pert}}}\, \frac{\NCZ_{\gamma\delta}}{\NCZ_{\text{pert}}} \simeq \frac{\NCZ_{\gamma\delta}}{\NCZ_{\text{pert}}}\, \frac{\rmi}{2\pi}\, \log \abs{\frac{\sin \left(\frac{\pi n_1}{2k-1}\right)-\sin \left(\frac{\pi n_2}{2k-1}\right)}{\sin \left(\frac{\pi n_1}{2k-1}\right)+\sin \left(\frac{\pi n_2}{2k-1}\right)}} + \cdots.
\end{align}
\end{enumerate}

\subsection{Multiple ZZ- and Negative-Tension ZZ-Instantons}\label{subsec:ZZ+negativeZZ}

Having gathered calculational evidence for the role and corresponding amplitudes associated to both ZZ- and negative-tension ZZ-branes in Liouville BCFT, let us next address the case of multiple \textit{pure} (negative-tension) ZZ D-instantons. We recover results in the literature concerning multiple ZZ-instantons, and enlarge this class to also include negative-tension ZZ-instantons.

Starting off with the integral for $\ell$ multiple ZZ D-instantons we first consider\footnote{As in the $(2,0)$ case, we include a combinatorial factor accounting for all FZZT branes on the same sheets $\alpha\beta$.}
\begin{align}
\label{eq:multi-FZZT-integral}
\frac{\NCZ_{\alpha \beta, \ldots,\alpha \beta}}{\NCZ_{\text{pert}}} &\simeq \frac{1}{\ell!}\int_{\CC_{mn}} \prod_{i=1}^{\ell}\frac{\text{d}x_i}{2\pi}\, \prod_{i<j} \left(x_i-x_j\right)^2 \times \\
&\times \exp \left( \sum_{i=1}^{\ell} \CA^{[-1]}_{\alpha \beta}(x_i) + \sum_{i=1}^{\ell}  \CA^{[0]}_{\alpha \beta}(x_i) + \sum_{i=1}^{\ell} \sum_{j=i+1}^{\ell} \widehat{\CA}^{[0]}_{\alpha \beta, \alpha \beta}(x_i, x_j) + \cdots \right). \nonumber
\end{align} 
\noindent
Notice now the appearance of a more general Vandermonde-like determinant, of dimension $\ell$, as a direct generalization of splitting the annulus contributions in ``regular'' and ``logarithmic'' parts as in equations \eqref{eq:split-regular-term-vandermonde} and \eqref{eq:regular-split-11}. This saddle-point integration is straightforward to perform, albeit for completeness we fill in some of the steps below. Firstly, expanding around $x_i=x^{\star}+\lambda_i \sqrt{g_{\text{s}}}$ we have at lowest $g_{\text{s}}$ orders
\begin{align}
\frac{\NCZ_{\alpha \beta, \ldots,\alpha \beta}}{\NCZ_{\text{pert}}} &\simeq \frac{g_{\text{s}}^{\frac{\ell^2}{2}}}{\left(2\pi\right)^{\ell} \ell!}\, \exp \Bigg( \ell \left( \CA^{[-1]}_{\alpha \beta}(x_{nm}) + \CA^{[0]}_{\alpha \beta}(x_{nm}) \right) + \frac{1}{2} \left( \ell^2-\ell \right) \widehat{\CA}^{[0]}_{\alpha \beta, \alpha \beta}(x_{nm}, x_{nm}) \Bigg) \times \\
&\times \int \prod_{i=1}^{\ell} \text{d}\lambda_i \prod_{i<j} \left(\lambda_i-\lambda_j\right)^2 \rme^{\frac{1}{2} g_{\text{s}} \sum\limits_{i=1}^{\ell} \lambda^2_i \partial^2_x \CA^{[-1]}_{\alpha \beta}(x_{nm})} + \cdots. \nonumber
\end{align}
\noindent
Recalling that the relation between $\CA^{[0]}_{\alpha \beta}$ and $\widehat{\CA}^{[0]}_{\alpha \beta, \alpha \beta}$ at lowest $g_{\text{s}}$-order is simply a factor of two, we perform the saddle-point integration to find 
\begin{equation}
\label{eq:multi-POS-ZZs}
\frac{\NCZ_{\alpha \beta, \ldots,\alpha \beta}}{\NCZ_{\text{pert}}} \simeq \frac{G_2 \left(\ell+1\right)}{\left(2\pi\right)^{\ell/2}} \left( \frac{\cot^2 \left(\frac{\pi m}{p}\right) - \cot^2 \left(\frac{\pi n}{q}\right)}{4\, \mathsf{A}_{\text{D}}(m,n) \left(p^2-q^2\right)}\right)^\frac{\ell^2}{2} \rme^{\ell\, \mathsf{A}_{\text{D}}(m,n)} + \cdots,
\end{equation} 
\noindent
where $G_2 (x)$ is the Barnes $G$-function. This agrees exactly with well-known results across the literature; \textit{e.g.}, \cite{akk03, hhikkmt04, st04, iky05, msw08, emms22b}. As usual, we specialize the above result to the $(2,2k-1)$ minimal-string case and find
\begin{equation}
\label{eq:22km1multipleinst}
\left.\frac{\NCZ_{\alpha \beta, \ldots,\alpha \beta}}{\NCZ_{\text{pert}}}\right|_{(2,2k-1)} \simeq \frac{G_{2} \left(\ell+1\right)}{\left(2\pi\right)^{\ell/2}}\, \left( g_{\text{s}}\, \frac{(-1)^{n+k} \cot\frac{n\pi}{2k-1}}{32 \left(2k-1\right) \sin^2\frac{n\pi}{2k-1}}\right)^{\frac{\ell^2}{2}} \rme^{\ell\, \mathsf{A}_{\text{D}}(1,n)} + \cdots.
\end{equation}

Next, let us extend the above result for the case of multiple negative-tension ZZ-branes. This just amounts to exchanging the sheets in \eqref{eq:multi-FZZT-integral} with each other (\textit{i.e.}, swapping $\alpha$ and $\beta$). Thus we consider the integral
\begin{align}
\frac{\NCZ_{\beta\alpha, \ldots, \beta\alpha}}{\NCZ_{\text{pert}}} &\simeq \frac{1}{\ell!} \int_{\bar{\CC}_{mn}} \prod_{i=1}^{\ell}\frac{\text{d}x_i}{2\pi}\, \prod_{i<j} \left(x_i-x_j\right)^2 \times \\
&\times \exp \left( \sum_{i=1}^{\ell} \CA^{[-1]}_{\beta \alpha }(x_i) + \sum_{i=1}^{\ell}  \CA^{[0]}_{\beta \alpha }(x_i) + \sum_{i=1}^{\ell} \sum_{j=i+1}^{\ell} \widehat{\CA}^{[0]}_{\beta \alpha, \beta \alpha}(x_i, x_j) + \cdots \right), \nonumber
\end{align} 
\noindent
where $\bar{\CC}_{mn}$ denotes the steepest-\textit{ascent} contour through the saddle-point $x_{mn}$ instead of the steepest-\textit{descent} contour $\CC_{mn}$. As may already be familiar to the reader, when one finds purely physical-sheet or purely non-physical-sheet ZZ-brane configurations under consideration, the resulting contributions are simple to evaluate. Noting that $\CA^{[-1]}_{\beta \alpha}(x_{nm}) = - \CA^{[-1]}_{\alpha \beta}(x_{nm})$, we find by comparison with the previous integral: 
\begin{equation}
\label{eq:multi-NEG-ZZs}
\frac{\NCZ_{\beta\alpha, \ldots, \beta\alpha}}{\NCZ_{\text{pert}}} \simeq \frac{G_2 \left(\ell+1\right)}{\left(2\pi\right)^{\ell/2}} \left( - \frac{\cot^2 \left(\frac{\pi m}{p}\right) - \cot^2 \left(\frac{\pi n}{q}\right)}{4\, \mathsf{A}_{\text{D}}(m,n) \left(p^2-q^2\right)}\right)^\frac{\ell^2}{2} \rme^{-\ell\, \mathsf{A}_{\text{D}}(m,n)} + \cdots.
\end{equation} 
\noindent
Once again specializing to the $(2,2k-1)$ case we have
\begin{equation}
\label{eq:22km1multipleantiinst}
\left.\frac{\NCZ_{\beta\alpha, \ldots, \beta\alpha}}{\NCZ_{\text{pert}}}\right|_{(2,2k-1)} \simeq \frac{G_{2} \left(\ell+1\right)}{\left(2\pi\right)^{\ell/2}}\, \left( - g_{\text{s}}\, \frac{(-1)^{n+k} \cot\frac{n\pi}{2k-1}}{32 \left(2k-1\right) \sin^2\frac{n\pi}{2k-1}}\right)^{\frac{\ell^2}{2}}\rme^{-\ell\, \mathsf{A}_{\text{D}}(1,n)} + \cdots,
\end{equation}
\noindent
We see that we find a contribution which only differs from the previously known ones essentially by minus signs\footnote{We stress yet once again that although the minus sign in the exponent only trivially alters contributions of purely negative-tension configurations, its effect in mixed sectors is the furthest from trivial: it introduces poles in the integrands and therefore a completely novel set of predictions emerged.}. Ensemble, these two formulae \eqref{eq:multi-POS-ZZs} and \eqref{eq:multi-NEG-ZZs} are an example of a more general case of the resonant behavior alluded to in subsection~\ref{subsec:bcft-resonant-pairs}. It is also interesting to compare these formulae \eqref{eq:multi-POS-ZZs} and \eqref{eq:multi-NEG-ZZs} with the corresponding multi-instanton formulae for pure eigenvalue tunneling and for pure anti-eigenvalue tunneling in \cite{mss22}, where the Barnes function appears in the same fashion and where the string-coupling also scales quadratically. This is a well-known phenomenon that is sometimes dubbed backward-forward relation \cite{bssv22, mss22}.

\subsection{Boundary CFT for Resurgent FZZT Branes}\label{subsec:FZZT-BCFT}

Having understood ZZ D-instantons from a BCFT perspective we can ask if this also gives us some leverage on negative-tension FZZT-brane calculations. A natural question to pose would be if resonance also naturally extends to those contributions, possibly due to the multi-sheeted nature of the FZZT moduli space. Such nonperturbative contributions would appear for example in the study of the large-order behavior of correlation functions of the type
\begin{equation}
W_{h} \left( x_1, \ldots, x_{h} \right) = \ev{\mathcal{O}(x_1) \cdots \mathcal{O}(x_{h})},
\end{equation}
\noindent
where $\mathcal{O}(x)$ denotes a macroscopic-loop operator (\textit{i.e.}, a worldsheet partition-function with fixed boundary length) or a resolvent operator for example. This type of FZZT nonperturbative contributions and their large-order effects may be found in \cite{eggls23}, in the JT-gravity context.

Let us start with the discussion of a single FZZT-brane with boundary parameter on the sheet $\alpha$. The argument is the same as before: the combinatorics of boundaries leads to an exponentiation of the disconnected surfaces \cite{p94} and we thus find the contribution\footnote{Recall that the annulus contribution herein does not need to be regularized.}
\begin{align}
\label{eq:1-inst-FZZT-contribution}
W^{\text{FZZT}}_1 \left(x\right) &\simeq \exp \left( \DiscFZZT{\upzeta_{\alpha}} + \frac{1}{2}\,\, \AnnulusFZZT{\upzeta_{\alpha}}{\upzeta_{\alpha}} + \cdots \right) = \nonumber \\
&\simeq \exp \Bigg( \mathsf{A}_{\text{D}} \left(\upzeta_{\alpha}(x)\right) + \frac{1}{2} \mathsf{A}_{\text{A}} \left(\upzeta_{\alpha}(x), \upzeta_{\alpha}(x)\right) + \cdots \Bigg),
\end{align}
\noindent
where the disk and annulus were already computed in \eqref{eq:FZZT-disk} and \eqref{eq:FZZT-annulus-same-sheet}. It is interesting to study the above expression in more detail. There are two cases to consider: 
\begin{enumerate}
\item We start with $(p,q)=(2,2k-1)$ for simplicity, as in this case the FZZT moduli space is just double-sheeted. Those sheets are labeled by $\alpha \in \lbrace{0,1\rbrace}$ where $\alpha=0$ is the canonical choice for the physical sheet. Calculating the D-brane tension we find (similarly to our earlier ZZ story) a resonant pair associated to the two-sheets (this is in some sense the single disk-amplitude analogue of formula \eqref{eq:differences-FZZT-disks})
\begin{align}
\mathsf{A}_{\text{D}} \left(\zeta_1(x)\right) = \mu^{\frac{p+q}{2p}} \int^{x(\zeta_0)} \text{d}x \left(-1\right) y(x) = -\mathsf{A}_{\text{D}} \left(\zeta_0(x)\right).
\end{align}
\noindent
This implies that for $(p,q)=(2,2k-1)$ minimal string theories the nonperturbative structure of FZZT-brane contributions is quite similar to the structure of ZZ contributions.
\item The story changes when considering $p>2$. Then the moduli space has more sheets and the relation between the disk amplitudes on the different sheets will be more complicated (even though it still is resonant, it will be so in a more intricate way). Here the nonperturbative structure of FZZT contributions will fundamentally differ from the ZZ story outlined above.
\end{enumerate}

The hermitian matrix model analogue of the above calculation is well-known and we follow \cite{msw07, mss22}. We will come back to this in the next section~\ref{sec:NPminimalstring}, but for the moment consider $M$ a $N\times N$ hermitian matrix, $V(x)$ the matrix model potential, $V_{\text{h;eff}}(x)$ the corresponding holomorphic effective potential, and with $A_{0;2}(x, x)$ to be found in \cite{mss22}; so as to write
\begin{align}
\label{eq:Matrix-ell-instanton}
\rme^{-\frac{1}{2g_{\text{s}}} V(x)} \left\langle\det\left(x-M\right)\right\rangle \simeq \exp\left(-\frac{1}{2g_{\text{s}}} V_{\text{h;eff}}(x) + A_{0;2}(x, x) + \cdots \right).
\end{align}
\noindent
Upon double-scaling, \eqref{eq:1-inst-FZZT-contribution} and \eqref{eq:Matrix-ell-instanton} agree and a detailed discussion may be found in \cite{sss19}.

Having established the resurgent resonance of a single FZZT-brane, we can ask for the general case. With all we have said above, it is not complicated to arrive at 
\begin{align}
\label{eq:General-FZZT-Brane-Contribution}
W^{\text{FZZT}}_{h} \left( x_1, \ldots, x_{h} \right) \simeq \exp \left( \sum\limits_{i=1}^{h} \mathsf{A}_{\text{D}}\left(\zeta_{\alpha_i}(x_i)\right) + \frac{1}{2}\, \sum\limits_{i=1}^{h}\sum\limits_{j=1}^{h} \mathsf{A}_{\text{A}}\left(\upzeta_{\alpha_i}(x_i), \upzeta_{\alpha_j}(x_j)\right) + \cdots \right).
\end{align}
\noindent
Let us quickly comment on this result. The one obvious difference between \eqref{eq:General-FZZT-Brane-Contribution} and the FZZT amplitudes in earlier subsections is that we are now no-longer writing \textit{differences} of FZZT-branes. Those differences are convenient to use as the choice of base-point for the FZZT integration in, for example, \eqref{eq:FZZTdisc} drops out. Herein we are interested in exactly one FZZT brane and so we have to deal with this intricacy\footnote{Writing differences of FZZT amplitudes in comparison to single ones will contribute a factor of $2$ in the result for the minimal string branes: in the latter case we consider half-cycles (single FZZT branes) while in the former case we have a full-cycle (difference of two FZZT branes). It would be interesting to compare the two approaches further. Note how the distinction in these two approaches is responsible for the well-known difference of a factor of 2 between matrix-model instanton actions and minimal-string ones (for example, compare to formula \eqref{eq:Matrix-ell-instanton}).}. More on this question may be found in \cite{mmss04, sss19}. In addition, for $x_i\neq x_j$ the above annulus contributions are all well-defined and there is no need of regularization (up to the infinite constant terms which can appear \cite{sss19}). Finally, depending on the choice of sheets characterized by $\lbrace \alpha_i\rbrace$, \eqref{eq:General-FZZT-Brane-Contribution} will calculate different nonperturbative contributions. 

Let us consider some instructive examples of formula \eqref{eq:General-FZZT-Brane-Contribution}. We have already seen the one-brane contribution above. The generalization to $\ell$ branes of the same type (\textit{i.e.}, with $x_1 = x_2 = \cdots = x_{\ell} \equiv x$ and $\alpha_1 = \alpha_2 = \cdots = \alpha_{\ell} \equiv \alpha$) is straightforward and reads
\begin{equation}
\label{eq:identical-FZZT-branes}
W^{\ell\text{-FZZT}}_{1} \left(x\right) \simeq \frac{1}{\ell!}\, \exp \left( \ell\,\mathsf{A}_{\text{D}}\left(\zeta_{a}(x)\right) + \frac{\ell^2}{2}\, \mathsf{A}_{\text{A}}\left(\upzeta_{\alpha}(x), \upzeta_{\alpha}(x)\right) + \cdots \right),
\end{equation}
\noindent
where we have included a symmetry factor of $1/{\ell!}$, and the annulus contribution needs no regularization and is computed via \eqref{eq:FZZT-annulus-same-sheet}. Let us mention in passing that for $(p,q)=(2,2k-1)$ a resonant pair ($\alpha=0$ and $\alpha=1$) will emerge from \eqref{eq:identical-FZZT-branes} exactly as described in and below formula \eqref{eq:1-inst-FZZT-contribution}. Further, it is interesting to understand the FZZT analogue of the $(1,1)$ contribution earlier discussed for ZZ-branes. Explicitly choosing $(p,q)=(2,2k-1)$, and $\alpha_1=0$ and $\alpha_2=1$, we find
\begin{align}
W^{(1,1)\text{-FZZT}}_{2} \left(x_1,x_2\right) &\simeq \exp \Bigg( \mathsf{A}_{\text{D}}\left(\zeta_{0}(x_1)\right) + \mathsf{A}_{\text{D}}\left(\zeta_{1}(x_2)\right) + \frac{1}{2}\, \mathsf{A}_{\text{A}}\left(\upzeta_{0}(x_1), \upzeta_{0}(x_1)\right) + \nonumber \\
&+ \frac{1}{2}\, \mathsf{A}_{\text{A}}\left(\upzeta_{1}(x_2), \upzeta_{1}(x_2)\right) + \mathsf{A}_{\text{A}}\left(\upzeta_{0}(x_1), \upzeta_{1}(x_2)\right) + \cdots \Bigg).
\end{align}
\noindent
Here an interesting question arises: what happens as $x_2\to x_1$? To this end, let us spell out this formula explicitly
\begin{align}
W^{(1,1)\text{-FZZT}}_{2} \left(x_1,x_2\right) &\simeq \frac{1}{4p^2} \left(\frac{1}{\upzeta_1(x_2)}-\frac{1}{\upzeta_0(x_1)}\right) \frac{1}{x_1-x_2}\, \rme^{\mathsf{A}_{\text{D}} \left(\zeta_{0}(x_1)\right) - \mathsf{A}_{\text{D}} \left(\zeta_{0}(x_2)\right)} + \cdots.
\end{align}
\noindent
First, as $x_2\to x_1$ the disk contributions in the exponential cancel each other. This discussion is analogous to the one in (and under) formula \eqref{eq:11-contribution-minimal-strings} (just that here we do not perform integrations over FZZT moduli as in the earlier ZZ case). Further, from the annulus contributions we find a divergence as $x_2$ approaches $x_1$. This is again in line with the discussion in subsection~\ref{subsec:generic-ZZ-BCFT}. Physically, the term $1/(x_1-x_2)$ implies that FZZT-branes on different sheets behave a lot like the eigenvalue--anti-eigenvalue pairs discussed in \cite{mss22}. Two FZZT branes attract, and if they do so close to a pinch of the moduli space they form a bound state---exactly as discussed in the previous subsection. On the other hand if they meet somewhere else on moduli space, they just annihilate. Last but not least let us comment on a curious fact: the above $(1,1)$ nonperturbative contribution appears for example as a nonperturbative correction to the correlator $W_2(x_1, x_2) = \frac{1}{(x_1-x_2)^2} \left( \cdots \right)$ \cite{msw07}, which itself is intricately related to the Bergmann kernel that plays a prominent role in the topological recursion construct \cite{eo07a} (in appropriate coordinates it is just a double-pole as $x_2\to x_1$). In fact, the Bergmann kernel sets up the computation of perturbative contributions in the recursion, and it is hence quite interesting how the above $(1,1)$ correction appears to show a similar coordinate behavior. This could hint that this kernel may yet also play some relevant role in the computation of nonperturbative corrections via the topological recursion (see \cite{eggls23}).

\section{All Instantons of Nonperturbative Minimal Strings}\label{sec:NPminimalstring}

Let us now turn to the matrix model analysis. This will first consist of a multi-pinched (degenerate hyperelliptic) generalization of \cite{mss22}, which can then be applied to the double-scaled minimal-string spectral curves of \cite{ss03} so as to find a match against\footnote{We mainly address the one-matrix model throughout this section, albeit we will comment on possible future research venues to fully extend our results to the two-matrix model in subsection~\ref{subsec:2matrixmodel}.} the $(p,q)=(2,2k-1)$ BCFT results across the previous section. Such results also allow for a study of JT gravity \cite{sss19}.

Setting the matrix-model stage following, \textit{e.g.}, \cite{m04, msw07, msw08}, we first recall the partition function in diagonal gauge
\be
\mathcal{Z}_{N} = \frac{1}{N!} \int \prod\limits_{i=1}^{N} \frac{\text{d}\lambda_i}{2\pi}\, \Delta^2 (\lambda)\, \text{e}^{-\frac{1}{g_{\text{s}}} \sum\limits_{i=1}^{N} V(\lambda_i)}.
\ee
\noindent
Exponentiation of the Vandermonde determinant $\Delta (\lambda)$ naturally leads to the introduction of the holomorphic effective potential $V_{\text{h;eff}}(x)$ from where the large-$N$ (at fixed $t = g_{\text{s}} N$ 't~Hooft coupling \cite{th74}) spectral-curve $y(x)$ follows as $V_{\text{h;eff}}'(x) = y(x)$ \cite{bipz78, ackm93}. This is the prime input for the topological recursion \cite{eo07a}, which then yields the matrix-model free energy $\CF = \log \CZ$ genus-by-genus as $\CF_g(t)$, as an asymptotic perturbative expansion in the string coupling.

If the matrix-model potential $V(\lambda)$ has $s$ critical points, the most general eigenvalue configuration will have support on a multi-cut configuration $\NCC = [a_1,a_2] \cup [a_3,a_4] \cup \cdots \cup [a_{2s-1},a_{2s}]$, with corresponding hyperelliptic spectral curve 
\be
\label{eq:hyperelliptic-spectral-curve}
y^2 (x) = M^2 (x)\, \prod_{i=1}^{2s} \left( x-a_{i} \right),
\ee
\noindent
where $M(x)$ is the moment function. Distributing the total $N$ eigenvalues across these $s$-cuts, it is natural to associate partial 't~Hooft couplings $t_{i} = g_{\text{s}} N_{i}$ with each cut. These are moduli for the matrix model, albeit fixed to adding up to the total 't~Hooft coupling $t$. We will be interested in the minimal-string double-scaling limit, with all degenerate cuts but for the perturbative cut, hence geometrically corresponding to a multi-pinched spectral curve.

In particular, we are interested in the spectral curve description of $(p,q)=(2,2k-1)$ minimal string theory following \cite{ss03}. These models are described by
\be
\label{eq:minimal-string-Tp-Tq}
T_{p} (y) = T_{q} (x),
\ee
\noindent
where recall $T_{p} \left( \cos \theta \right) = \cos p \theta$ are Chebyshev polynomials of the first kind. This geometry describes a $p$-sheeted covering of the complex $x$-plane, which is obtained from a degenerate hyperelliptic spectral curve with $\frac{1}{2} \left( p-1 \right) \left( q-1 \right)$ pinched $A$-cycles in the double-scaling limit---hence resulting in a genus-zero Riemann surface, \textit{i.e.}, we find a double-scaled single-cut along $\NCC = (-\infty,1]$. This is essentially the Riemann surface $\Sigma_{p,q}$ of section~\ref{sec:ZZminimalstring} \cite{mmss04}, where the eigenvalue complex $x$-plane is identified with $\mu_{\text{B}}$ as in \eqref{eq:uniformization-minimal-string-x}. In our one-matrix model scenario this simplifies, as $T_{2} (y) = 2y^2-1$, and the double-scaled spectral curve becomes
\be
\label{eq:dsl-spectral-curve-MS}
y^2 = \frac{1}{2} \left( 1 + T_{2k-1} (x) \right) = 2^{2k-3} \left( x+1 \right) \prod_{n=1}^{k-1} \left( x-x_{n}^{\star} \right)^2,
\ee
\noindent
with $x_{n}^{\star} = - \cos \frac{2\pi n}{2k-1}$ for $1 \le n \le k-1$. This is clearly of the form \eqref{eq:hyperelliptic-spectral-curve} albeit for a single infinite cut. The moment function can be extracted from the above expression as\footnote{Note that there is a minus-sign difference between conventions herein and the ones used in \cite{gs21}. The reason for this is so that in the following subsections we match notation against the matrix model results in \cite{mss22}, where eigenvalue tunneling to the physical sheet is related to the saddles of $V_{\text{h;eff}} (x)$. This is also consistent with the difference in the sign of the exponent between sections~\ref{sec:ZZminimalstring} and~\ref{sec:NPminimalstring} (recall footnote~\ref{footnote:ExplainingMinusSign1} in subsection~\ref{subsec:bcft-resonant-pairs}).}
\be
M(x) = -\frac{1}{\sqrt{2}}\, \frac{T_{k}(x)+T_{k-1}(x)}{x+1},
\ee
\noindent
whose zeroes are precisely at the pinched $A$-cycles. These $x_{n}^{\star}$ are of course the relevant points in the definition of ZZ-branes from FZZT-branes in \eqref{eq:ZZ-from-FZZTs}; \textit{i.e.}, they are where the ZZ-branes sit (on the physical sheet). Finally, the corresponding holomorphic effective potential is given by 
\be
V_{\text{h;eff}} (x) = -\frac{1}{2k+1}\, T_{\frac{2k+1}{2}} (x) + \frac{1}{2k-3}\, T_{\frac{2k-3}{2}} (x)
\ee
\noindent
(see, \textit{e.g.}, \cite{gs21} for a recent discussion of these topics).

\subsection{Arbitrary Double-Scaled Spectral Geometry}\label{subsec:double-scaled-geometry}

We begin by reconsidering the matrix-model predictions for various transseries sectors in \cite{mss22} and extend them to our multi-pinched, double-scaled configurations. 

\paragraph{On Transseries Sectors from Matrix Models:}

Let us immediately specialize to the one-cut setting with $s=1$ in \eqref{eq:hyperelliptic-spectral-curve} and $\NCC = [a,b]$,
\begin{equation}
\label{eq:1-cut-matrix-sc}
y(x) = M(x) \sqrt{(x-a)(x-b)},
\end{equation}
\noindent
where the moment function $M(x)$ has multiple zeros $x_1^{\star}$, $x_2^{\star}$, $\ldots$, as nonperturbative saddle points. There is one additional subtlety appearing in this multi-saddle-point case in comparison to the single-pinch considered in \cite{mss22}: depending on the locations of the $x_i^{\star}$, (anti) eigenvalue-tunneling may now also happen in-between  nonperturbative saddles. For example, a previously perturbative eigenvalue might first tunnel to $x_1^{\star}$ and only then, upon crossing the next Stokes line, will it tunnel to $x_2^{\star}$---where the saddle-point $x_2^{\star}$ would otherwise never be directly reachable by tunneling straight from the perturbative-cut. For this reason, the configuration of saddle points is crucial and here we shall only consider the two saddles closest to the cut ($x_1^{\star}$ and $x_2^{\star}$). The generic case with an arbitrary number of nonperturbative saddles may then be obtained straightforwardly building on our ensuing discussion. 

\begin{figure}
\centering
\includegraphics[scale=0.7]{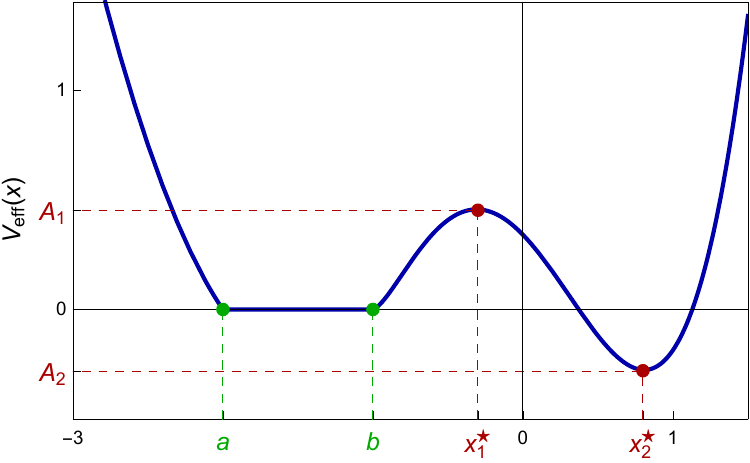}
\caption{Schematic plot of the real-part of the matrix-model holomorphic effective-potential (on the physical sheet) with two real saddle-points $x_1^{\star}$ and $x_2^{\star}$. The saddle $x_1^{\star}$ is closer to the cut with endpoints $a$ and $b$. The corresponding instanton actions $A_1$ (to $x_1^{\star}$) and $A_2$ (to $x_2^{\star}$) are real and resonant. Compare this plot to the double-scaled $(2,5)$ minimal-string plot in figure \ref{fig:minimal-string-resonance-check}.}
\label{fig:Matrix-Potential-Configuration}
\end{figure}

More specifically, focus upon the following setting: both saddles lie on the real line, with $x_1$ siting closer to the cut than $x_2$. In addition we require the instanton actions to be real and labeled analogously to the saddles; such that $A_1$ is associated to $x_1$ and $A_2$ to $x_2$. Furthermore we require that $A_1$ and $A_2$, although themselves resonant, do not resonate between each other. One such configuration is depicted in figure~\ref{fig:Matrix-Potential-Configuration}. The fact that the instanton actions are both real and resonant \cite{mss22} implies that the problem we are studying has two Stokes lines: a forward one along the positive real axis, and a backwards one across the negative real axis (see figure~\ref{fig:matrix-eigenvalue-tunneling-25-potential}).

Following the notation introduced in \cite{mss22}, we denote the corresponding partition-function and free-energy matrix-model nonperturbative contributions as
\begin{equation}
\label{eq:Z-and-F-(anti)ev-labels}
\mathcal{Z}^{(\ell_1|\bar{\ell}_1)(\ell_2|\bar{\ell}_2)}, \qquad \mathcal{F}^{(\ell_1|\bar{\ell}_1)(\ell_2|\bar{\ell}_2)},
\end{equation}
\noindent
where $\ell_1$ ($\bar{\ell}_1$) (anti) eigenvalues sit at $x_1^{\star}$ and likewise for $x_2^{\star}$ (generically $\ell_1 \neq \bar{\ell}_1 \neq \ell_2 \neq \bar{\ell}_2$). The above matrix-integral quantities may alternatively be computed out of string equations (more on this in appendices~\ref{app:minimal-string-equation-setup} and~\ref{app:minimal-string-25}). Again following the notation in \cite{mss22}, string-equation nonperturbative transseries-sectors are now denoted with standard\footnote{Recap notation: transseries as $Z$ and $F$, matrix-integral as $\CZ$ and $\CF$, and BCFT as $\NCZ$ and $\NCF$.} ``roman notation'', and the above configuration with two resonant saddle-points translates to a four-parameter resonant resurgent-transseries of the form (see, \textit{e.g.}, \cite{abs18, gs21})
\begin{align}
\label{eq:generic-4parameter-TS}
F \left(g_{\text{s}}, \boldsymbol{\sigma}\right) = \sum_{\boldsymbol{n} \in \BN_{0}^{4}} \boldsymbol{\sigma}^{\boldsymbol{n}}\, \rme^{-(n_1-m_1)\frac{A_1}{g_{\text{s}}}-(n_2-m_2)\frac{A_2}{g_{\text{s}}}}\, F^{(\boldsymbol{n})} (g_{\text{s}}),
\end{align}
\noindent
where $\boldsymbol{\sigma}$ is the four-dimensional vector of transseries parameters, $A_1$ and $A_2$ are resurgent instanton actions, and $F^{(\boldsymbol{n})} (g_{\text{s}})$ are nonperturbative transseries sectors. These sectors are usually labeled as $\boldsymbol{n} = (n_1,m_1,n_2,m_2)$ to distinguish from the (anti) eigenvalue labeling in \eqref{eq:Z-and-F-(anti)ev-labels}, albeit they are of course in correspondence as we discuss below \cite{mss22}. In short, (anti) eigenvalue tunneling is associated to instantons on the (non) physical sheet, producing exponentially suppressed $\sim \rme^{-\ell A/g_{\text{s}}}$ (enhanced $\sim \rme^{+\bar{\ell} A/g_{\text{s}}}$) contributions associated to action $A$. For a particular single $A$, these contributions span a two-dimensional lattice of nonperturbative contributions labeled by $(\ell|\bar{\ell})$ where the two directions are then in resonant correspondence with the suppressed (enhanced) instantons in the transseries. The general case would include an arbitrary number of resonant instanton actions; \textit{e.g.}, \cite{abs18, gs21}.

\begin{figure}
\centering
\begin{tikzpicture}[scale=0.7]
\draw[line width=1pt] (-11,5.5) -- (12,5.5);
\draw[line width=1pt] (-11,0) -- (12,0);
\draw[line width=1pt] (-10.1, 6.5) -- (-10.1, -5.5);
\node[rotate=90] at (-10.5, 2.75) {$\text{Disc}_{0}$};
\draw[line width=1pt] (3.6, 6.5) -- (3.6, -5.5);
\node[rotate=90] at (-10.5, -2.75) {$\text{Disc}_{\pi}$};
\node at (-3.3, 6) {Contour Deformations};
\node at (8, 6) {Resurgent Borel Plane};
\node at (-2.85, 4.6) {\footnotesize $\mathcal{Z}^{(0|0)(0|0)}(\abs{g_{\text{s}}}\rme^{\rmi 0^{+}})-\mathcal{Z}^{(0|0)(0|0)}(\abs{g_{\text{s}}}\rme^{\rmi 0^{-}}) \to \mathcal{Z}^{(1|0)(0|0)}(g_{\text{s}})$};
\node at (-3.2, -0.8) {\footnotesize $\mathcal{Z}^{(1|0)(0|0)}(\abs{g_{\text{s}}}\rme^{\rmi \pi^{+}})-\mathcal{Z}^{(1|0)(0|0)}(\abs{g_{\text{s}}}\rme^{\rmi \pi^{-}}) \to \mathcal{Z}^{(0|0)(1|0)}(\abs{g_{\text{s}}}\rme^{\rmi \pi}) + \cdots$};
\node at (7.2, 4.6) {\footnotesize $\text{Disc}_0 Z^{(0,0,0,0)}=\rme^{-\frac{A_1}{g_{\text{s}}}}\times$};
\node at (8.4, 3.6) {\footnotesize $\times\mathsf{S}_{(0,0,0,0)\to(1,0,0,0)}Z^{(1,0,0,0)}$};
\node at (7.3, -0.8) {\footnotesize $\text{Disc}_{\pi} Z^{(1,0,0,0)}=\rme^{-\frac{A_2-A_1}{g_{\text{s}}}}\times$};
\node at (8.4, -1.8) {\footnotesize $\times\mathsf{S}_{(1,0,0,0)\to(0,0,1,0)}Z^{(0,0,1,0)}$};
	\begin{scope}[scale=0.6, shift={({-13},{3.5})}]
		\draw[ForestGreen, line width=2pt] (-2, 0) -- (0, 0);
		\draw[ForestGreen, fill=ForestGreen] (-2,0) circle (1.1ex);
		\node at (-2, -0.7) {$a$};
		\draw[ForestGreen, fill=ForestGreen] (0,0) circle (1.1ex);
		\draw[cornellred, fill=cornellred] (3,0) circle (1.1ex);
		\node at (0, -.7) {$b$};
		\draw[cornellred, fill=cornellred, line width=1pt] (6,0) circle (1.1ex);
		\node at (6, -0.7) {$x^{\star}_2$};
		\node at (3, -0.7) {$x^{\star}_1$};
		\draw[blue, line width=1.5pt] plot [smooth, tension=0.4] coordinates{(-3, 0)(2, 0.1)(2.8,0.4)(2.9, 1)(3,2)};
		\draw[blue, line width=1.5pt] plot [smooth, tension=0.3] coordinates{(-3, 0)(2,-0.1)(2.8,-0.4)(2.9, -1)(3, -2)};
		\node at (8.2,0.4) {\Large $\xrightarrow{\boldsymbol{\epsilon\,\rightarrow\,0}}$};
		\node[blue] at (1.8, 1.1) {$\CC^{+}$};
		\node[blue] at (1.6, -1.1) {$\CC^{-}$};
	\end{scope}
	\begin{scope}[scale=0.6, shift={({-0.8},{3.5})}]
		\draw[ForestGreen, line width=2pt] (-2, 0) -- (0, 0);
		\draw[ForestGreen, fill=ForestGreen] (-2,0) circle (1.1ex);
		\node at (-2, -0.7) {$a$};
		\draw[ForestGreen, fill=ForestGreen] (0,0) circle (1.1ex);
		\draw[cornellred, fill=cornellred] (3,0) circle (1.1ex);
		\node at (0, -.7) {$b$};
		\draw[cornellred, fill=cornellred, line width=1pt] (6,0) circle (1.1ex);
		\node at (6, -0.7) {$x^{\star}_2$};
		\node at (3, -0.7) {$x^{\star}_1$};
		\draw[blue, line width=1.5pt] plot [smooth, tension=0.4] coordinates{(3, -2)(2.95, 0)(3,2)};
		\node[blue] at (2, 1) {$\CC^{\star}_1$};
	\end{scope}	
	\begin{scope}[scale=0.6, shift={({-14},{-5.4})}]
		\draw[ForestGreen, line width=2pt] (-2, 0) -- (0, 0);
		\draw[ForestGreen, fill=ForestGreen] (-2,0) circle (1.1ex);
		\node at (-2, -0.7) {$a$};
		\draw[ForestGreen, fill=ForestGreen] (0,0) circle (1.1ex);
		\draw[cornellred, fill=cornellred] (3,0) circle (1.1ex);
		\node at (0, -.7) {$b$};
		\draw[cornellred, fill=cornellred, line width=1pt] (6,0) circle (1.1ex);
		\node at (6, -0.7) {$x^{\star}_2$};
		\node at (3, -0.7) {$x^{\star}_1$};
		\draw[blue, line width=1.5pt] plot [smooth, tension=0.4] coordinates{(-1.5, -2)(-0.5, -1)(0, -0.3)(3,0)(5.8, 0.3)(6.2,1)(7,2)};
		\draw[blue, line width=1.5pt] plot [smooth, tension=0.3] coordinates{(-1.5, 2)(-0.5, 1)(0, 0.3)(3,0)(5.8, -0.3)(6.2,-1)(7,-2)};
		\node at (8.3,0.4) {\Large $\xrightarrow{\boldsymbol{\epsilon\,\rightarrow\,0}}$};
		\node[blue] at (4.8, 1.1) {$\CC^{\star}_{1, +}$};
		\node[blue] at (4.6, -1.1) {$\CC^{\star}_{1, -}$};
	\end{scope}
	\begin{scope}[scale=0.6, shift={({-1.45},{-5.4})}]
		\draw[ForestGreen, line width=2pt] (-2, 0) -- (0, 0);
		\draw[ForestGreen, fill=ForestGreen] (-2,0) circle (1.1ex);
		\node at (-2, -0.7) {$a$};
		\draw[ForestGreen, fill=ForestGreen] (0,0) circle (1.1ex);
		\draw[cornellred, fill=cornellred] (3,0) circle (1.1ex);
		\node at (0, -.7) {$b$};
		\draw[cornellred, fill=cornellred, line width=1pt] (6,0) circle (1.1ex);
		\node at (6, -0.7) {$x^{\star}_2$};
		\node at (3, -0.7) {$x^{\star}_1$};
		\draw[blue, line width=1.5pt] plot [smooth, tension=0.4] coordinates{(7, -2)(6.5, -1.2)(6,0)(6.5, 1.2)(7,2)};
		\draw[blue, line width=1.5pt] plot [smooth, tension=0.4] coordinates{(7, -2)(6.5, -1.2)(6,0)(6.5, 1.2)(7,2)};
		\draw[blue, line width=1.5pt] plot [smooth, tension=0.4] coordinates{(-1, -2)(-.5, -1.2)(0,0)(-.5, 1.2)(-1,2)};
		\node[blue] at (5.4, 1.4) {$\CC^{\star}_2$};
	\end{scope}	
	\begin{scope}[scale=0.6, shift={({12.5},{2.7})}]
\draw[line width=1.5pt, ->] (-6,0) -- (6, 0);
\draw[line width=1.5pt, ->] (0,-1.5) -- (0, 1.5);
		\draw[cornellred, fill=cornellred] (4.5,0) circle (1.2ex);
		\draw[cornellred, fill=cornellred] (-4.5,0) circle (1.2ex);
\node[black] at (4.5, -1.3) {$A_1$};
\draw[blue, line width=1.5pt] (0,0) -- (5.5, 0.7);
\draw[blue, line width=1.5pt] (0,0) -- (5.5, -0.7);
\draw[->, color=blue, line width=1.5pt] (6, -0.5) to[bend right =30] (6, 0.5);
	\node[color=blue] at (6.9, 0.6) {$\underline{\mathfrak{S}}_{0}$};
	\end{scope}	
	\begin{scope}[scale=0.6, shift={({13.8},{-6.5})}]
\draw[line width=1.5pt, ->] (-6,0) -- (6, 0);
\draw[line width=1.5pt, ->] (0,-1.5) -- (0, 1.5);
		\draw[cornellred, fill=cornellred] (-2.6,0) circle (1.2ex);
		\draw[cornellred, fill=cornellred] (2.6,0) circle (1.2ex);
\draw[orange, line width=1.5pt] (0,0) -- (-5.5, 0.7);
\draw[orange, line width=1.5pt] (0,0) -- (-5.5, -0.7);
\draw[->, color=orange, line width=1.5pt] (-6, 0.5) to[bend right =30] (-6, -0.5);
\node[black] at (-2.8, -1.3) {$A_2-A_1$};
	\node[color=orange] at (-6.9, -0.7) {$\underline{\mathfrak{S}}_{\pi}$};		
	\end{scope}
	\end{tikzpicture}
\caption{Schematic visualization of tunneling one eigenvalue from the perturbative-cut to the saddle $x_2^{\star}$, using the matrix-model potential in figure~\ref{fig:Matrix-Potential-Configuration}. First (upper row), undergoing a forward discontinuity with the perturbative configuration tunnels an eigenvalue from the perturbative-cut to $x_1^{\star}$ (here the cut of the effective potential is visualized in green; the nonperturbative saddles are shown in red; the eigenvalue steepest-descent contours are schematically plotted in blue; and $\epsilon$ denotes the offset of the argument of $g_{\text{s}}$ coming from the discontinuity). The corresponding Borel plane picture is shown on the right, where the residue attached to the action $A_1$ is picked up (we only show the singularity relevant for our transition). Second (lower row), we undergo the backwards discontinuity starting from the $(1|0)(0|0)$ configuration. Because we are employing the backwards discontinuity the phase of $g_{\text{s}}$ has shifted by $\pi$. We observe how the backwards discontinuity indeed populates the second saddle $x_2^{\star}$ in this way. Notice also the appearance of an additive perturbative term in this transition which has been studied in \cite{mss22}. Again on the right-hand side we show the Borel plane with the singularity shown in red.}
\label{fig:matrix-integration-contours}
\end{figure}
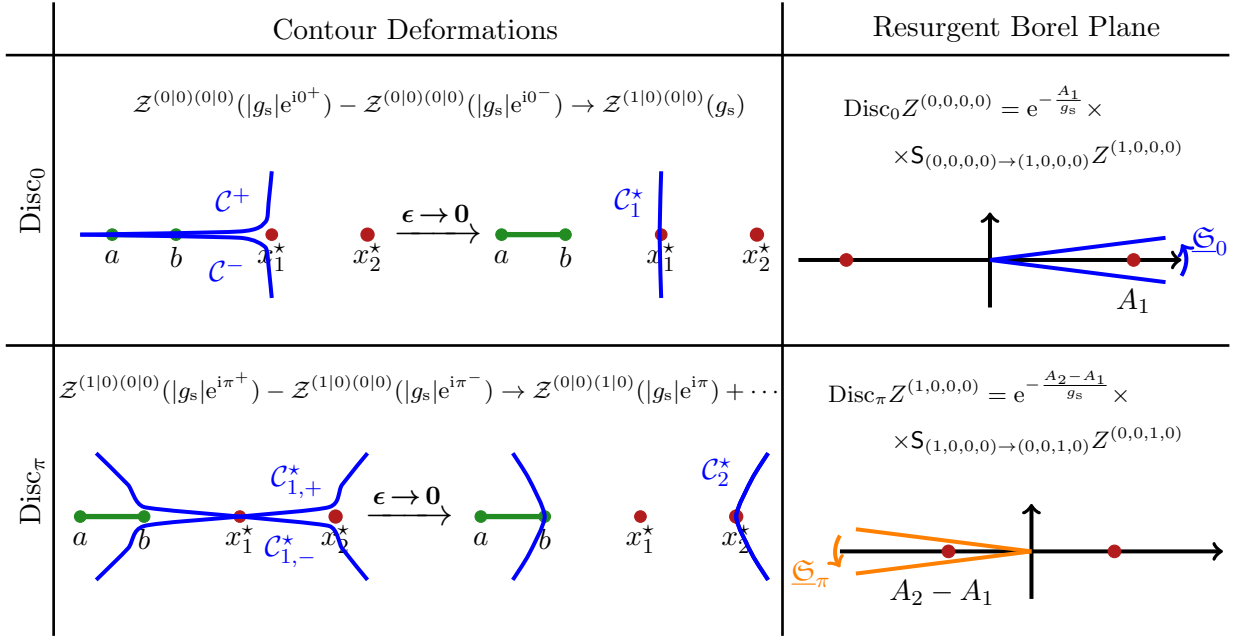

Let us start by understanding the intricacies of eigenvalue-tunneling in the setting described above. It has been thoroughly established how to populate the saddle $x_1^{\star}$ in \cite{mss22}. Namely, starting with the perturbative-sector we can undergo a forward discontinuity (this is the one at $\arg g_{\text{s}} = 0$) which tunnels eigenvalues to the saddle $x_1^{\star}$; or we can undergo a backward discontinuity (the one at $\arg g_{\text{s}} = \pi$) which tunnels anti-eigenvalues to $x_1^{\star}$ (see figure~\ref{fig:matrix-eigenvalue-tunneling-25-potential}). Interestingly, both options leave the second saddle-point $x_2^{\star}$ \textit{empty}. Given that there are only forward and backward Stokes lines, this implies that there is no \textit{direct} tunneling from the perturbative-cut to $x_2^{\star}$. On the other hand, having first populated $x_1^{\star}$, it is now possible to tunnel eigenvalues and anti-eigenvalues from there to $x_2^{\star}$. In other words, if we wish to populate $x_2^{\star}$, we first need to populate $x_1^{\star}$ and only thereafter tunnel to $x_2^{\star}$. Let us illustrate this procedure with the example of a single eigenvalue. The first step, where an eigenvalue has left the cut in favor of $x_{1}^{\star}$, is well-known \cite{msw07, mss22} (see figure~\ref{fig:matrix-integration-contours}). On the matrix model side this eigenvalue-tunneling contour-deformation reads
\begin{equation}
\mathcal{Z}^{(1|0)(0|0)}(t, g_{\text{s}}) = \frac{1}{2\pi}\, \mathcal{Z}^{(0|0)(0|0)}(t-g_{\text{s}}, g_{\text{s}}) \int_{\CC_1^{\star}}\text{d}x\, \rme^{-\frac{1}{g_{\text{s}}}V(x)}\left\langle\det\left(x-M\right)^2\right\rangle,
\end{equation}
\noindent
where we follow the notation from \cite{mss22}, and $\CC_1^{\star}$ is the steepest-descent contour associated to $x_1^{\star}$. Furthermore we have the well-studied ``bridge'' relation \cite{d91, msw07, mss22}
\begin{equation}
\label{eq:bridge-relation-1000}
\rme^{-\frac{A_1}{g_{\text{s}}}}\, \mathsf{S}_{(0,0,0,0)\to(1,0,0,0)}\, \frac{Z^{(1,0,0,0)}(t, g_{\text{s}})}{Z^{(0,0,0,0)}(t, g_{\text{s}})} \simeq \frac{\mathcal{Z}^{(1|0)(0|0)}(t, g_{\text{s}})}{\mathcal{Z}^{(0|0)(0|0)}(t, g_{\text{s}})}.
\end{equation}
\noindent
Let us now tunnel the eigenvalue sitting at $x_1^{\star}$ to the furthest saddle $x_2^{\star}$. This contour deformation occurs at the backwards discontinuity as illustrated in figure~\ref{fig:matrix-integration-contours}, and we focus only on the $(0|0)(1|0)$ contribution of the full discontinuity\footnote{Notice how the contour deformation picks up an additional, additive perturbative term. This phenomenon was discussed in \cite{mss22}. Herein we are only interested in the $(0|0)(1|0)$ contribution and we will ignore such term.}. We arrive at
\begin{equation}
\mathcal{Z}^{(0|0)(1|0)}(t, g_{\text{s}}) = \frac{1}{2\pi}\, \mathcal{Z}^{(0|0)(0|0)}(t-g_{\text{s}}, g_{\text{s}}) \int_{\CC_2^{\star}}\text{d}x\, \rme^{-\frac{1}{g_{\text{s}}}V(x)} \left\langle\det\left(x-M\right)^2\right\rangle.
\end{equation}
\noindent
On the resurgent-transseries side, the backward discontinuity of the $(1,0,0,0)$ contribution at leading exponential damping amounts to
\begin{align}
\rme^{-\frac{A_1}{g_{\text{s}}}}\, \mathsf{S}_{(0,0,0,0)\to(1,0,0,0)}\, \text{Disc}_{\pi} Z^{(1,0,0,0)} = \rme^{-\frac{A_2}{g_{\text{s}}}}\, \mathsf{S}_{(0,0,0,0)\to(1,0,0,0)}\, \mathsf{S}_{(1,0,0,0)\to(0,0,1,0)}\, Z^{(0,0,1,0)} + \cdots.
\end{align}
\noindent
Now comparing the matrix-integral calculation with the resurgent transseries, in the spirit of \cite{mss22}, we find the ``bridge'' relation
\begin{equation}
\label{eq:bridge-relation-0010}
\rme^{-\frac{A_2}{g_{\text{s}}}}\, \mathsf{S}_{(0,0,0,0)\to(1,0,0,0)}\, \mathsf{S}_{(1,0,0,0)\to(0,0,1,0)}\, \frac{Z^{(0,0,1,0)}(t, g_{\text{s}})}{Z^{(0,0,0,0)}(t, g_{\text{s}})} \simeq \frac{\mathcal{Z}^{(0|0)(1|0)}(t, g_{\text{s}})}{\mathcal{Z}^{(0|0)(0|0)}(t, g_{\text{s}})}.
\end{equation}
\noindent
It is interesting to compare this result with the one for the $(1,0,0,0)$ sector in \eqref{eq:bridge-relation-1000}. At first one might have expected the Borel residue $\mathsf{S}_{(0,0,0,0)\to(0,0,1,0)}$ to appear in \eqref{eq:bridge-relation-0010}---but this Borel residue actually vanishes as there is no direct tunneling between the cut and the $x_2^{\star}$ saddle. This explains the appearance of a more complicated combination of Borel residues in \eqref{eq:bridge-relation-0010} (as we have to walk two ``resurgent steps'' to arrive at the $(0,0,1,0)$ sector). On the other hand it is interesting to note how the $(0,0,1,0)$ transseries sector is still computed around the nonperturbative saddle $x_2^{\star}$. In fact, only the structure of its Borel residue pre-factor has changed.

We are now in a position to briefly outline the full tunneling mechanics for the example of the matrix-model potential depicted in figure~\ref{fig:Matrix-Potential-Configuration}. Eigenvalues (anti-eigenvalues) sitting in the perturbative cut tunnel to the nonperturbative saddle $x_1^{\star}$ at the forward (backward) discontinuity. This is exactly the scenario outlined in \cite{mss22}. Now, including the second nonperturbative saddle-point $x_2^{\star}$, this picture needs generalization. Tunneling directly from the cut to $x_2^{\star}$ is impossible, as outlined above. On the other hand tunneling between the two nonperturbative saddles can happen, albeit with a twist: now eigenvalue tunneling occurs at the \textit{backwards} discontinuity, whereas anti-eigenvalues tunnel at the \textit{forward} discontinuity. This is illustrated in figure~\ref{fig:matrix-eigenvalue-tunneling-25-potential}. 

\definecolor{deepsaffron}{rgb}{1.0, 0.8, 0.6}
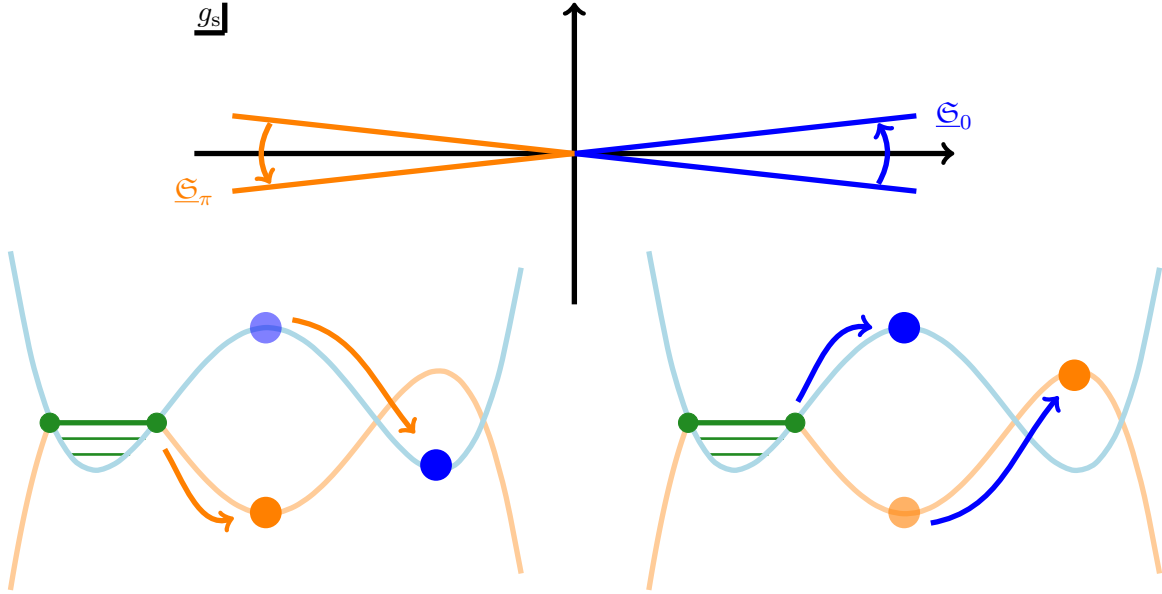
\begin{figure}
\centering
	\begin{tikzpicture}
	\draw[->, line width=2pt] (-5,0) -- (5, 0);
	\draw[->, line width=2pt] (0,-2) -- (0, 2);
	\draw[line width=2pt] (-5, 1.6) -- (-4.6, 1.6);
	\draw[line width=2pt] (-4.6, 1.6) -- (-4.6, 2);
	\node at (-4.8, 1.8) {$g_{\text{s}}$};
	\draw[blue, line width=2pt] (0,0) -- (4.5, -0.5);
	\draw[blue, line width=2pt] (0,0) -- (4.5, 0.5);
	\draw[->, blue, line width=2pt] (4, -0.4) to[bend right =30] (4, 0.4);
	\node[blue] at (5, 0.5) {$\underline{\mathfrak{S}}_0$};
	\draw[color=orange, line width=2pt] (0,0) -- (-4.5, -0.5);
	\draw[color=orange, line width=2pt] (0,0) -- (-4.5, 0.5);
	\draw[->, color=orange, line width=2pt] (-4, 0.4) to[bend right =30] (-4, -0.4);
	\node[color=orange] at (-5, -0.5) {$\underline{\mathfrak{S}}_{\pi}$};
	\begin{scope}[scale=0.7, shift={({-9},{-6})}]
	\draw[color=ForestGreen, line width=2pt] (-0.88,0.9) -- (1.15,0.9);
	\draw[color=ForestGreen, line width=1pt] (-0.7,0.6) -- (0.94,0.6);
	\draw[color=ForestGreen, line width=1pt] (-0.51,0.3) -- (0.65,0.3);
	 \draw[scale=2, domain=0.6:4, smooth, variable=\x, deepsaffron, line width=2pt] plot ({\x}, {-2*1.04*\x*\x + 2*0.645*\x*\x*\x - 0.2*\x*\x*\x*\x+0.94});
	 \draw[scale=2, domain=-0.8:-0.44, smooth, variable=\x, deepsaffron, line width=2pt] plot ({\x}, {-2*1.04*\x*\x + 2*0.645*\x*\x*\x - 0.2*\x*\x*\x*\x+0.94});
	 \draw[scale=2, domain=-0.8:4, smooth, variable=\x, LightBlue, line width=2pt] plot ({\x}, {2*1.04*\x*\x - 2*0.645*\x*\x*\x + 0.2*\x*\x*\x*\x});
	 \fill[orange, line width=2pt] (3.2,-0.8) circle (1.8ex);
	 \draw[orange, line width=2pt, ->] (1.3, 0.4) to[out=300, in=200] (2.6, -0.9);
	 \fill[blue, line width=2pt] (6.4,0.1) circle (1.8ex);
	 \fill[blue, line width=2pt, opacity=0.5] (3.2,2.7) circle (1.8ex);
	 \draw[orange, line width=2pt, ->] (3.7, 2.85) to[out=350, in=130] (6, 0.7);
	\draw[ForestGreen, fill=ForestGreen] (-0.86,0.9) circle (1.1ex);
	 \draw[ForestGreen, fill=ForestGreen] (1.15,0.9) circle (1.1ex);
	\end{scope}
	\begin{scope}[scale=0.7, shift={({3},{-6})}]
	\draw[color=ForestGreen, line width=2pt] (-0.88,0.9) -- (1.15,0.9);
	\draw[color=ForestGreen, line width=1pt] (-0.7,0.6) -- (0.94,0.6);
	\draw[color=ForestGreen, line width=1pt] (-0.51,0.3) -- (0.65,0.3);
	 \draw[scale=2, domain=0.6:4, smooth, variable=\x, deepsaffron, line width=2pt] plot ({\x}, {-2*1.04*\x*\x + 2*0.645*\x*\x*\x - 0.2*\x*\x*\x*\x+0.94});
	 \draw[scale=2, domain=-0.8:-0.44, smooth, variable=\x, deepsaffron, line width=2pt] plot ({\x}, {-2*1.04*\x*\x + 2*0.645*\x*\x*\x - 0.2*\x*\x*\x*\x+0.94});
	 \draw[scale=2, domain=-0.8:4, smooth, variable=\x, LightBlue, line width=2pt] plot ({\x}, {2*1.04*\x*\x - 2*0.645*\x*\x*\x + 0.2*\x*\x*\x*\x});
	 \fill[orange, line width=2pt, opacity=0.6] (3.2,-0.8) circle (1.8ex);
	 \draw[blue, line width=2pt, ->] (1.2, 1.3) to[out=60, in=170] (2.6, 2.7);
	 \fill[orange, line width=2pt] (6.4,1.8) circle (1.8ex);
	 \fill[blue, line width=2pt] (3.2,2.7) circle (1.8ex);
	 \draw[blue, line width=2pt, ->] (3.7, -1) to[out=10, in=230] (6.1, 1.4);
	\draw[ForestGreen, fill=ForestGreen] (-0.86,0.9) circle (1.1ex);
	 \draw[ForestGreen, fill=ForestGreen] (1.15,0.9) circle (1.1ex);
	\end{scope}
	\end{tikzpicture}
\caption{Schematic visualization of the mechanics of (anti) eigenvalue tunneling, for the holomorphic effective potential shown in figure~\ref{fig:Matrix-Potential-Configuration}. The physical sheet is plotted in blue and the non-physical one is orange, while the cut if shown in green. Whereas the tunneling to the saddle closest to the cut ($x_1^{\star}$) is exactly the same as discussed in \cite{mss22} the picture needs generalization for the second saddle ($x_2^{\star}$), which cannot be directly reached from the perturbative cut. Instead, it gets populated by tunneling starting from the closer nonperturbative saddle ($x_1^{\star}$). This comes with an inversion relative to the standard picture, as now eigenvalues (blue) tunnel at the \textit{backward} discontinuity while anti-eigenvalues (orange) tunnel at the \textit{forward} one.}
\label{fig:matrix-eigenvalue-tunneling-25-potential}
\end{figure}

Armed with the full tunneling-mechanics for the doubly-pinched case, let us expand on the above example with the (novel) calculation of the $(1|1)(1|0)$ free-energy contribution (this will add to our upcoming non-trivial checks). Following \cite{mss22} and the arguments outlined above we arrive at this sector by undergoing a forward discontinuity, picking up the $(1|0)(0|0)$ contribution, and then undergoing a backwards discontinuity. This result contains the $(0|1)(1|0)$ contribution, which we pick out. Then we are left to undergo a final forward Stokes transition to reach $(1|1)(1|0)$, and this will be our focus in the following. Carrying out this procedure carefully for both resurgent-transseries and matrix-integral quantities described above, yields the ``bridge'' relation between both as (we refer the reader to \cite{mss22} for details on this type of calculations)
\begin{align}
\label{eq:bridge-relation-1110}
\rme^{-\frac{A_2}{g_{\text{s}}}}\, \mathsf{S}_{(0,0,0,0)\to (1,0,0,0)}\, \mathsf{S}_{(1,0,0,0)\to(0,1,1,0)}\, \mathsf{S}_{(0,1,1,0)\to(1,1,1,0)}\, \frac{Z^{(1,1,1,0)}(t, g_{\text{s}})}{Z^{(0,0, 0,0)}(t, g_{\text{s}})} \simeq \frac{\mathcal{Z}^{(1|1)(1|0)} (t, g_{\text{s}})}{\mathcal{Z}^{(0|0)(0|0)} (t, g_{\text{s}})}.
\end{align}
\noindent
Evaluation of the above right-hand side requires solving the triple integral (which we shall do up to order $\sqrt{g_{\text{s}}}$) \cite{mss22} 
\newcommand{\pvint}{\,\text{PV}\hspace{-4pt}\int}
\begin{align}
\frac{\mathcal{Z}^{(1|1)(1|0)} (t, g_{\text{s}})}{\mathcal{Z}^{(0|0)(0|0)} (t, g_{\text{s}})}  &= \frac{1}{2}\, \frac{\mathcal{Z}^{(0|0)(0|0)} (t-g_{\text{s}}, g_{\text{s}})}{\mathcal{Z}^{(0|0)(0|0)} (t, g_{\text{s}})}\, \pvint_{\CC_{1}^{\text{res}} + \CC^{\star}_1} \frac{\text{d}x_1}{2\pi}\, \pvint_{\CC^{\star}_2} \frac{\text{d}x_2}{2\pi}\, \pvint_{\bar{\CC}^{\star}_1} \frac{\text{d}\bar{x}_1}{2\pi}\, \times \\
&
\hspace{-50pt}
\times \frac{\left(x_1-x_2\right)^2}{\left(x_1-\bar{x}_1\right)^2 \left(x_2-\bar{x}_1\right)^2}\, \rme^{- \frac{1}{g_{\text{s}}} \left( V(x_1) + V(x_2) - V(\bar{x}_1) \right)} \ev{\frac{\det \left(x_1-M\right)^{2} \det \left(x_2-M\right)^{2}}{\det \left(\bar{x}_1-M\right)^{2}}}_{N-1}, \nonumber
\end{align}
\noindent
where the contour $\CC_1^{\star}$ ($\CC_2^{\star}$) is the eigenvalue steepest-descent contour associated to $x_1^{\star}$ ($x_2^{\star}$), the contour $\CC_1^{\text{res}}$ is the residue contour connecting the saddle-point $x_1^{\star}$ to the positive endpoint of the cut, $b$, and the contour $\bar{\CC}_1^{\star}$ is the anti-eigenvalue steepest-descent contour associated with $x_1^{\star}$. For details on the contours and on the following calculation we must again refer the reader to \cite{mss22}. Let us observe that the lowest $g_{\text{s}}$-order of the $(1,1,1,0)$ free-energy sector appears at order $\sqrt{g_{\text{s}}}$ in the above calculation\footnote{Note how for the $(1,1,1,0)$ free-energy sector the overall lowest-appearing order is $\sqrt{g_{\text{s}}}$, whereas for the matrix-integral contribution $(1|1)(1|0)$ the lowest-order appears at $1/\sqrt{g_{\text{s}}}$. This order is assembled by the product of the $(1|1)(0|0)$ and $(0|0)(1|0)$ contributions---sectors which have been calculated in \cite{mss22}. Although this is a good check on our earlier results, it yields no new information. Therefore we can cancel those terms on both sides of \eqref{eq:bridge-relation-1110}.}. Isolating this contribution we find the result\footnote{Amplitude definitions are in formula (3.5) of \cite{mss22}, where $\mathcal{A}_0(x,\bar{x})=2A_{0;2}(x,x)+2A_{0;2}(\bar{x}, \bar{x})-4A_{0;2}(x,\bar{x})$.}
\begin{align}
\rme^{-\frac{A_2}{g_{\text{s}}}}\, \mathsf{S}_{(0,0,0,0)\to (1,0,0,0)}\, \mathsf{S}_{(1,0,0,0)\to(0,1,1,0)}\, \mathsf{S}_{(0,1,1,0)\to(1,1,1,0)}\, F^{(1,1,1,0)} &\simeq \\
&
\hspace{-300pt}
\simeq\frac{\mathcal{Z}^{(1|1)(1|0)} (t, g_{\text{s}})}{\mathcal{Z}^{(0|0)(0|0)} (t, g_{\text{s}})} - \frac{\mathcal{Z}^{(1|1)(0|0)} (t, g_{\text{s}})}{\mathcal{Z}^{(0|0)(0|0)} (t, g_{\text{s}})}\, \frac{\mathcal{Z}^{(0|0)(1|0)} (t, g_{\text{s}})}{\mathcal{Z}^{(0|0)(0|0)} (t, g_{\text{s}})} = \frac{\rmi}{(2\pi)^{3/2}}\, \rme^{-\frac{1}{g_{\text{s}}} \left( V_{\text{eff}}(x^{\star}_2)-V_{\text{eff}}(b) \right)} \sqrt{\frac{ g_{\text{s}}}{V^{\prime\prime}_{\text{h,eff}}(x_2^{\star})}} \times \nonumber \\
&
\hspace{-300pt}
\times \frac{b-a}{8 \left(x_2^{\star}-a\right)\left(x_2^{\star}-b\right)} \int_{b}^{x_1^{\star}} \text{d}\bar{x}_1\, \Bigg\lbrace \partial_{x_1} \left. \left( \frac{(x_2^{\star}-x_1)^2}{(x_2^{\star}-\bar{x}_1)^2}\, \rme^{\mathcal{A}_{0}(x_1,\bar{x}_1)+4A_{0;2}(x_1,x^{\star}_2)-4A_{0;2}(\bar{x}_1,x^{\star}_2)}\right) \right|_{x_1=\bar{x}_1} \Bigg\rbrace + o(g_{\text{s}}), \nonumber
\end{align}
\noindent
which we can evaluate to yield
\begin{align}
\label{eq:result-1110-sector}
\rme^{-\frac{A_2}{g_{\text{s}}}}\, \mathsf{S}_{(0,0,0,0)\to (1,0,0,0)}\, \mathsf{S}_{(1,0,0,0)\to(0,1,1,0)}\, \mathsf{S}_{(0,1,1,0)\to(1,1,1,0)}\, F^{(1,1,1,0)} &\simeq \frac{\rmi}{(2\pi)^{3/2}}\, \rme^{-\frac{1}{g_{\text{s}}} \left( V_{\text{eff}}(x^{\star}_2)-V_{\text{eff}}(b) \right)} \times \nonumber\\
&
\hspace{-300pt}
\times \sqrt{\frac{ g_{\text{s}}}{V^{\prime\prime}_{\text{h,eff}}(x_2^{\star})}}\, \frac{1}{\left(x_2^{\star}-a\right)\left(x_2^{\star}-b\right)}\, \frac{b-a}{8}\, \log \Bigg\lbrace \frac{\left(2 \sqrt{(x_1^{\star}-a) (x_1^{\star}-b)}+a+b-2 x_1^{\star}\right)^2}{\left(a-b\right)^4 \left(x_2^{\star}-x_1^{\star}\right)^2} \times \\
&
\hspace{-300pt}
\times \left( \left(a+b\right)\left(x_2^{\star}+x_1^{\star}\right) - 2 \left(\sqrt{(a-x_2^{\star}) (a-x_1^{\star}) (x_2^{\star}-b) (x_1^{\star}-b)}+x_2^{\star} x_1^{\star} + a b \right) \right)^2 \Bigg\rbrace + o(g_{\text{s}}).\nonumber
\end{align}
\noindent
One interesting feature of this result is the appearance of a logarithmic contribution which is \text{not} directly associated to resonance (\textit{e.g.}, compare with the logarithmic contributions in \cite{mss22}). Furthermore, note how the pre-factor of the logarithm is just the usual $(0|0)(1|0)$ contribution.

Many other sectors have been recently computed in \cite{mss22}, to where we refer the reader for more details on the match between transseries and matrix-integrals nonperturbative data.

\paragraph{Double-Scaling Limit:}

In order to reach minimal-string or JT-gravity formulae, the above (off-critical) matrix model results have to be double-scaled (\textit{e.g.}, see the recent discussions in \cite{gs21}). In practice one starts from a one-cut multi-pinched spectral curve---as in \eqref{eq:hyperelliptic-spectral-curve} with $s=1$ or, equivalently, as in \eqref{eq:1-cut-matrix-sc}---and zooms-in on the endpoint of the cut $a_1$, effectively producing a single cut along $\NCC = (-\infty, a_1]$ (compare with formula \eqref{eq:dsl-spectral-curve-MS}). On what concerns nonperturbative sectors, this procedure was described in \cite{gs21, mss22} but which we next briefly recall. Starting from the one-cut setting, introduce a scaling parameter $\upvarepsilon$ and take\footnote{This scaling is simple: the one-cut double-scaled curve $y(x) = M(x) \sqrt{x-a}$ on $\NCC=(-\infty, a]$ may be rewritten as on a finite-cut $\NCC_{\text{finite}}=[a-\upvarepsilon,a]$ by replacing $x$ with $x+\frac{(x-a)^2}{\upvarepsilon}$. The limit $\upvarepsilon\to+\infty$ then recovers $\NCC_{\text{finite}} \to \NCC$.}
\begin{align}
a_1 &\to a_{\text{dsl}}, & a_2 \to a_{\text{dsl}}-\upvarepsilon, \\
M(x) &\to \frac{M_{\text{dsl}}(x)}{\sqrt{\upvarepsilon}}, & M^{\prime}(x) \to \frac{M_{\text{dsl}}^{\prime}(x)}{\sqrt{\upvarepsilon}},
\end{align}
\noindent
in the limit $\upvarepsilon\to+\infty$. As we only address double-scaled quantities from now on, we drop the ``dsl'' labeling of all quantities. This leads to the multi-pinched double-scaled spectral curve
\begin{equation}
\label{eq:general-double-scaled-sc}
y(x) = M(x)\sqrt{x-a},
\end{equation}
\noindent
with $k$ pinches---which we take located at $x_n^{\star}$ with $n=1,\ldots,k$.

Consider the free energy which is computed from that spectral curve\footnote{This free energy may be computed from the spectral curve for example via the topological recursion \cite{eo07a}.} where $g_{\text{s}}$ is the expansion parameter now in the double-scaled theory. Focus on the nonperturbative contributions to that free-energy when only one pinch, say, $x^{\star}_1$, of the spectral curve \eqref{eq:general-double-scaled-sc} is populated. These are transseries sectors of the type $(n_1,m_1,0,\ldots,0)$ which will be associated to (anti) eigenvalue configurations of the type $(\ell_1|\bar{\ell}_1)(0|\bar{0})\cdots(0|\bar{0})$, with some pre-factor of Borel residues that depends strongly on the resurgent path under consideration (as outlined above). But even if the Borel residues (resurgent path) that one has to employ in order to arrive at a resurgent sector changes, the (anti) eigenvalue configuration associated to the sector will not change---we are still computing the same nonperturbative contributions, just with different Borel residues attached to it. What this effectively implies is that for all practical purposes this case may be treated as if a single-pinch problem of the type in \cite{mss22}, ignoring all other zero-labelings, and we shall directly use the results therein in the following. These results yield the double-scaled formulae that we present below, where we will keep the single-pinch labeling of \cite{mss22} for simplicity of notation. 

Nonperturbative contributions from the saddle $x^{\star}_1$ span the two-dimensional lattice $(\ell_1|\bar{\ell}_1)$, and we begin at its edges spelling out partition-function ratios---yielding combinations of free-energy sectors in the double-scaled theory. For the $(\ell_1|0)$ and $(0|\bar{\ell}_1)$ sectors associated to the saddle point $x^{\star}_1$ we find (we compute here solely the lowest-appearing order in $g_{\text{s}}$)
\begin{align}
\label{eq:dsl-l0-sector}
\frac{\mathcal{Z}^{(\ell_1|0)}(g_{\text{s}})}{\mathcal{Z}^{(0|0)}(g_{\text{s}})} &\simeq \frac{G_{2}(\ell_1+1)}{\left(2\pi\right)^{\ell_1/2}}\, \rme^{-\frac{\ell_1}{g_{\text{s}}} \left(V_{\text{h;eff}}(x^{\star}_1)-V_{\text{h;eff}}(a)\right)} \left(\frac{g_{\text{s}}}{16 M^{\prime}(x^{\star}_1) \left(x^{\star}_1-a\right)^{5/2}}\right)^{\frac{\ell_1^2}{2}} + \cdots, \\
\label{eq:dsl-0lb-sector}
\frac{\mathcal{Z}^{(0|\bar{\ell}_1)}(g_{\text{s}})}{\mathcal{Z}^{(0|0)}(g_{\text{s}})} &\simeq \frac{G_{2}(\bar{\ell}_1+1)}{\left(2\pi\right)^{\bar{\ell}_1/2}}\, \rme^{\frac{\bar{\ell}_1}{g_{\text{s}}} \left(V_{\text{h;eff}}(x^{\star}_1)-V_{\text{h;eff}}(a)\right)} \left(-\frac{g_{\text{s}}}{16 M^{\prime}(x^{\star}_1) \left(x^{\star}_1-a\right)^{5/2}}\right)^{\frac{\bar{\ell}_1^2}{2}} + \cdots.
\end{align}
\noindent
Similarly, we next obtain the $(1|1)$ contribution around the saddle $x^{\star}_1$, finding
\begin{align}
\label{eq:dsl-11-sector}
&\frac{\mathcal{Z}^{(1|1)}(g_{\text{s}})}{\mathcal{Z}^{(0|0)}(g_{\text{s}})} \simeq \frac{1}{g_{\text{s}}}\, \frac{\rmi}{2\pi} \left(V_{\text{h;eff}}(x_1^{\star}) - V_{\text{h;eff}}(a)\right) + 0 + g_{\text{s}}\, \frac{\rmi}{24\pi}\, \Bigg\lbrace\frac{3 \left(a-x^{\star}_{1}\right) M'(a) + M(a)}{M(a)^2 \left(x^{\star}_{1}-a\right)^{3/2}} - \\
&- \frac{4 \left(a-x^{\star}_{1}\right)^2 M''(x^{\star}_{1})^2 - 2 \left(a-x^{\star}_{1}\right) M'(x^{\star}_{1}) \left\{ 2 \left(a-x^{\star}_{1}\right) M^{(3)}(x^{\star}_{1}) + M''(x^{\star}_{1}) \right\} + 19 M'(x^{\star}_{1})^2}{8 \left(x^{\star}_{1}-a\right)^{5/2}  M'(x^{\star}_{1})^3} \Bigg\rbrace + o(g_{\text{s}}^2). \nonumber
\end{align} 
\noindent
Notice how this contribution has no associated exponential transmonomials, as the resonant actions annihilate each other. Still, \textit{this is} a nonperturbative contribution; see, \textit{e.g.}, \cite{gikm10, asv11, as13, abs18, gs21, mss22}. Last but not least we turn to the $(2|1)$ configuration around the saddle $x^{\star}_1$, where we find
\begin{align}
\label{eq:dsl-21-sector}
\frac{\mathcal{Z}^{(2|1)} (g_{\text{s}})}{\mathcal{Z}^{(0|0)} (g_{\text{s}})} - \frac{\mathcal{Z}^{(1|0)} (g_{\text{s}})}{\mathcal{Z}^{(0|0)} (g_{\text{s}})}\, \frac{\mathcal{Z}^{(1|1)} (g_{\text{s}})}{\mathcal{Z}^{(0|0)} (g_{\text{s}})} &\simeq \rme^{-\frac{1}{g_{\text{s}}} \left(V_{\text{h;eff}}(x_1^{\star}) - V_{\text{h;eff}}(a)\right)} \times \\
&
\hspace{-100pt}
\times \frac{\rmi}{8 \left(2\pi\right)^2}\, \sqrt{\frac{2\pi g_{\text{s}}}{M^{\prime}(x^{\star}_1) \left(x^{\star}_1-a\right)^{5/2}}} \left\{ 2 \gamma_{\text{E}} + \log \left( \frac{2^8}{g_{\text{s}}^2}\, M^{\prime}(x^{\star}_1)^2 \left(x^{\star}_1-a\right)^{5} \right) \right\} + o(g_{\text{s}}^{3/2}). \nonumber
\end{align}
\noindent
Note that we are here computing a special combination of partition-function ratios. That this combination is the interesting one to consider is related to the very special resonant structure that spectral curves of the type \eqref{eq:general-double-scaled-sc} inherently possess. Further details can be found in \cite{mss22} and we will get back to this in the upcoming subsection~\ref{subsec:22k-1minstring} .

Having understood the case where just one of the pinches (herein $x_1^{\star}$) is populated, let us next move-on to also populate a second pinch, say $x_2^{\star}$---as in the off-critical matrix-model calculation we started-off with. In this case, we find that \eqref{eq:result-1110-sector} double-scales to
\begin{align}
\label{eq:dsl-result-1110-sector}
\frac{\mathcal{Z}^{(1|1)(1|0)} (g_{\text{s}})}{\mathcal{Z}^{(0|0)(0|0)} (g_{\text{s}})} - \frac{\mathcal{Z}^{(1|1)(0|0)} (g_{\text{s}})}{\mathcal{Z}^{(0|0)(0|0)} (g_{\text{s}})}\,  \frac{\mathcal{Z}^{(0|0)(1|0)} (g_{\text{s}})}{\mathcal{Z}^{(0|0)(0|0)} (g_{\text{s}})} &\simeq \frac{\rmi}{(2\pi)^{3/2}}\, \rme^{-\frac{1}{g_{\text{s}}} \left(V_{\text{eff}}(x^{\star}_2)-V_{\text{eff}}(b) \right)} \times \\
&
\hspace{-150pt}
\times \sqrt{\frac{ g_{\text{s}}}{16 M^{\prime}(x_2^{\star})}}\, \frac{1}{\left(x_2^{\star}-a\right)^{\frac{5}{4}}}\, \log \left\lbrace \frac{2a-x_2^{\star}+2\sqrt{\left(x_2^{\star}-a\right) \left(x_1^{\star}-a\right)}-x_1^{\star}}{\left(x_2^{\star}-x_1^{\star}\right)} \right\rbrace + o(g_{\text{s}}). \nonumber
\end{align}

The lingering question at this stage is: can the above results be reproduced via BCFT?

\subsection{$(2,2k-1)$ Minimal Strings: BCFT Matches the Matrix Model}\label{subsec:22k-1minstring}

Let us now use the double-scaled one-matrix model results derived above and apply them to the $(2,2k-1)$ minimal-string spectral curve \eqref{eq:dsl-spectral-curve-MS}. When comparing minimal-string and (double scaled) matrix-model results, it is well known how tunneled eigenvalues correspond to ZZ-branes; \textit{e.g.}, \cite{d92, akk03, ss03, hhikkmt04, st04, iky05, msw07, emms22b}. What we can now establish by direct comparison is that the tunneled anti-eigenvalues in \cite{mss22} correspond to the negative-tension ZZ-branes proposed in section~\ref{sec:ZZminimalstring}. This is established by comparing BCFT results for different free-energy nonperturbative sectors from section~\ref{sec:ZZminimalstring} against our present double-scaled matrix-model formulae.

We start by spelling out formulae \eqref{eq:dsl-l0-sector}-\eqref{eq:dsl-0lb-sector}-\eqref{eq:dsl-11-sector}-\eqref{eq:dsl-21-sector}-\eqref{eq:dsl-result-1110-sector} for the specific case of the $(2,2k-1)$ minimal string. The instanton action reads\footnote{Also recall that there is a minus-sign difference between our conventions and the ones used in \cite{gs21}.} \cite{ss03, st04}
\begin{equation}
\label{eq:22km1MinimalStringActions}
A_{n,k} = (-1)^{k+n+1} \left(\frac{1}{2k+1}+\frac{1}{2k-3}\right) \sin\frac{2\pi n}{2k-1}.
\end{equation}
\noindent
Besides the minus-sign difference which has already been addressed in footnote~\ref{footnote:ExplainingMinusSign1}, notice the difference of a factor of $2$ to the action computed from BCFT in formula \eqref{eq:22km1-bcft-Instanton-Action}. This factor goes back to the the fact that the BCFT calculation considers full cycles on the moduli space of FZZT branes while the matrix model and string equation calculations work with half cycles. Upon comparison of the various results this factor needs to be taken into account\footnote{This difference of conventions can be resolved for example by rescaling the spectral curve appropriately.}. Furthermore a useful identity to use is 
\be
M^{\prime}_k(x_{n,k}^{\star}) \left(x_{n,k}^{\star}-a\right)^{5/2} = (-1)^{k+n} \left(2k-1\right) \tan \frac{n\pi}{2k-1}\, \sin^2 \frac{n\pi}{2k-1}.
\ee
\noindent
For the multiple ZZ and negative-tension ZZ contributions, we hence find to leading order
\begin{align}
\label{eq:MS-l0-sector}
\frac{\mathcal{Z}^{(\ell|0)}_{n,k}(g_{\text{s}})}{\mathcal{Z}^{(0|0)}_{n,k}(g_{\text{s}})} &\simeq \frac{G_{2} \left(\ell+1\right)}{\left(2\pi\right)^{\ell/2}}\, \rme^{-\frac{\ell}{g_{\text{s}}} A_{n,k}} \left( g_{\text{s}}\, \frac{(-1)^{n+k} \cot\frac{n\pi}{2k-1}}{16 \left(2k-1\right) \sin^2\frac{n\pi}{2k-1}}\right)^{\frac{\ell^2}{2}} + \cdots,\\
\label{eq:MS-0lb-sector}
\frac{\mathcal{Z}^{(0|\bar{\ell})}_{n,k}(g_{\text{s}})}{\mathcal{Z}^{(0|0)}_{n,k}(g_{\text{s}})} &\simeq \frac{G_{2} \left(\bar{\ell}+1\right)}{\left(2\pi\right)^{\bar{\ell}/2}}\, \rme^{\frac{\bar{\ell}}{g_{\text{s}}} A_{n,k}} \left(g_{\text{s}}\, \frac{(-1)^{n+k+1} \cot\frac{n\pi}{2k-1}}{16 \left(2k-1\right) \sin^2\frac{n\pi}{2k-1}}\right)^{\frac{\bar{\ell}^2}{2}} + \cdots.
\end{align}
\noindent
Up to the factor of $2$ we just alluded to (which appears inside the brackets in the formulae above), these expressions \textit{exactly match} the corresponding BCFT calculations specialized to the $(2,2k-1)$ case, which we have given in formulae \eqref{eq:22km1multipleinst} and \eqref{eq:22km1multipleantiinst}. Proceeding with the $(1|1)$ sector we find next (and which we now compute to one higher order than was possible via BCFT---recall the comment below \eqref{eq:22km1-cft-result-11})
\begin{align}
\label{eq:MS-11-sector}
\frac{\mathcal{Z}^{(1|1)}(g_{\text{s}})}{\mathcal{Z}^{(0|0)}(g_{\text{s}})} &= \frac{1}{g_{\text{s}}}\, \frac{\rmi}{2\pi}\, A_{n,k} + 0 + \nonumber \\
&
+ g_{\text{s}}\, \frac{\rmi (-1)^k}{192 \pi \left(2k-1\right) \sin^3 \left(\frac{2\pi n}{2k-1}\right)} \left\{ 48 (-1)^n \cos \left(\frac{2\pi n}{2k-1}\right) - 8 \cos \left(\frac{3\pi n}{2k-1}\right) + \right. \nonumber \\
&
+ 8 \cos \left(\frac{\pi n}{2k-1}\right) \left( k \left(k-1\right) \left[ \cos \left(\frac{4\pi n}{2k-1}\right) - 1 \right] - 3 \right) + \\
&
\left. 
+ (-1)^{n+1} \left( \left[ 4 k \left(k-1\right) - 7 \right] \cos \left(\frac{4\pi n}{2k-1}\right) - 4 k \left(k-1\right) - 25 \right) \right\} + o(g_{\text{s}}^2). \nonumber
\end{align}
\noindent
For this sector we now need to compare against the BCFT result in \eqref{eq:22km1-cft-result-11}, and to the first two lowest orders---where again we find \textit{precise agreement}. The next sector we consider is $(2|1)$, where we now have 
\begin{align}
\label{eq:MS-21-sector}
&\quad\quad\frac{\mathcal{Z}^{(2|1)}_{n,k} (g_{\text{s}})}{\mathcal{Z}^{(0|0)}_{n,k} (g_{\text{s}})} - \frac{\mathcal{Z}^{(1|0)}_{n,k} (g_{\text{s}})}{\mathcal{Z}^{(0|0)}_{n,k} (g_{\text{s}})}\, \frac{\mathcal{Z}^{(1|1)}_{n,k} (g_{\text{s}})}{\mathcal{Z}^{(0|0)}_{n,k} (g_{\text{s}})} \simeq \rme^{-\frac{1}{g_{\text{s}}} A_{n,k}}\times\\
&\times \frac{\rmi}{16\pi}\, \sqrt{\frac{\,g_{\text{s}}(-1)^{n+k}\cot\frac{n\pi}{2k-1}}{2\pi (2k-1)\sin^2\frac{n\pi}{2k-1}}}\, \left\{ 2 \gamma_{\text{E}} + \log \left( \frac{2^8}{g_{\text{s}}^2}\, \frac{(2k-1)^2\sin^4\frac{n\pi}{2k-1}}{\cot^2\frac{n\pi}{2k-1}}\right) \right\} + o(g_{\text{s}}^{3/2}).\nonumber
\end{align}
\noindent
Once again this result \textit{precisely matches} its BCFT counterpart in \eqref{eq:BCFT21final} (once one correctly adjusts a factor of $2$ for every occurrence of the action in the above formula). Last but not least the same holds for the $(1|1)(1|0)$ contribution. For the saddle-points $x_n^{\star}$ and $x_m^{\star}$, with $x_n^{\star}$ further away from the cut, we explicitly find the relation
\begin{align}
\frac{\mathcal{Z}^{(1|1)(1|0)} (g_{\text{s}})}{\mathcal{Z}^{(0|0)(0|0)} (g_{\text{s}})} - \frac{\mathcal{Z}^{(1|1)(0|0)} (g_{\text{s}})}{\mathcal{Z}^{(0|0)(0|0)} (g_{\text{s}})}\, \frac{\mathcal{Z}^{(0|0)(1|0)} (g_{\text{s}})}{\mathcal{Z}^{(0|0)(0|0)} (g_{\text{s}})} &\simeq \frac{\rmi}{(2\pi)^{3/2}}\, \rme^{-\frac{1}{g_{\text{s}}}\, A_{n,k}} \times \\
&
\hspace{-100pt}
\times \sqrt{\frac{g_{\text{s}} \left(-1\right)^{k+n} \cot \frac{n\pi}{2k-1}}{16 \left (2k-1\right) \sin^2 \frac{n\pi}{2k-1}}}\, \log \abs{ \frac{\sin \left(\frac{\pi m}{2k-1}\right) - \sin \left(\frac{\pi n}{2k-1}\right)}{\sin \left(\frac{\pi m}{2k-1}\right) + \sin \left(\frac{\pi n}{2k-1}\right)} } + o(g_{\text{s}}^{3/2}). \nonumber
\end{align}
\noindent
Indeed this result also \textit{perfectly matches} the corresponding BCFT calculation given in \eqref{eq:BCFT1110final}, upon the inclusion of the familiar factor of $2$ in the appropriate places. Having \textit{precisely matched} many BCFT and double-scaled matrix-model results, supports both our results and our interpretation. Nonetheless, we proceed with further checks---now against string-equation results and via numerical large-order matches.

\subsection{BCFT/Matrix-Model Results Match String-Equation Results}\label{subsec:BCFT-MM-stringeqs}

Albeit comparison of BCFT versus matrix-model results has already established a rather non-trivial consistency check of our proposal, let us move-on and further check our results against string equations\footnote{To be fully precise, string equations really arise from matrix-model orthogonal polynomials. Hence what we truly imply is that our earlier ``matrix model'' results are actually ``spectral curve'' or ``matrix integral'' results.}. Appendix~\ref{app:minimal-string-equation-setup} outlines the general set-up of minimal-string string-equations, and we refer the reader to \cite{gs21} for another recent discussion precisely along the lines of the ensuing discussion. Comparisons between matrix-model and string-equation results were also extensively conducted in \cite{mss22}, as string-equations iteratively yield the complete resurgent transseries solutions including all possible nonperturbative sectors. Further note how this ``triple check'' is useful also if to go forward with calculations: while BCFT or matrix model computations at high-instanton or high-genera are very slow, the same via string equations may be quite fast and efficient. Once the match has been established, one may then mainly rely on string equations to produce high-instanton and high-order data, also establishing Stokes data in the process.

\begin{figure}
\hspace{-20pt}
\begin{tabular}{p{0.50\textwidth} p{0.50\textwidth}}
  \vspace{0pt} \includegraphics[width=0.52\textwidth]{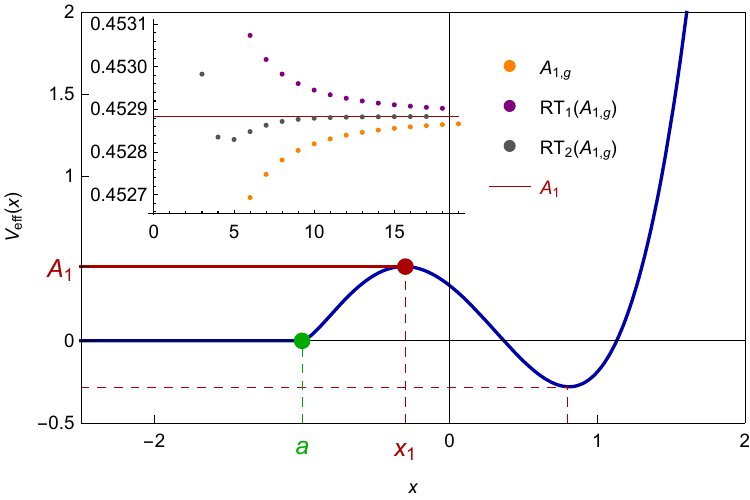}&
  \vspace{0pt} \includegraphics[width=0.50\textwidth]{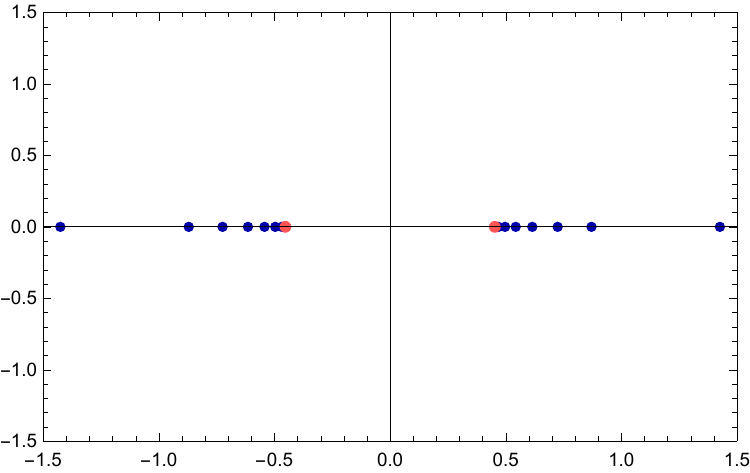}
\end{tabular}
\caption{Numerical tests for the perturbative sector of the specific-heat of the $(2,5)$ minimal string. On the left-hand side we plot the holomorphic effective potential, together with large-order checks of the perturbative sector. Here $A_{1,g}=2 g \sqrt{u_{g}/u_{g+1}}$ and $\text{RT}_{k}$ denotes the $k$th Richardson transform \cite{abs18}. We observe that, as expected from the discussion in subsection~\ref{subsec:double-scaled-geometry}, the leading asymptotics only ``see'' the instanton action $A_1$ which in the matrix-model computations is associated to the saddle-point closest to the cut. On the right-hand side we plot the poles of the Borel--Pad\'e approximation \cite{abs18} to the perturbative specific-heat, up to genus $40$. The symmetry of the Borel singularities is evident, thus supporting that minimal strings are resonant.}
\label{fig:minimal-string-resonance-check}
\end{figure}

Let us start with a quick numerical exploration of the resonant structure of minimal strings using string-equation data for the $(p,q)=(2,5)$ example. Here we check the perturbative specific-heat up to genus $40$. The explicit results are shown in figure~\ref{fig:minimal-string-resonance-check} and they support how minimal strings are indeed resonant (analytically shown in \cite{gs21}). Notice how the asymptotic checks in figure~\ref{fig:minimal-string-resonance-check} only see the leading instanton action $A_1$---and not the smaller action $A_2$. This is exactly in line with the matrix-model discussion in subsection~\ref{subsec:double-scaled-geometry}, and reflects how eigenvalues cannot directly tunnel to the saddle-point $x_2^{\star}$ (note how moving between free-energy and specific-heat via double-integration preserves the instanton actions and the Borel residues up to global rescaling---as such, matrix-model free-energy features also reflect in the specific-heat).

We now focus on the $(1,1)$ sector at leading-order in genus. It turns out that this contribution may be obtained analytically for arbitrary $k$ from direct use of the minimal-string string-equations. In appendix~\ref{app:minimal-string-equation-setup} we derive the formula for the lowest-genus contribution to the $(1,1)$ sector of the string specific-heat, which is given explicitly in formula \eqref{eq:MC11Sector} for the multicritical hierarchy and in \eqref{eq:MS11Sectoru} for the minimal-string hierarchy. Then, recalling the ``bridge'' relation, our results can be written in terms of free energy via simple exponentiation as \cite{mss22}
\begin{align}
\label{eq:oneonesectortransseries}
\mathsf{S}_{(0,0)\rightarrow (0,1)}\, \mathsf{S}_{(0,1)\rightarrow (1,1)}\, \frac{Z^{(1,1)} (g_{\text{s}})}{Z^{(0,0)} (g_{\text{s}})} &= \\
&
\hspace{-50pt}
= \mathsf{S}_{(0,0)\rightarrow (0,1)}\, \mathsf{S}_{(0,1)\rightarrow (1,1)} \left\{ F^{(1,1)} (g_{\text{s}}) + F^{(1,0)} (g_{\text{s}})\, F^{(0,1)} (g_{\text{s}}) \right\}, \nonumber
\end{align}
\noindent
and all that remains to be done is translate the prediction of \eqref{eq:MS-11-sector} from free energy $F$ to specific heat $u$. We do so by recalling the relation\footnote{Herein we are focusing on the string-equation variable $z$-dependence, see appendix~\ref{app:minimal-string-equation-setup}.}
\begin{equation}
g_{\text{s}}^2 F^{\prime\prime} (z) = - \frac{1}{2} u (z).
\end{equation} 
\noindent
Thus, noting that at lowest genus only the $(1,1)$ free-energy sector contributes, it turns out that this equality becomes equivalent to the prediction that \eqref{eq:MS11Sectoru} is  proportional\footnote{Note that this is a proportionality relation as the particular equality depends on the Borel residues of the problem, which are themselves normalization-dependent and could hence rescale the $(1,1)$ sector; see, \textit{e.g.}, \cite{asv11}.} to the $(1,1)$ specific-heat sector as
\begin{equation}
\label{eq:u0MSk=A''}
u_{0,{\text{MS-}}k}^{(1,1)}(z) \propto A^{\prime\prime}(z).
\end{equation}
\noindent
The simplicity of this formula is surprising at first sight, given the complicated structure of $u_{0,{\text{MS-}}k}^{(1,1)}(z)$. However, below we will show that the two expressions exactly coincide \textit{for all} $k$.

Firstly, one needs to use prior results obtained in \cite{gs21}, which we have explicitly included in the appendix~\ref{app:minimal-string-equation-setup}. Then, differentiating \eqref{eq:instActionODEMS} we can solve for $A^{\prime\prime}(z)$ to find 
\begin{equation}
\label{eq:2ndDerivInstActionMS}
2 A^{\prime\prime}(z) = -\frac{\sum\limits_{n=0}^{k-1} \left( \sum\limits_{\ell=k-2\left[\frac{k-1}{2}\right]}^{k} \left(\ell-n-1\right)\, \ell\, t_{\ell} (k)\, \frac{\alpha_{\ell,\ell-n}}{\alpha_{\ell,\ell}}\, u_0^{\ell-n-2}(z)\, u_0^{\prime}(z) \right) A'(z)^{2n}}{\sum\limits_{n=0}^{k-1} \left( \sum\limits_{\ell=k-2\left[\frac{k-1}{2}\right]}^{k} n\, \ell\, t_{\ell} (k)\, \frac{\alpha_{\ell,\ell-n}}{\alpha_{\ell,\ell}}\, u_0^{\ell-n-1}(z) \right) A'(z)^{2n-1}}.
\end{equation}
\noindent
In fact we are almost done: first we note that by solving the $k$th minimal-string string-equation at lowest genus, one can find $u^{\prime}_0(z)$ as 
\begin{equation}
u^{\prime}_0(z) = \left( \sum_{p=0}^{[\frac{k-1}{2}]} t_{k-2p}(k) \left(k-2p\right) u_0^{k-2p-1}(z) \right)^{-1}.
\end{equation}
\noindent
We immediately recognize this factor as the denominator of $u_{0,{\text{MS-}}k}^{(1,1)}(z)$ in \eqref{eq:MS11Sectoru}. Further, we recognize the denominator (up to a proportionality constant) of the expression for $A^{\prime\prime}(z)$ above to be precisely\footnote{Due to the backward-forward relations discussed in \cite{bssv22}, these two sectors are in fact the same at lowest genus.} the \textit{inverse} of $u_{0,{\text{MS-}}k}^{(1,0)}(z) \times u_{0,{\text{MS-}}k}^{(0,1)}(z) = u_{0,{\text{MS-}}k}^{(1,0)}(z)^2$; see \eqref{eq:MS10Sectoru}. Finally, upon relabelling the sum appearing in the numerator of \eqref{eq:2ndDerivInstActionMS} as $\ell \to k-2p$, one sees that this term precisely matches the double-sum in the numerator of \eqref{eq:MS11Sectoru}; the first of which is associated to the minimal string KdV times, and the second belonging to the ``multicritical factors'' $u_{0,{\text{MC-}}k}^{(1,1)}(z)$ given in \eqref{eq:MC11Sector}. All of the above ingredients add up so as to precisely produce \eqref{eq:2ndDerivInstActionMS} above, thus demonstrating the equality in \eqref{eq:u0MSk=A''}.

Let us perform another check, albeit less analytic, for the results for the $(2,0)$ sector. In particular, we now wish to verify formula \eqref{eq:MS-l0-sector} for $\ell=2$ against string equation results. Notice that the starting genus for that equation is in fact of order $g^2_{\text{s}}$. In other words, at order $g_{\text{s}}$ the right-hand side of this equation should equal zero and, translating to free energies, this formula actually tells us that the lowest contributions should cancel. Therefore, at order $g_{\text{s}}$ we find
\begin{equation}
\label{eq:MS-l=2,0-sector}
\frac{\mathcal{Z}^{(2|0)}_{n,k}(g_{\text{s}})}{\mathcal{Z}^{(0|0)}_{n,k}(g_{\text{s}})}  = \mathsf{S}_{(0,0)\rightarrow (0,2)}\, \rme^{-\frac{2}{g_{\text{s}}} A_{n,k}} \left( F_0^{(2,0)} (g_{\text{s}}) + \frac{1}{2} \left( F_0^{(1,0)} (g_{\text{s}}) \right)^2 \right) = 0.
\end{equation}
\noindent
To check the validity of this result directly from the string equation, we again first translate the above to a statement for the specific heat as
\begin{equation}
u_{0,{\text{MS-}}k}^{(2,0)}(z) = \left( \frac{u_{0,{\text{MS-}}k}^{(1,0)}(z)}{A^{\prime}(z)}\right)^2.
\end{equation}
\noindent
This equality may be shown to hold \textit{for all} $k$, directly from the \eqref{eq:MS10Sectoru} $(1,0)$ and \eqref{eq:MS20SectorSpecificHeat} $(2,0)$ lowest-genus coefficients. In fact, a factor of $\left( u_{0,{\text{MS-}}k}^{(1,0)}(z) \right)^2$ already appears implicitly in equation  \eqref{eq:MS20SectorSpecificHeat} for the $(2,0)$ sector, and it is then straightforward to see that the remaining terms precisely reproduce \eqref{eq:instActionODEMS}, the defining equation for $A^{\prime}(z)$ in terms of the perturbative coefficient $u_0(z)$. Since this is an easier calculation than the one performed for the $(1,1)$ sector above, and as it is not particularly illuminating, we omit the details. Instead, we proceed to verify \eqref{eq:MS-l0-sector} at next-to-leading order in $g_{\text{s}}$. In full generality this would require generic higher-order coefficients in the $(2,0)$ and $(1,0)$ sectors which are not available as expressions for arbitrary $k$. Nonetheless, we can of course make explicit checks at various fixed values of $k$. First evaluate \eqref{eq:MS-l0-sector} with $\ell=2$ at order $g_{\text{s}}^2$,
\begin{equation}
\label{eq:20sectorStokesExpression}
\frac{\mathcal{Z}^{(2|0)}_{n,k}(g_{\text{s}})}{\mathcal{Z}^{(0|0)}_{n,k}(g_{\text{s}})} = \frac{1}{2\pi}\, \rme^{-\frac{2}{g_{\text{s}}} A_{n,k}} \left( g_{\text{s}}\, \frac{(-1)^{n+k+1} \cot\frac{n\pi}{2k-1}}{16 \left(2k-1\right) \sin^2\frac{n\pi}{2k-1}}\right)^{2}.
\end{equation}
\noindent
By using the ``bridge'' relation established in \cite{mss22} we can link the transseries partition function to our matrix model calculation. Rewriting quantities in terms of the free energy we find
\begin{equation}
\label{eq:20sectorfreeEnergy}
\frac{\mathcal{Z}^{(2|0)}_{n,k}(g_{\text{s}})}{\mathcal{Z}^{(0|0)}_{n,k}(g_{\text{s}})} \simeq \mathsf{S}_{(0,0)\rightarrow (0,2)}\, \rme^{-\frac{2}{g_{\text{s}}} A_{n,k}} \left( F_1^{(2,0)} (g_{\text{s}}) + F_1^{(1,0)} (g_{\text{s}})\, F_0^{(1,0)} (g_{\text{s}}) \right) + \cdots.
\end{equation}
\noindent
To make the comparison with the string equations, all we have left to do is relate the free-energy coefficients to those of the specific heat. Similarly to what was done in \cite{gs21} for the lowest two orders in $g_{\text{s}}$, we find that these free energy sectors can be rewritten as
\begin{equation}
F_1^{(2,0)} (g_{\text{s}}) = - \frac{1}{8 \left(A^{\prime}(z)\right)^2} \left( u_{1,{\text{MS-}}k}^{(2,0)}(z) + \frac{\partial_z u_{0,{\text{MS-}}k}^{(2,0)}(z)}{A^{\prime}(z)} - \frac{3}{2} \frac{A^{\prime\prime}(z)}{\left(A^{\prime}(z)\right)^2}\, u_{0,{\text{MS-}}k}^{(2,0)}(z) \right)
\end{equation}
\noindent
for the $(2,0)$ sector, and as
\begin{equation}
F_1^{(1,0)} (g_{\text{s}}) = - \frac{1}{2 \left(A^{\prime}(z)\right)^2} \left( u_{1,{\text{MS-}}k}^{(1,0)}(z) + 2 \frac{\partial_z u_{0,{\text{MS-}}k}^{(1,0)}(z)}{A^{\prime}(z)} - 3 \frac{A^{\prime\prime}(z)}{\left(A^{\prime}(z)\right)^2}\, u_{0,{\text{MS-}}k}^{(1,0)}(z) \right)
\end{equation}
\noindent
for the $(1,0)$ sector. As already mentioned, a check at arbitrary $k$ requires closed-forms for the coefficients $u_{1,{\text{MS-}}k}^{(2,0)}(z)$ and $u_{1,{\text{MS-}}k}^{(1,0)}(z)$, which would in turn need more knowledge than we currently have concerning the coefficients of the Gel'fand--Dikii polynomials (see appendix~\ref{app:minimal-string-equation-setup} for a more detailed explanation). However, we can of course still check this explicitly for several values of $k$. We have done so for $k=2,3,4,5$ and we obtained precise agreements between the above equations \eqref{eq:20sectorStokesExpression} and \eqref{eq:20sectorfreeEnergy}. Alongside the earlier arbitrary-$k$ check on the order-$g_{\text{s}}$ $(2,0)$ and $(0,2)$ sectors (the latter being given by the former via backward forward symmetry; see for example \cite{bssv22}) and order-$g^{-1}_{\text{s}}$ $(1,1)$ sector, we now have very compelling evidence for the validity of the BCFT and matrix-integral results fully matching against string-equation results. 

Finally, let us check the above double-scaled matrix-model results against the string-equation transseries for $(p,q)=(2,5)$ (discussed in appendix~\ref{app:minimal-string-25}). For the following comparison we will not fix to the conformal background, so as to understand how the matrix model computation actually reproduces the functional dependence on the string-equation variable $z$. The $(2,5)$ spectral curve away from the conformal background reads \cite{gs21}
\begin{equation}
y_{(2,5)}(x) = \left( 8 x^2 - 4 u_0(z)\, x + 3 u_0(z)^2 - 5\right) \sqrt{x+u_0(z)}.
\end{equation}
\noindent
Some preliminary checks on the canonical Borel residue were already conducted in \cite{gs21} and we will not repeat them herein, but still follow notation and computations therein. This Borel residue result can be found in table~\ref{tab:BorelMS251}. Let us start with the non-trivial $(2|1)$ contribution. Plugging-in the spectral-curve data into \eqref{eq:dsl-21-sector} yields
\begin{equation}
\frac{\mathcal{Z}^{(2|1)} (g_{\text{s}})}{\mathcal{Z}^{(0|0)} (g_{\text{s}})} - \frac{\mathcal{Z}^{(1|0)} (g_{\text{s}})}{\mathcal{Z}^{(0|0)} (g_{\text{s}})}\, \frac{\mathcal{Z}^{(1|1)} (g_{\text{s}})}{\mathcal{Z}^{(0|0)} (g_{\text{s}})} \simeq \rme^{-\frac{1}{g_{\text{s}}}A_{1,3}}\, \frac{\sqrt{g_{\text{s}}} \left(U^2+5\right)^{7/8} \left( 2 \gamma_{\text{E}} + \log \left( \frac{500 \sqrt{10} \left(5-U\right)^5 U^2}{g_{\text{s}}^2 \left(U^2+5\right)^{7/2}} \right) \right)}{4\, \sqrt[8]{2} \cdot 5^{7/8}\, \pi^{3/2} \left(5-U\right)^{5/4} \sqrt{U}} + \cdots,
\end{equation}
\noindent
where $U=\frac{\sqrt{5}}{u_0}\sqrt{2-u_0^2}$. Comparing with expression \eqref{eq:compare25fe21} then allows for the computation of a non-trivial Borel residue. Namely, we find
\begin{equation}
\mathsf{S}_{(1,1)\to(1,0)} = -\frac{\rmi}{\sqrt{\pi}}\left(\frac{5}{2}\right)^{3/8}\left(\gamma_{\text{E}}-\log\left(-10\cdot 2^{1/4}\cdot 5^{3/4}\right)\right)
\end{equation}
\noindent
This result has exactly the features that we would expect from previous studies: we find the $\gamma_{\text{E}}$ number in addition to a logarithmic contribution \cite{bssv22, mss22}. Similarly we can treat the $(1|1)(1|0)$ contribution where \eqref{eq:dsl-result-1110-sector} yields 
\begin{align}
\rme^{-\frac{A_{2,3}}{g_{\text{s}}}}\, \mathsf{S}_{(0,0,0,0)\to (1,0,0,0)}\, \mathsf{S}_{(1,0,0,0)\to(0,1,1,0)}\, \mathsf{S}_{(0,1,1,0)\to(1,1,1,0)}\, F^{(1,1,1,0)} &\simeq \rme^{-\frac{1}{g_{\text{s}}}A_{2,3}} \times \\
&
\hspace{-200pt}
\times \frac{-\rmi}{2 \cdot 2^{1/8} \cdot 5^{7/8} \pi^{3/2}}\, \frac{\left(U^2+5\right)^{7/8}}{U^{1/2} \left(5+U\right)^{5/4}}\, \tanh^{-1} \sqrt{\frac{10}{5+U}-1} + \cdots. \nonumber
\end{align}
\noindent
We can again compare with the transseries structure of the $(2,5)$ minimal string. Plugging-in $F^{(1,1,1,0)}$ from \eqref{eq:compare25fe1110}, we find perfect agreement for all $z$. Moreover, using the fact that the Borel residues $\mathsf{S}_{(0,0, 0,0)\to(1,0,0,0)}\, \mathsf{S}_{(0,1, 1,0)\to(1,1,1,0)} = -\frac{1}{\pi}\left(\frac{2}{5}\right)^{3/4}$ are known (to be the square of the canonical Borel residue, as dictated by resurgence) we can predict one further (so far unknown) Borel residue, 
\begin{equation}
\mathsf{S}_{(1,0,0,0)\to(0,1,1,0)} = -\frac{\rmi}{2\sqrt{\pi}}\left(\frac{2}{5}\right)^{\frac{3}{8}}.
\end{equation}
\noindent
We summarize our results for the Borel residues of the $(2,5)$ minimal string in tables~\ref{tab:BorelMS251} and~\ref{tab:BorelMS252}. A thorough large-order check of these results would be interesting, but because the transseries study of $(2,5)$ is rather intricate we leave it for future work.

\begin{table}
\centering
\begin{tabular}{c|c|c|c}
$\mathsf{S}_{(0,0)\to(1,0)}$ & $\mathsf{S}_{(0,0)\to(0,1)}$ & $\mathsf{S}_{(0,1)\to(1,1)}$ & $\mathsf{S}_{(1,1)\to(1,0)}$ \\\hline 
$-\frac{\rmi}{\sqrt{\pi}}\left(\frac{2}{5}\right)^{3/8}$ & $\frac{1}{\sqrt{\pi}}\left(\frac{2}{5}\right)^{3/8}$  & $-\frac{\rmi}{\sqrt{\pi}}\left(\frac{2}{5}\right)^{3/8}$ & $ -\frac{\rmi}{\sqrt{\pi}}\left(\frac{5}{2}\right)^{3/8}\left(\gamma_{\text{E}}-\log\left(-10\cdot 2^{1/4}\cdot 5^{3/4}\right)\right)$
\end{tabular}
\caption{List of a few Borel residues for the $(2,5)$ minimal string. Here we compute only Borel residues associated to the saddle closest to the cut (see figure \ref{fig:minimal-string-resonance-check}).}
\label{tab:BorelMS251}
\end{table}

\begin{table}
\centering
\begin{tabular}{c|c}
$\mathsf{S}_{(1,0,0,0)\to(0,0,1,0)}$ & $\mathsf{S}_{(1,0,0,0)\to(0,1,1,0)}$ \\\hline 
$\rmi$ & $-\frac{\rmi}{2\sqrt{\pi}}\left(\frac{2}{5}\right)^{3/8}$
\end{tabular}
\caption{List of a few Borel residues for the $(2,5)$ minimal string. Here we now compute Borel residues which appear when both saddles are populated.}
\label{tab:BorelMS252}
\end{table}

\subsection{The Case of Jackiw--Teitelboim Gravity}\label{subsec:JTgravity}

The $(2,2k-1)$ minimal-string series has a well-defined $k \to +\infty$ limit which has seen great interest in recent years: this is the double-scaled matrix model \cite{sss19} describing quantum JT-gravity \cite{t83, j85, ap14, ms16, msy16, sw17}. At strict nonperturbative level, this limit was recently addressed in \cite{gs21}, and the resurgent asymptotics of JT gravity studied in detail in \cite{eggls23}---at least on what concerns the non-resonant part of the JT transseries. What we can now add, given our present results, is what happens at nonperturbative \textit{resonant} level. For that, we just apply our earlier machinery to the JT-gravity double-scaled spectral curve \cite{eo07b, sss19},
\be
\label{eq:JT-spectral-curve}
y(x) = \frac{1}{2\pi} \sin \left( 2\pi \sqrt{x} \right).
\ee
\noindent
Note how running the topological recursion on this JT spectral curve is quite straightforward, \textit{e.g.}, \cite{sss19, gs21, eggls23}. The same holds true for our nonperturbative resurgent formulae. The zeros of this spectral curve are located at $\sqrt{x_n^{\star}} = \upzeta_n^{\star}=\frac{n}{2}$, with $n\in\mathbb{N}^{+}$. Furthermore we find the (to become useful below) expressions for the moment function $M^{\prime}(x_n^{\star}) = (-1)^n \frac{2}{n^2}$ and the instanton action (see \cite{eggls23} for further details)
\begin{equation}
A_{n} = (-1)^{n+1}\, \frac{n}{4\pi^2}.
\end{equation}

\begin{figure}
\centering
\includegraphics[scale=0.63]{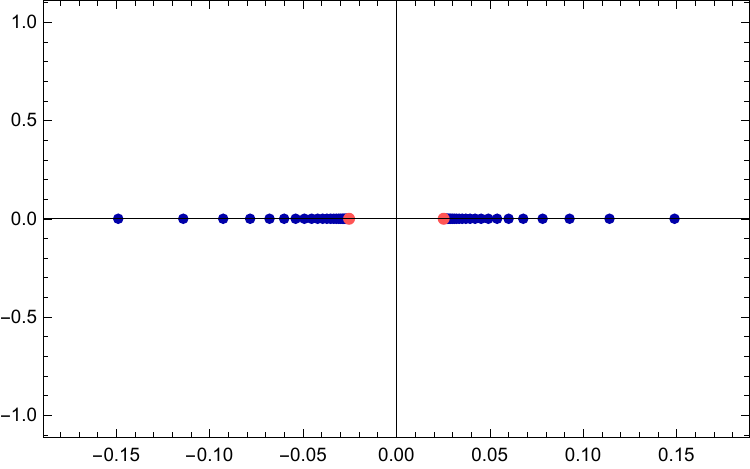}
\includegraphics[scale=0.63]{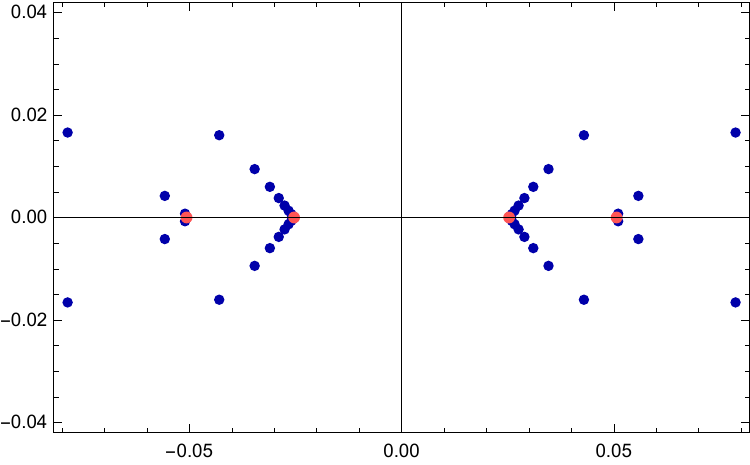}
\caption{Plot of poles from the Borel--Pad\'e approximant to the perturbative-sector data (up to $g_{\text{s}}$-order $122$ \cite{eggls23}). The poles are shown in blue, and the analytic locations of the instanton actions in red. On the left-hand side we show a direct Borel--Pad\'e approximation whereas on the right we have enhanced it using a conformal transformation (also visible on the typical doubling of the ``branch-cuts''). This allows us to distinctively see the ``branch-cuts'' associated to higher instanton singularities on the Borel plane. Symmetry of these singularities is evident.}
\label{fig:JT-pert-borel-pade}
\end{figure}

\begin{figure}
\centering
\includegraphics[scale=0.65]{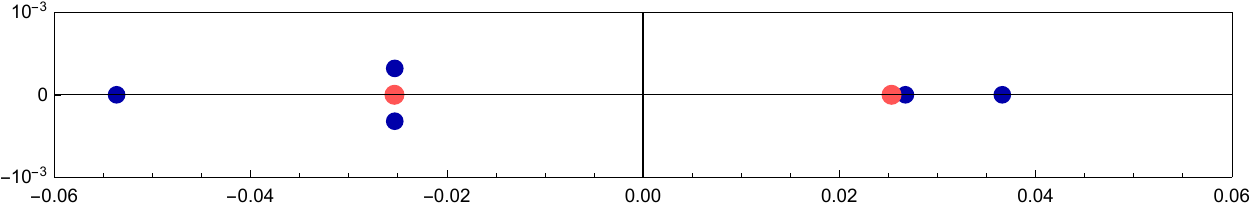}
\caption{Borel--Pad\'e approximant to the one-instanton sector data of JT gravity. Generating nonperturbative data from the topological recursion is slow and it has only been computed up to 12th order \cite{eggls23}. As such, the Pad\'e approximants do not produce a lot of poles. In spite of this, both symmetric instanton contributions associated to resonance are clearly visible in the plot.}
\label{fig:JT-non-pert-borel-pade}
\end{figure}

The resonant nature of JT gravity has been established at the level of its string-equation instanton-actions in \cite{gs21}. We herein start by checking how resonance is further supported numerically. For this we have performed a Borel--Pad\'e approximation of the JT perturbative sector (see, \textit{e.g.}, \cite{abs18}) using its first 122 coefficients (and using data from \cite{eggls23}). The results are shown in figure~\ref{fig:JT-pert-borel-pade}. A similar calculation can be performed for the one-instanton contribution, just with much less numerical accuracy (we have used the whole 12 coefficients obtained via the nonperturbative topological recursion in \cite{eggls23} for the one-instanton sector). The result is now visualized in figure~\ref{fig:JT-non-pert-borel-pade}. It is clear from the symmetry of the Borel singularities, alongside the instanton-action calculations in \cite{gs21}, how resonance is indeed a feature of JT-gravity.

With all this evidence we are equipped to spell out the nonperturbative contributions outlined in subsection~\ref{subsec:double-scaled-geometry} for the case of JT gravity. Straightforward application of our formulae to saddle $x_n^{\star}$ yields the results 
\begin{align}
\label{eq:JT-l0-sector}
\frac{\mathcal{Z}^{(\ell|0)}(g_{\text{s}})}{\mathcal{Z}^{(0|0)}(g_{\text{s}})} &\simeq \frac{G_{2} \left(\ell+1\right)}{\left(2\pi\right)^{\ell/2}}\, \rme^{-\frac{\ell}{g_{\text{s}}} A_n} \left( g_{\text{s}}\, \frac{(-1)^n}{n^3}\right)^{\frac{\ell^2}{2}} + \cdots, \\
\label{eq:JT-0lb-sector}
\frac{\mathcal{Z}^{(0|\bar{\ell})}(g_{\text{s}})}{\mathcal{Z}^{(0|0)}(g_{\text{s}})} &\simeq \frac{G_{2} \left(\bar{\ell}+1\right)}{\left(2\pi\right)^{\bar{\ell}/2}}\,  \rme^{\frac{\bar{\ell}}{g_{\text{s}}} A_n} \left(-g_{\text{s}}\, \frac{(-1)^n}{n^3}\right)^{\frac{\bar{\ell}^2}{2}} + \cdots,
\end{align}
\noindent
as well as
\begin{align}
\label{eq:JT-11-sector}
\frac{\mathcal{Z}^{(1|1)}(g_{\text{s}})}{\mathcal{Z}^{(0|0)}(g_{\text{s}})} \simeq \frac{1}{g_{\text{s}}}\, \frac{\rmi\, (-1)^{n+1}\, n}{8\pi^3} + 0 + g_{\text{s}}\, \frac{\rmi}{12\pi  n^3}\, \Big\{ 4 + 2 \pi^2 n^2 + (-1)^{n+1} \left( n^2\pi^2+10 \right) \Big\} + o(g_{\text{s}}).
\end{align} 
\noindent
In addition we have the non-trivial $(2|1)$ contribution,
\begin{align}
\label{eq:JT-21-sector}
\frac{\mathcal{Z}^{(2|1)} (g_{\text{s}})}{\mathcal{Z}^{(0|0)} (g_{\text{s}})} - \frac{\mathcal{Z}^{(1|0)} (g_{\text{s}})}{\mathcal{Z}^{(0|0)} (g_{\text{s}})}\, \frac{\mathcal{Z}^{(1|1)} (g_{\text{s}})}{\mathcal{Z}^{(0|0)} (g_{\text{s}})} \simeq \rme^{-\frac{1}{g_{\text{s}}} A_n}\, \frac{1}{4\pi}\, \sqrt{\frac{g_{\text{s}}\,(-1)^{n+1}}{2\pi n^3}} \left\{ 2 \gamma_{\text{E}} + \log \left( \frac{n^6}{g_{\text{s}}^2} \right) \right\} + o(g_{\text{s}}^{3/2}).
\end{align}
\noindent
Finally, taking two distinct saddles, $x^{\star}_n$ and $x^{\star}_m$, with $x^{\star}_n$ further from the cut, we may also write down the $(1|1)(1|0)$ contribution as
\begin{equation}
\frac{\mathcal{Z}^{(1|1)(1|0)} (g_{\text{s}})}{\mathcal{Z}^{(0|0)(0|0)} (g_{\text{s}})} - \frac{\mathcal{Z}^{(1|1)(0|0)} (g_{\text{s}})}{\mathcal{Z}^{(0|0)(0|0)} (g_{\text{s}})}\, \frac{\mathcal{Z}^{(0|0)(1|0)} (g_{\text{s}})}{\mathcal{Z}^{(0|0)(0|0)} (g_{\text{s}})} \simeq \rme^{-\frac{1}{g_{\text{s}}} A_{n}}\, \frac{1}{2\pi}\, \sqrt{\frac{g_{\text{s}}\,(-1)^{n+1}}{2\pi n^3}}\, \log \left(\frac{m-n}{m+n}\right) + o(g_{\text{s}}^{3/2}).
\end{equation}

Checks on the validity of the above formulae against BCFT or string equations are essentially out of reach at the moment (but see for the JT string-equation discussions in \cite{os19, gs21}). One non-trivial check we can do, however, is to go back to the construction of JT-gravity as the $k \to +\infty$ limit of the $(2,2k-1)$ minimal-string \cite{sss19}. This limit received nonperturbative support in \cite{gs21}, upon using the adequate rescalings of nonperturbative quantities as 
\begin{align}
& y(x) = \frac{(-1)^{k-1}}{2\pi}\, T_{2k-1} \left(\frac{2\pi\sqrt{x}}{2k-1}\right)  & \xrightarrow{k\to+\infty} \qquad & \qquad \frac{\sin\left(2\pi\sqrt{x}\right)}{2\pi}, \\
& x_{n}^{\star} = \frac{\left(2k-1\right)^2}{8\pi^2} \left(1-\cos\frac{2\pi n}{2k-1}\right)  & \xrightarrow{k\to+\infty} \qquad & \qquad \frac{n^2}{4}, \\
& A_{n,k} = (-1)^{n+1}\frac{(2k-1)^2}{16\pi^3}\frac{4k-2}{(2k+1)(2k-3)}\sin\frac{2\pi n}{2k-1}  & \xrightarrow{k\to+\infty} \qquad & \qquad \frac{(-1)^{n+1}n}{4\pi^2}, \\
& M^{\prime}(x_n^{\star}) \left(x_n^{\star}-a\right)^{5/2} = \frac{4\pi (-1)^{n+1}\cot\frac{\pi n}{2k-1}}{(2k-1)^3\sin\left(\frac{\pi\,n}{2k-1}\right)^2} & \xrightarrow{k\to+\infty} \qquad & \qquad \frac{(-1)^n}{32\pi n^3}.
\end{align}
\noindent
Applying these rescalings and corresponding $k \to +\infty$ limits to our $(2,2k-1)$ minimal-string formulae from subsection~\ref{subsec:22k-1minstring}---which themselves have already been triple-checked against BCFT and string equation results---we have explicitly checked\footnote{Notice that the check of the third $g_{\text{s}}$-order of the $(1|1)$ contribution needs some more terms, such as $M^{\prime\prime}(x_n^{\star})$, but even though the computation is long it is straightforward.} that they yield \textit{precise matches} with the above JT results. In this sense, at the very least our nonperturbative JT-gravity results are fully consistent with the large $k$ limit of the minimal string.

Finally let us fix a normalization for the JT-gravity transseries, and compute Borel residues explicitly from our above ``transseries reparametrization invariant'' quantities \cite{asv11}, using the ``bridge'' relations given in \cite{mss22}. For the free energy, we choose to normalize sectors around the saddle $x_n^{\star}$ as
\begin{align}
\label{eq:JTgravityFreeEnergyConventions-a}
F^{(1,0)} &= \sqrt{g_{\text{s}}}\, \sqrt{\frac{(-1)^{n+1}}{n^3}} + o(g_{\text{s}}^{3/2}), & F^{(0,1)} &= \sqrt{g_{\text{s}}}\, \sqrt{\frac{(-1)^{n+1}}{n^3}} + o(g_{\text{s}}^{3/2}), \\
\label{eq:JTgravityFreeEnergyConventions-b}
F^{(2,1)} &= \sqrt{g_{\text{s}}}\, \sqrt{\frac{(-1)^{n+1}}{n^3}}\, \alpha\, \log\frac{n^6}{g_{\text{s}}^2} + o(g_{\text{s}}^{3/2}), & \alpha &= -\frac{1}{4}.
\end{align}
\noindent
This immediately yields a (small) number of Borel residues which can be calculated for the saddle-point closest to the cut, and we list them in table~\ref{tab:BorelJT}.

\begin{table}
\centering
\begin{tabular}{c|c|c|c}
$\mathsf{S}_{(0,0)\to(1,0)}$ & $\mathsf{S}_{(0,0)\to(0,1)}$ & $\mathsf{S}_{(0,1)\to(1,1)}$ & $\mathsf{S}_{(1,1)\to(1,0)}$ \\\hline 
$-\frac{\rmi}{\sqrt{2\pi}}$ & $\frac{1}{\sqrt{2\pi}}$  & $-\frac{\rmi}{\sqrt{2\pi}}$ & $-\frac{\rmi}{\sqrt{2\pi}}\, \frac{\gamma_{\text{E}}}{2}$  
\end{tabular}
\caption{List of a few Borel residues for JT gravity, for the saddle $n=1$ and using the free energy normalization given in \eqref{eq:JTgravityFreeEnergyConventions-a}-\eqref{eq:JTgravityFreeEnergyConventions-b}. We can calculate the canonical forward Borel residue (see as well \cite{eggls23}) but also its backward resonant sibling. Moreover we find the non-trivial Borel residue $\mathsf{S}_{(1,1)\to(1,0)}$ which sits inside the resonant sum of the $(2|1)$ contribution.}
\label{tab:BorelJT}
\end{table}

\subsection{Towards the Two-Matrix Model?}\label{subsec:2matrixmodel}

Having focused on the one-matrix model up to this stage, and making contact with the $(p,q)=(2,2k-1)$ minimal string results of section~\ref{sec:ZZminimalstring}, we may now ask if our formulae may be extended to the full-fledged two-matrix model---hence if they may make contact with the complete $(p,q)$ BCFT formulae in section~\ref{sec:ZZminimalstring}. Let us immediately warn the reader that this subsection does not include a complete calculation, but rather a motivating set-up to build-upon in future work.

\paragraph{Two-Matrix Model:}

Begin with the two-matrix model \cite{iz80, dkk93, e02, kk04, iky05, ceo06}, now with partition function
\begin{equation}
\label{eq:twomatrixmodel}
\mathcal{Z}_{N} = \frac{1}{\text{vol}\left(\text{U}(N)\right)} \iint \rmd M_1 \rmd M_2\, \text{e}^{-\frac{1}{g_{\text{s}}}\, \text{Tr}\, \big\{ V_1(M_1) + V_2(M_2) - M_1  M_2 \big\}},
\end{equation}
\noindent
where both matrices are hermitian $N\times N$, and where the polynomial potentials have degrees $\deg V_1^{\prime} = d_1$, $\deg V_2^{\prime} = d_2$. As for the one-matrix model, the large-$N$ 't~Hooft limit is described by a spectral curve, encoded in the algebraic (polynomial) equation \cite{e02, ceo06, eo07a}
\begin{equation}
\label{eq:twomatrixmodelspectralcurve}
\CE (x,y) = 0, \qquad \text{ with} \qquad \deg_x \CE = d_1+1, \quad \deg_y \CE = d_2+1,
\end{equation}
\noindent
from where one wishes to obtain $y \equiv Y(x)$, albeit this is now a bit more subtle. Consider a point $p \in \Sigma$ on the Riemann surface $\Sigma$ described by \eqref{eq:twomatrixmodelspectralcurve} and the corresponding maps $x = \CX (p)$ and $y = \CY (p)$. More specifically, for fixed $x$ this geometry describes a $(d_2+1)$-sheeted covering of the complex $x$-plane. Label these sheets\footnote{Notice that the labeling of sheets that we are using for the two-matrix model spectral curve differs slightly from the one used in section \ref{sec:ZZminimalstring}. Nevertheless they are consistent and related (see formula \eqref{eq:saddle-zeta}-\eqref{eq:saddle-zeta-beta}).} by $i=0,\ldots,d_2$, and let $p_i (x)$ be the points on each of these which project down to $x$ via $\CX (p_i) = x$. Then, \eqref{eq:twomatrixmodelspectralcurve} is satisfied with this $x$ and the corresponding\footnote{Equivalently, for fixed $y$, \eqref{eq:twomatrixmodelspectralcurve} describes a $(d_1+1)$-sheeted covering of the complex $y$-plane. Label these sheets by $j=0,\ldots,d_1$, and let $\widetilde{p}_j (y)$ be the points on each of these which project down to $y$ via $\CY (\widetilde{p}_j) = y$. Then, \eqref{eq:twomatrixmodelspectralcurve} is satisfied with this $y$ and the corresponding
\be
x_j \equiv X_j (y) = \CX (\widetilde{p}_j (y)).
\ee}
\be
y_i \equiv Y_i (x) = \CY (p_i (x)).
\ee
\noindent
There is one (unique) special sheet, the \textit{physical} $0$-th sheet, where\footnote{Respectively, where $X_0(y) = V_2^{\prime} (y) - \frac{N}{y} + o(1/y^2)$ for $y \to +\infty$.} for $x \to +\infty$
\be
Y_0(x) = V_1^{\prime} (x) - \frac{N}{x} + o(1/x^2).
\ee
\noindent
But moreover, we find branch-points whenever two of the sheets of $\Sigma$ meet, \textit{i.e.}, whenever $Y_{n} (x) = Y_{m} (x)$ (or $X_{n} (y) = X_{m} (y)$) for $n \neq m$. We are interested in branch-points away from the endpoints of the cuts, \textit{i.e.}, in the nonperturbative saddles.

Consider one such nonperturbative saddle, labeled by $(x_{mn}, y_{mn})$. As two sheets touch, this is a double-point of the spectral curve, and we are (at least locally) in the framework of \cite{mss22}---\textit{i.e.}, these are two distinct points in the uniformization cover, and as such both must be taken into account when computing nonperturbative contributions. This is in fact considered when computing the corresponding instanton action as in \cite{kk04}, where one finds
\begin{equation}
\label{eq:effectiveactiontwomatrixmodel}
A_{mn} = \oint_{B_{mn}} \text{d}x\, y = \int^{\infty}_{x_{mn}} \text{d}x\, Y_{n}(x) - \int^{\infty}_{x_{nm}} \text{d}x\, Y_{m}(x).
\end{equation}
\noindent
Herein $B_{mn}$ is the $B$-cycle through the two points, and recall that multiple sheets meet at $\infty$ \cite{kk04}. This set-up is illustrated in figure~\ref{fig:twomatrixexample23}.

\definecolor{modebeige}{rgb}{0.59, 0.44, 0.09}
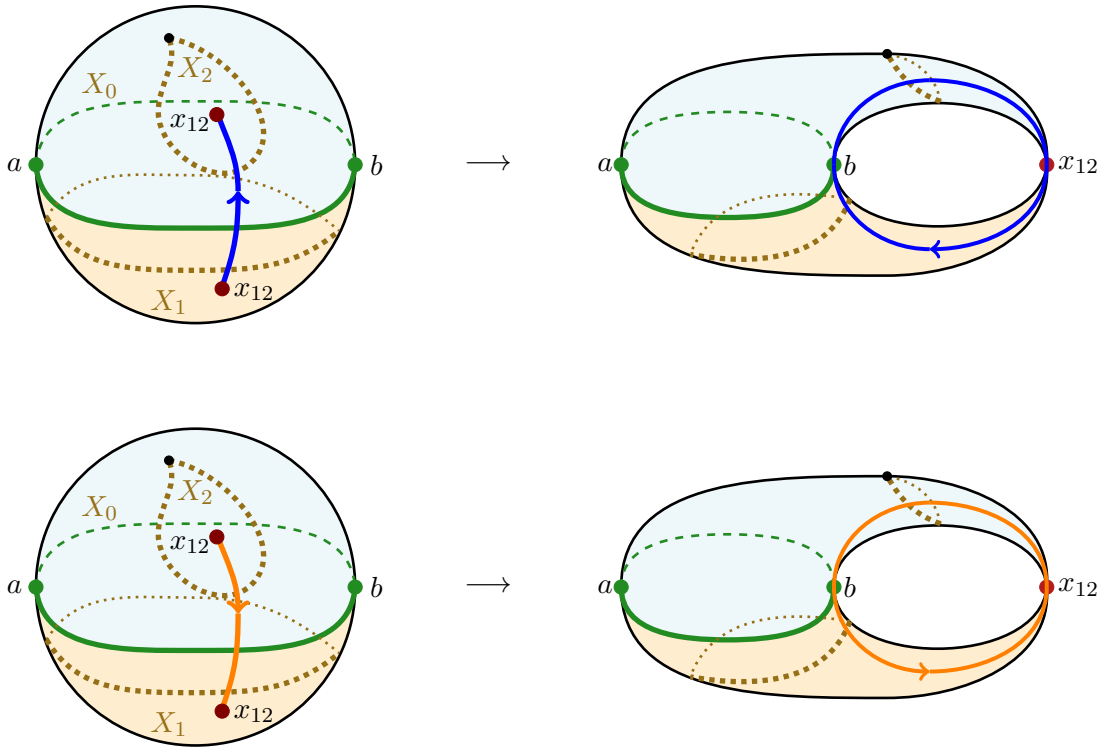
\begin{figure}
\centering
\begin{tikzpicture}
\begin{scope}[scale=0.7,  shift={({-6},{0})}]
\draw[fill=LightBlue,fill opacity=0.2, line width=1pt] (-3,0) to [out=90,in=180] (0,3)
	to [out=0,in=90] (3,0)
    to [out=265,in=0] (0,-1.2)
    to [out=180, in=275] cycle;
    \draw[fill=darktangerine,fill opacity=0.2, line width=1pt] (-3,0) to [out=270,in=180] (0,-3)
	to [out=0,in=270] (3,0)
    to [out=265,in=0] (0,-1.2)
    to [out=180, in=275] cycle;
    \draw[color=ForestGreen, line width=2pt] (-3,0) to [out=275, in=180] (0, -1.2)
    to [out=0, in=265] (3,0);
    \draw[dashed, color=ForestGreen, line width=1pt] (-3,0) to [out=85, in=180] (0, 1.2)
    to [out=0, in=95] (3,0);
    \draw[ForestGreen, fill=ForestGreen] (-3,0) circle (.8ex);
\draw[ForestGreen, fill=ForestGreen] (3,0) circle (.8ex);
\draw[dotted, modebeige, line width=2pt] (-0.5, 2.4) to[out=350, in=40] (1, 0)
to[out=220, in=270] (-0.7, 1)
to[out=90, in=290] cycle;
\draw[dotted, modebeige, line width=2pt] (-2.8, -1) to[out=300, in=180] (0,-2)
to[out=0, in=220] (2.7, -1.25);
\draw[dotted, modebeige, line width=1pt] (-2.8, -1) to[out=60, in=180] (0,-0.2)
to[out=0, in=140] (2.7, -1.25);
\draw[blue, line width=2pt, ->] (0.5, -2.3) to[out=70, in=270] (0.8, -0.5);
\draw[blue, line width=2pt] (0.8, -0.5) to[out=90, in=290] (0.4, 1);
\draw[Maroon, fill=Maroon] (0.5,-2.35) circle (.8ex);
\draw[Maroon, fill=Maroon] (0.4,0.95) circle (.8ex);
\node at (-3.4, 0) {$a$};
\node at (3.4, 0) {$b$};
\node[modebeige] at (-0.5, -2.6) {$X_1$};
\node[modebeige] at (-1.8, 1.5) {$X_0$};
\node[modebeige] at (0, 1.8) {$X_2$};
\node at (-0.1, 0.8) {$x_{12}$};
\node at (1.1, -2.4) {$x_{12}$};
\node at (5.5, 0) {$\longrightarrow$};
\draw[black, fill] (-0.5,2.4) circle (.5ex);
\end{scope}
\begin{scope}[scale=0.7,  shift={({6},{0})}]
	\draw[fill=LightBlue,fill opacity=0.2, line width=1pt] (0,0) to [out=90,in=95] (4,0)
	to [out=85,in=0] (1,2.1)
    to [out=180,in=90] (-4,0)
    to [out=270, in=180] (-2, -1)
    to [out=0, in=270] cycle;
    \draw[fill=darktangerine,fill opacity=0.2, line width=1pt] (-4,0)
    to [out=270,in=180] (1,-2.1)
    to [out=0,in=275] (4,0)
    to [out=265,in=270] (0,0)
    to [out=270, in=0] (-2, -1)
    to [out=180, in=270] cycle;
    \draw[color=ForestGreen, line width=2pt] (-4,0) to [out=270, in=180] (-2, -1)
    to [out=0, in=270] (0,0);
    \draw[dashed, color=ForestGreen, line width=1pt] (-4,0) to [out=90, in=180] (-2, 1)
    to [out=0, in=90] (0,0);
\draw[ForestGreen, fill=ForestGreen] (-4,0) circle (.8ex);
\draw[ForestGreen, fill=ForestGreen] (0,0) circle (.8ex);
\draw[cornellred, fill=cornellred] (4,0) circle (.8ex);
\node at (-4.3, 0) {$a$}; 
\node at (0.3, 0) {$b$};
\node at (4.6, 0) {$x_{12}$}; 
\draw[dotted, modebeige, line width=2pt] (2, 1.2) to[out=160, in=300] (1,2.1);
\draw[dotted, modebeige, line width=1pt] (2, 1.2) to[out=100, in=350] (1,2.1);
\draw[dotted, modebeige, line width=1pt] (0.3, -0.65) to[out=170, in=70] (-2.7,-1.7);
\draw[dotted, modebeige, line width=2pt] (0.3, -0.65) to[out=250, in=350] (-2.7,-1.7);
    \draw[blue, line width=1.5pt] (0,0) to [out=90, in=180] (1.8, 1.6);
    \draw[blue, line width=1.5pt] (1.8, 1.6) to [out=0, in=90] (4, 0);
    \draw[blue, line width=1.5pt] (0,0) to [out=270, in=180] (1.8, -1.6);
    \draw[blue, line width=1.5pt, <-] (1.8, -1.6) to [out=0, in=270] (4, 0);
    \draw[black, fill] (1,2.1) circle (.5ex);
\end{scope} 
\begin{scope}[scale=0.7,  shift={({-6},{-8})}]
\draw[fill=LightBlue,fill opacity=0.2, line width=1pt] (-3,0) to [out=90,in=180] (0,3)
	to [out=0,in=90] (3,0)
    to [out=265,in=0] (0,-1.2)
    to [out=180, in=275] cycle;
    \draw[fill=darktangerine,fill opacity=0.2, line width=1pt] (-3,0) to [out=270,in=180] (0,-3)
	to [out=0,in=270] (3,0)
    to [out=265,in=0] (0,-1.2)
    to [out=180, in=275] cycle;
    \draw[color=ForestGreen, line width=2pt] (-3,0) to [out=275, in=180] (0, -1.2)
    to [out=0, in=265] (3,0);
    \draw[dashed, color=ForestGreen, line width=1pt] (-3,0) to [out=85, in=180] (0, 1.2)
    to [out=0, in=95] (3,0);
    \draw[ForestGreen, fill=ForestGreen] (-3,0) circle (.8ex);
\draw[ForestGreen, fill=ForestGreen] (3,0) circle (.8ex);
\draw[dotted, modebeige, line width=2pt] (-0.5, 2.4) to[out=350, in=40] (1, 0)
to[out=220, in=270] (-0.7, 1)
to[out=90, in=290] cycle;
\draw[dotted, modebeige, line width=2pt] (-2.8, -1) to[out=300, in=180] (0,-2)
to[out=0, in=220] (2.7, -1.25);
\draw[dotted, modebeige, line width=1pt] (-2.8, -1) to[out=60, in=180] (0,-0.2)
to[out=0, in=140] (2.7, -1.25);
\draw[orange, line width=2pt] (0.5, -2.3) to[out=70, in=270] (0.8, -0.5);
\draw[orange, line width=2pt, <-] (0.8, -0.5) to[out=90, in=290] (0.4, 1);
\draw[Maroon, fill=Maroon] (0.5,-2.35) circle (.8ex);
\draw[Maroon, fill=Maroon] (0.4,0.95) circle (.8ex);
\node at (-3.4, 0) {$a$};
\node at (3.4, 0) {$b$};
\node[modebeige] at (-0.5, -2.6) {$X_1$};
\node[modebeige] at (-1.8, 1.5) {$X_0$};
\node[modebeige] at (0, 1.8) {$X_2$};
\node at (-0.1, 0.8) {$x_{12}$};
\node at (1.1, -2.4) {$x_{12}$};
\node at (5.5, 0) {$\longrightarrow$};
\draw[black, fill] (-0.5,2.4) circle (.5ex);
\end{scope}
\begin{scope}[scale=0.7,  shift={({6},{-8})}]
	\draw[fill=LightBlue,fill opacity=0.2, line width=1pt] (0,0) to [out=90,in=95] (4,0)
	to [out=85,in=0] (1,2.1)
    to [out=180,in=90] (-4,0)
    to [out=270, in=180] (-2, -1)
    to [out=0, in=270] cycle;
    \draw[fill=darktangerine,fill opacity=0.2, line width=1pt] (-4,0)
    to [out=270,in=180] (1,-2.1)
    to [out=0,in=275] (4,0)
    to [out=265,in=270] (0,0)
    to [out=270, in=0] (-2, -1)
    to [out=180, in=270] cycle;
    \draw[color=ForestGreen, line width=2pt] (-4,0) to [out=270, in=180] (-2, -1)
    to [out=0, in=270] (0,0);
    \draw[dashed, color=ForestGreen, line width=1pt] (-4,0) to [out=90, in=180] (-2, 1)
    to [out=0, in=90] (0,0);
\draw[ForestGreen, fill=ForestGreen] (-4,0) circle (.8ex);
\draw[ForestGreen, fill=ForestGreen] (0,0) circle (.8ex);
\draw[cornellred, fill=cornellred] (4,0) circle (.8ex);
\node at (-4.3, 0) {$a$}; 
\node at (0.3, 0) {$b$};
\node at (4.6, 0) {$x_{12}$}; 
\draw[dotted, modebeige, line width=2pt] (2, 1.2) to[out=160, in=300] (1,2.1);
\draw[dotted, modebeige, line width=1pt] (2, 1.2) to[out=100, in=350] (1,2.1);
\draw[dotted, modebeige, line width=1pt] (0.3, -0.65) to[out=170, in=70] (-2.7,-1.7);
\draw[dotted, modebeige, line width=2pt] (0.3, -0.65) to[out=250, in=350] (-2.7,-1.7);
    \draw[orange, line width=1.5pt] (0,0) to [out=90, in=180] (1.8, 1.6);
    \draw[orange, line width=1.5pt] (1.8, 1.6) to [out=0, in=90] (4, 0);
    \draw[orange, line width=1.5pt, ->] (0,0) to [out=270, in=180] (1.8, -1.6);
    \draw[orange, line width=1.5pt] (1.8, -1.6) to [out=0, in=270] (4, 0);
    \draw[black, fill] (1,2.1) circle (.5ex);
\end{scope} 
	\end{tikzpicture}
\caption{Spectral curve of the two-matrix model with $d_1=2$, $d_2=1$. On the left we plot the spectral curve via uniformization on the sphere, where the two points are identified with each other, \textit{i.e.}, $x_{12}=\mathcal{X}(p)$. Their connecting $B$-cycles are plotted in analogy with \cite{mss22}, and the different sheets are also illustrated (either by color or by name). Swapping $Y(x)$ physical ({\color{LightBlue}blue}) and non-physical ({\color{darktangerine}orange}) sheets as in \eqref{eq:effectiveactiontwomatrixmodel} effectively corresponds to reversing the integration direction of the cycle (recall that the integration direction which is consistent with the matrix model is opposed to the BCFT one). More complicated configurations will have a multitude of sheets, albeit nonperturbative saddles always lie at double-points and (at least locally) the discussion in \cite{mss22} applies. These present plots build on figures first appearing in \cite{kk04}.}
\label{fig:twomatrixexample23}
\end{figure}

It is now clear how the constructions in \cite{mss22} should translate to the two-matrix model case. In spite of now finding a multitude of sheets (rather than just two), the generic structure at the nonperturbative saddle has not changed: in fact, therein, one still finds two distinct sheets coming together. In addition, swapping sheets effectively reverses the integration direction in \eqref{eq:effectiveactiontwomatrixmodel} leading to an overall minus sign, and eventually to the subsequent transseries resonant structures as computed in the one-matrix model case \cite{mss22}. The difficulty now, of course, is making this explicit for all possible nonperturbative saddles, which we leave for future work.

As a warm-up calculation, let us herein focus on the one-instanton contribution and its resonant sibling. The two-matrix model one-instanton contribution has been investigated in detail in \cite{iky05}. Moreover the full double-scaled minimal-string result was computed therein and it is consistent with the computations in this paper (\textit{e.g.}, compare formulae (27) and (35) in section~4 of \cite{iky05} to our formula \eqref{eq:1-instanton-result}). Indeed the instanton action computed in \cite{iky05} is exactly of the type \eqref{eq:effectiveactiontwomatrixmodel}. Therefore we expect resonance to be a clear feature of the two-matrix model. Furthermore let us briefly investigate the chemical potential associated to the one-instanton. It reads \cite{iky05, emms22b}
\begin{equation}
\label{eq:two-matrix-chemical-potential}
\frac{\mathcal{Z}^{(1|0)}}{\mathcal{Z}^{(0|0)}} \equiv \frac{\mathcal{Z}_{m n}}{\mathcal{Z}} = \sqrt{\frac{g_{\text{s}}}{2\pi}}\, \frac{1}{\left(\partial\CX(p_m)\partial\CY(p_n) - \partial\CX(p_n)\partial\CY(p_m)\right)^{1/2}}\, \frac{1}{p_m-p_n}\, \rme^{-\frac{A_{mn}}{g_{\text{s}}}} + \cdots,
\end{equation}
\noindent
where in the first identification we are making contact with the notation in section~\ref{sec:ZZminimalstring}---and hence label nonperturbative contributions by the sheets that their $B$-cycles connect. Let us now understand how the chemical potential changes when comparing an instanton to a negative-instanton. This amounts to swapping $n$ with $m$ in the above \eqref{eq:two-matrix-chemical-potential}. This produces a factor of $\rmi$ which is exactly consistent with our results in subsection~\ref{subsec:bcft-resonant-pairs}. Furthermore, it agrees with the general expectation in resonant problems \cite{asv11, gs21, mss22}. We leave a careful study of the full two-matrix model nonperturbative contributions for future work.

\paragraph{$(p,q)$ Minimal String Theory:}

The double-scaled version of the previous discussion is of course simpler, and it further matches to the $(p,q)$ BCFT minimal string discussion in section~\ref{sec:ZZminimalstring}, as well as reduces to the $(2,2k-1)$ one-matrix model minimal-string discussion of subsection~\ref{subsec:22k-1minstring}. Recall how the relevant Riemann surface $\Sigma_{p,q}$ is now genus-zero with $\frac{1}{2} \left( p-1 \right) \left( q-1 \right)$ pinched $A$-cycles, and is given by \eqref{eq:minimal-string-Tp-Tq}. These (pinched) singularities sit at \cite{ss03}
\begin{equation}
(x_{mn}, y_{mn}) = \left( (-1)^m \cos \frac{n \pi p}{q}, (-1)^n \cos \frac{m \pi q}{p} \right), \qquad \upzeta_{mn}^{\pm} = \cos \pi \left( \frac{m}{p} \pm \frac{n}{q} \right),
\end{equation}
\noindent
with $1 \le m \le p-1$, $1 \le n \le q-1$, $q m - p n > 0$, and where $\upzeta$ is the uniformization parameter. The instanton contributions are given by integrating the spectral curve over $B$-cycles across these singularities \cite{ss03}
\be
A_{mn} = \oint_{B_{mn}} \text{d}x\, y.
\ee
\noindent
As for the two-matrix model, swapping sheets at the pinched double-points will reverse integration directions and produce the minus-sign which gives rise to the resonant contributions. This is exactly our result in section~\ref{sec:ZZminimalstring}---switching sheets will reverse the integration direction of the $B_{mn}$ cycle as underlined in subsection~\ref{subsec:bcft-resonant-pairs} (and recall that the matrix-model integration-direction is reversed in comparison to BCFT as explained earlier).

\section{Towards Topological and Critical String Theories}\label{sec:topcritstrings}

Having understood how negative-tension D-branes are an unavoidable part of minimal string theory---in fact being \textit{required} by resurgence---the next question is how ubiquitous may they be across generic string theoretic backgrounds. In this section we begin the study of more intricate models, in topological and critical string theory. The results we report upon are not as exhaustive as in the minimal string example (due to the natural computational intricacies) but we nonetheless believe they amount to very clear and supporting evidence. Further work along these directions will be reported in the near future. On the topological string theory side, we focus upon one of the simplest toric Calabi--Yau geometries, the local curve \cite{m06}. This model may be solved with matrix-model methods and it hence builds upon \cite{mss22} and our earlier section~\ref{sec:NPminimalstring}. On the string theory side, we focus on D-branes in AdS, in particular the case of $\text{AdS}_3$ where BCFT methods come of help via the $\BH_3^+$--Liouville correspondence \cite{rt05, hr06, hs07} mapping the present computation to the Liouville analysis of section~\ref{sec:ZZminimalstring}. In particular, we address the analogous to the Liouville ZZ-calculations now for the case of discrete $\text{AdS}_2$ D-branes inside $\text{AdS}_3$ \cite{r05}.

\subsection{On Topological Strings on Toric Calabi--Yau Geometries}\label{subsec:top-strings}

Let us begin with topological string theory. It is a famous result that B-model topological strings on certain local Calabi--Yau backgrounds (the resolved $p=2$ local curve, more below) are equally described by hermitian matrix models \cite{dv02}. This is to say that they are described at resurgent, nonperturbative level by the results in \cite{mss22} and section~\ref{sec:NPminimalstring} herein. It is a fascinating story that this idea applies to more general toric Calabi--Yau geometries, in particular to toric varieties with enumerative geometry content---where making use of the topological recursion \cite{eo07a} on the \textit{mirror curve}\footnote{Recall how the mirror geometry to a toric Calabi--Yau threefold is essentially captured by a Riemann surface.} of the toric variety remarkably yields\footnote{There is one subtlety: this is not a \textit{verbatim} application of the topological recursion, as the mirror curve lives in $\BC^* \times \BC^*$ rather than $\BC \times \BC$; see \cite{bkmp07} for details on the required meromorphic-differential modification.} the complete B-model topological-string perturbative free-energy expansion, anywhere in moduli space \cite{m06, bkmp07} (in fact yielding both open- and closed-string B-model amplitudes). This further stands as an alternative computational procedure to the holomorphic anomaly equations \cite{bcov93}, free from any such ambiguities. Now, one clear feature of the results in \cite{mss22} alongside our matrix-model results in section~\ref{sec:NPminimalstring} is that they only depend upon large $N$ spectral-curve data. As such, they may also be (adequately) applied to a mirror curve and hence describe the resonant and resurgent nonperturbative content of these backgrounds \cite{msw07}. The one caveat is that our results only apply to one-cut models. Effectively, this reduces our range of topological string examples to local curves.

The local curve $X_p$ is a (family of) non-compact toric Calabi--Yau threefold described by the total space of a complex fiber bundle over a sphere (a direct sum of two line bundles; see, \textit{e.g.}, \cite{m04} for a review)
\be
X_p = \NCO (p-2) \oplus \NCO (-p) \to \BP^1.
\ee
\noindent
It is labeled by a single integer $p \in \BZ$, but as unchanged by $-p \leftrightarrow p-2$ one may restrict to $p \in \BN$. When $p=1$ this is the resolved conifold geometry \cite{gv95, gv98b}. When $p=2$ we obtain the aforementioned Dijkgraaf--Vafa geometries $\NCO (0) \oplus \NCO (-2) \to \BP^1$ related to hermitian matrix models \cite{dv02}. Herein we will explicitly use the non-trivial example of $p=3$. The free energies of A-model topological strings on $X_p$ were set-up in \cite{cgmps06}, and---albeit mirror symmetry in these geometries is subtle---a B-model-like computation was also set-up in this reference. Topological string amplitudes on this background have familiar (phase transition) singularities, but which are now in the $c=0$ critical class \cite{cgmps06} (where its associated double-scaled Painlev\'e~I equation was addressed in our present context in \cite{mss22}) rather than the usual conifold $c=1$ class \cite{gv95}.

The B-model-like computation of \cite{cgmps06} was revisited in \cite{msw07, csv16} and we will use the data computed therein (and refer the reader to those references for further details). This computation first yields the topological-string genus-$g$ free-energies on $X_p$ as $F_g \equiv F_g (\xi)$, where $\xi$ is a mirror-like coordinate relating to the  K\"ahler parameter $t$ (the size of the $\BP^1$) via the mirror map\footnote{A word on conventions: the mirror-like coordinate $w$ in \cite{cgmps06, csv16} relates to $\xi$ in \cite{m06, msw07} as $w=1-\xi$.}
\be
\label{eq:local-curve-mirror-map}
\rme^{-t} = \xi \left( 1-\xi \right)^{p\left(p-2\right)}.
\ee
\noindent
As already mentioned, one may also compute free energies for toric geometries using the topological recursion \cite{eo07a} on the appropriate curve---in this case on the mirror curve \cite{m06, bkmp07}. For the example we are interested in, it turns out that the curve will still be of the form \eqref{eq:hyperelliptic-spectral-curve} albeit now with non-polynomial moment function. More specifically, for the local curve one finds \cite{m06, bkmp07} (similar-looking spectral curves also hold for other toric mirrors)
\be
\label{eq:local-curve-spectral-curve}
y(x) = \frac{2}{x} \left( \tanh^{-1} \left( \frac{\sqrt{\left(x-a\right) \left(x-b\right)}}{x-\frac{1}{2} \left(a+b\right)} \right) - p\, \tanh^{-1} \left( \frac{\sqrt{\left(x-a\right) \left(x-b\right)}}{x+\sqrt{a b}} \right) \right),
\ee
\noindent
in which parametrization we can indeed apply the usual recursion \cite{eo07a}. Herein, the endpoints of the single cut $\NCC = [a,b]$ are located at
\be
a = \frac{\left( 1-\sqrt{\xi} \right)^2}{\left( 1-\xi \right)^p}, \qquad b = \frac{\left( 1+\sqrt{\xi} \right)^2}{\left( 1-\xi \right)^p}.
\ee

The nonperturbative study of topological strings on this background was initiated in \cite{m06} and then continued in \cite{msw07, csv16}. Nonperturbative saddles sit at the zeroes of the moment function, which for the explicit $p=3$ example yields
\be
\label{eq:localcurvepeq3saddle}
x^{\star} = \frac{4 a b}{(\sqrt{a}-\sqrt{b})^2}, \qquad p=3.
\ee
\noindent
This leads to its corresponding instanton action, albeit the expression is now very lengthy \cite{m06}. For completeness, first introduce the functions
\bea
f_1 (x) &=& \sqrt{\left(x-a\right) \left(x-b\right)} + x-\frac{1}{2} \left(a+b\right), \\
f_2 (x) &=& \sqrt{\left(x-a\right) \left(x-b\right)} + x+\sqrt{a b},
\eea
\noindent
which are then used in the main expression
\bea
F(x) &=& \frac{1}{2} p\, \log^2 x + p\, \log x\, \log^2 (\sqrt{a}+\sqrt{b}) - \log x\, \log \frac{\left( a-b \right)^2}{4} - 2\, \text{Li}_2 \left( \frac{- 2 f_1(x)}{(\sqrt{a}-\sqrt{b})^2} \right) - \nonumber \\
&&
- 2\, \text{Li}_2 \left( \frac{- 2 f_1(x)}{(\sqrt{a}+\sqrt{b})^2} \right) - 2p\, \text{Li}_2 \left( \frac{- f_2(x)}{2 \sqrt{ab}} \right) + 2p\, \text{Li}_2 \left( \frac{2 f_2(x)}{(\sqrt{a}+\sqrt{b})^2} \right) - \nonumber \\
&&
- \left\{ \log f_1(x) + \log \left( 1+\frac{2 f_1(x)}{(\sqrt{a}+\sqrt{b})^2} \right) - 2 \log \left( 1+\frac{2 f_1(x)}{(\sqrt{a}-\sqrt{b})^2} \right) \right\} \log f_1(x) - \nonumber \\
&&
- p \left\{ \log f_2(x) - \log \left( 1-\frac{2 f_2(x)}{(\sqrt{a}+\sqrt{b})^2} \right) + 2 \log \left( 1-\frac{f_2(x)}{2 \sqrt{a b}} \right) \right\} \log f_2(x).
\eea
\noindent
The instanton action finally follows from the usual ``difference of holomorphic effective potentials'' as
\be
A (\xi) = F(x^\star) - F(a).
\ee

\begin{figure}
\centering
\includegraphics[scale=0.65]{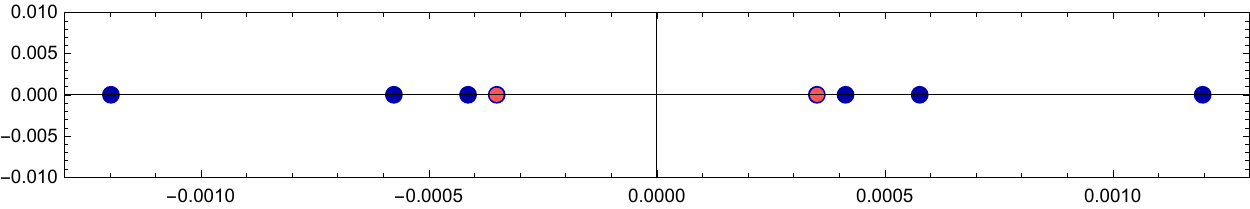}
\vspace{0.5cm}
\includegraphics[scale=0.65]{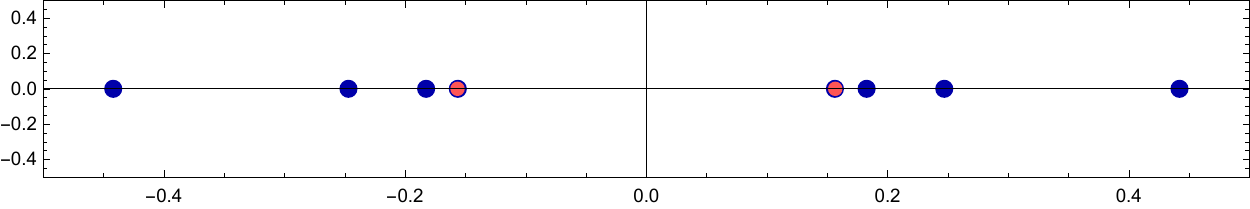}
\caption{Plot of poles from the Borel--Pad\'e approximant to the perturbative-sector data of the local curve with $p=3$ (using data up to $g_{\text{s}}$-order $18$ \cite{cgmps06, m06, msw07, csv16}). The upper plot is for $\xi=0.24$ while the lower one was plotted at $\xi=0.15$. The data points are shown in blue while the theoretical prediction for the resonant instanton actions is visualized in red. The inherent symmetry of Borel singularities in the plots fully supports our hypothesis of resonance.}
\label{fig:local-curve-borel-pade}
\end{figure}

For the explicit example of $p=3$, the instanton-action alongside one- and two-loop coefficients around the one-instanton sector were tested against large-order behavior in \cite{msw07}. Let us now briefly expand on these. Such tests are carried out in the B-model but use the computation of the free-energy coefficients which is obtained from A-model data following \cite{cgmps06, m06, csv16}. This A-model computation is based on the knowledge of high-degree Gromov--Witten invariants and is therefore limited in its computational capacity. Herein we have computed the first $18$ perturbative free-energy coefficients. The results of the corresponding Borel--Pad\'e analysis are visualized in figure~\ref{fig:local-curve-borel-pade} and they perfectly support the resonance hypothesis for the local curve.

On top of these tests, we can now directly use the matrix model results of \cite{mss22}, as well as our section~\ref{sec:NPminimalstring}, in order to compute several transseries sectors for the free energy of the topological string on the local curve $X_p$ (we shall occasionally specialize to the case $p=3$, where we can evaluate the saddle $x^{\star}$ explicitly in a simple form). As usual we begin with the $(\ell|0)$ sector. Curiously, even though the moment-function has a complicated functional form, its derivative evaluated on the saddle $x^{\star}$ simplifies considerably as
\begin{equation}
M'(x^{\star}) = \frac{\left(p-2\right) x^{\star} - p\, \sqrt{ab}}{\left(x^{\star}\right)^2 \left(a-x^{\star}\right) \left(x^{\star}-b\right)}.
\end{equation}
\noindent
Then, inserting the explicit expression for the $(\ell|0)$ sector given in \cite{mss22} (whose double-scaled version is given in \eqref{eq:dsl-l0-sector}) we find\footnote{The sign in front of the instanton-action in the exponent is consistent with the conventions in \cite{m06}. Furthermore, as we shall shortly see, it is also consistent with the sign-convention used in our minimal string actions \eqref{eq:22km1MinimalStringActions}.}
\begin{equation}
\label{eq:pLocalCurvel0Sector}
\frac{\mathcal{Z}_{X_p}^{(\ell|0)}(\xi, g_{\text{s}})}{\mathcal{Z}_{X_p}^{(0|0)}(\xi, g_{\text{s}})} \simeq \frac{G_{2}(\ell+1)}{\left(2\pi\right)^{\ell/2}}\, \rme^{-\frac{\ell}{g_{\text{s}}} A (\xi)} \left( \frac{g_{\text{s}} \left( x^{\star} \right)^2 \left(a-b\right)^2}{16 \left(x^{\star}-a\right)^{3/2} \left(x^{\star}-b\right)^{3/2} \left\{ p \sqrt{a b} - \left(p-2\right) x^{\star} \right\}} \right)^{\frac{\ell^2}{2}} + \cdots.
\end{equation}
\noindent
Since some of our explorations are done for the explicit case of $p=3$, we specialize the above expression to this value and evaluate explicitly on the saddle \eqref{eq:localcurvepeq3saddle} to find
\begin{align}
\label{eq:3LocalCurvel0Sector}
\frac{\mathcal{Z}_{X_3}^{(\ell|0)}(\xi, g_{\text{s}})}{\mathcal{Z}_{X_3}^{(0|0)}(\xi, g_{\text{s}})} &\simeq \frac{G_{2}(\ell+1)}{\left(2\pi\right)^{\ell/2}}\, \rme^{-\frac{\ell}{g_{\text{s}}} A (\xi)} \times \\
& 
\hspace{-50pt}
\times \left(\frac{g_{\text{s}}\, a^2 b^2 \left(a-b\right)^2 \left(\sqrt{a}-\sqrt{b}\right)^4}{\left( a b \left( 4a - \left(\sqrt{a}-\sqrt{b}\right)^2 \right) \left( 4b - \left(\sqrt{a}-\sqrt{b}\right)^2 \right) \right)^{3/2} \left( 3 \sqrt{a b} \left(\sqrt{a}-\sqrt{b}\right)^2 - 4 a b \right)} \right)^{\frac{\ell^2}{2}} + \nonumber \\
&
\hspace{-50pt}
+ \cdots. \nonumber
\end{align}
\noindent
We now move on to the case $(0|\bar{\ell})$. As is by now expected, this case only differs by the sign flip of the action, yielding
\begin{equation}
\frac{\mathcal{Z}_{X_p}^{(0|\bar{\ell})}(\xi, g_{\text{s}})}{\mathcal{Z}_{X_p}^{(0|0)}(\xi, g_{\text{s}})} \simeq \frac{G_{2}(\bar{\ell}+1)}{\left(2\pi\right)^{\bar{\ell}/2}}\, \rme^{+\frac{\bar{\ell}}{g_{\text{s}}} A (\xi)} \left( - \frac{g_{\text{s}} \left( x^{\star} \right)^2 \left(a-b\right)^2}{16 \left(x^{\star}-a\right)^{3/2} \left(x^{\star}-b\right)^{3/2} \left\{ p \sqrt{a b} - \left(p-2\right) x^{\star} \right\}} \right)^{\frac{\bar{\ell}^2}{2}} + \cdots. 
\end{equation}
\noindent
Specializing once more to $p=3$ we find
\begin{align}
\frac{\mathcal{Z}_{X_3}^{(0|\bar{\ell})}(\xi, g_{\text{s}})}{\mathcal{Z}_{X_3}^{(0|0)}(\xi, g_{\text{s}})} &\simeq \frac{G_{2}(\bar{\ell}+1)}{\left(2\pi\right)^{\bar{\ell}/2}}\, \rme^{+ \frac{\bar{\ell}}{g_{\text{s}}} A (\xi)} \times \\
&
\hspace{-50pt}
\times \left( - \frac{g_{\text{s}}\, a^2 b^2 \left(a-b\right)^2 \left(\sqrt{a}-\sqrt{b}\right)^4}{\left( a b \left( 4a - \left(\sqrt{a}-\sqrt{b}\right)^2 \right) \left( 4b - \left(\sqrt{a}-\sqrt{b}\right)^2 \right) \right)^{3/2} \left( 3 \sqrt{a b} \left(\sqrt{a}-\sqrt{b}\right)^2 - 4 ab \right)} \right)^{\frac{\bar{\ell}^2}{2}} + \nonumber \\
&
\hspace{-50pt}
+ \cdots. \nonumber
\end{align}
\noindent
The interesting non-trivial sectors are hidden in the ``bulk'' of the transseries. With a little bit more of work, we can compute the local curve $(1|1)$ transseries sector, up to and including order $g_{\text{s}}$ terms. This expression is however a bit messier, so below we only write the generic formula:
\begin{align}
\label{eq:pLocalCurve11Sectorgenus1}
\frac{\mathcal{Z}_{X_p}^{(1|1)}(\xi, g_{\text{s}})}{\mathcal{Z}_{X_p}^{(0|0)}(\xi, g_{\text{s}})} &= \frac{\rmi}{2\pi g_{\text{s}}}\,  A(\xi) - \\
&
\hspace{-55pt}
- \frac{\rmi g_{\text{s}}}{96\pi \left(x^{\star}-a\right)^{3/2} \left(x^{\star}-b\right)^{3/2} \left( p \sqrt{ab} - \left(p-2\right) x^{\star} \right)^3}\, \bigg\{ ab\, p^2 \left\{ 2 \left( x^{\star} \right)^2  \left(a^2+b^2\right) - 4 a^2 b^2 + 7 ab\, x^{\star} \left(a+b\right) \right\} + \nonumber \\
&
\hspace{-45pt}
+ \left(p-2\right)^2 \left( x^{\star} \right)^4 \left\{ 2 \left(a^2+b^2\right) + 7 x^{\star} \left(a+b\right) - 4 \left( x^{\star} \right)^2 \right\} + 2 p \left(p-2\right) x^{\star} \sqrt{ab}\, \times \nonumber \\
&
\hspace{-45pt}
\times \left\{ 5 a^2 b^2 - 9 \left( x^{\star} \right)^3 \left(a+b\right) - \left( x^{\star} \right)^2 \left( a - 10 \sqrt{ab} + b \right) \left( a + 3 \sqrt{ab} + b \right) - 9 a b\, x^{\star} \left(a+b\right) + 5 \left( x^{\star} \right)^4 \right\} + \nonumber \\
&
\hspace{-45pt}
+ 4 ab \left( x^{\star} \right)^2 \left\{ \left( 24 \left(p-1\right) - 7p^2 \right) \left( ab+\left( x^{\star} \right)^2 \right) - 20 a b \left(p-1\right) + 7 x^{\star} \left(a+b\right) \right\} \bigg\} - \nonumber \\
&
\hspace{-55pt}
- \frac{\rmi g_{\text{s}} \left( \left(p-2\right) \sqrt{ab} - p x^{\star} \right)}{48\pi \left(x^{\star}-a\right)^{3/2} \left(x^{\star}-b\right)^{3/2} \left( \left(p-1\right)^2 \left(\sqrt{a}-\sqrt{b}\right)^2 - \left(\sqrt{a}+\sqrt{b}\right)^2 \right)^2}\, \times \nonumber \\
&
\hspace{-45pt}
\times \left\{ \left( \left(p-1\right)^2 \left(\sqrt{a}-\sqrt{b}\right)^2 - \left(\sqrt{a}+\sqrt{b}\right)^2 \right) \left( \left( 3a + 2 \sqrt{ab} + 3b \right) \left( \left( x^{\star} \right)^2 + ab - \frac{2}{9} x^{\star} \left( 6a - \sqrt{ab} + 6b \right) \right) + \right. \right. \nonumber \\
&
\hspace{-45pt}
\left. \left.
+ \frac{32}{9} ab\, x^{\star} \right) + 4 \sqrt{ab} \left(\sqrt{a}+\sqrt{b}\right)^2 \left(a-x^{\star}\right) \left(b-x^{\star}\right) \right\} + \cdots. \nonumber
\end{align}
\noindent
Finally, the (non-trivial part of the) $(2|1)$ sector is given by
\begin{align}
\frac{\mathcal{Z}_{X_p}^{(2|1)}(\xi, g_{\text{s}})}{\mathcal{Z}_{X_p}^{(0|0)}(\xi, g_{\text{s}})} - \frac{\mathcal{Z}_{X_p}^{(1|0)}(\xi, g_{\text{s}})}{\mathcal{Z}_{X_p}^{(0|0)}(\xi, g_{\text{s}})}\, \frac{\mathcal{Z}_{X_p}^{(1|1)}(\xi, g_{\text{s}})}{\mathcal{Z}_{X_p}^{(0|0)}(\xi, g_{\text{s}})} &\simeq \rme^{-\frac{A(\xi)}{g_{\text{s}}}}\, \frac{\rmi x^{\star} \left(a-b\right) \sqrt{g_{\text{s}}\, \sqrt{\left(a-x^{\star}\right) \left(b-x^{\star}\right)}}}{8 \pi^{3/2} \left(a-x^{\star}\right) \left(x^{\star}-b\right) \sqrt{2 p \sqrt{a b} - 2 \left(p-2\right) x^{\star}}} \times \nonumber \\
&
\hspace{-140pt}
\times \left( 2 \gamma_{\text{E}} + \log \left( \frac{256 \left(a-x^{\star}\right)^3 \left(b-x^{\star}\right)^3 \left( p \sqrt{a b} - \left(p-2\right) x^{\star} \right)^2}{g_{\text{s}}^2\, \left( x^{\star} \right)^4 \left(a-b\right)^4} \right) \right) + \cdots.
\end{align}
\noindent
Specializing to the case $p=3$ we find
\begin{align}
\frac{\mathcal{Z}_{X_3}^{(2|1)}(\xi, g_{\text{s}})}{\mathcal{Z}_{X_3}^{(0|0)}(\xi, g_{\text{s}})} - \frac{\mathcal{Z}_{X_3}^{(1|0)}(\xi, g_{\text{s}})}{\mathcal{Z}_{X_3}^{(0|0)}(\xi, g_{\text{s}})}\, \frac{\mathcal{Z}_{X_3}^{(1|1)}(\xi, g_{\text{s}})}{\mathcal{Z}_{X_3}^{(0|0)}(\xi, g_{\text{s}})} &\simeq \rme^{-\frac{A(\xi)}{g_{\text{s}}}} \times \\
&
\hspace{-120pt}
\times \frac{\rmi \left(\sqrt{a}-\sqrt{b}\right)^3 \sqrt{g_{\text{s}} \left( a - 2 \sqrt{a b} + b \right) \sqrt{\frac{a b \left( 4 a \sqrt{a b} - 3 a^2 + 14 a b - 3 b^2 + 4 b \sqrt{a b} \right)}{\left( \sqrt{a}-\sqrt{b} \right)^4}}}}{2\pi^{3/2}\, \sqrt{6 a \sqrt{a b} - 20 a b + 6 b \sqrt{a b}} \left(\sqrt{a}-3\sqrt{b}\right) \left(3\sqrt{a}-\sqrt{b}\right) \left(\sqrt{a}+\sqrt{b}\right)} \times \nonumber \\
&
\hspace{-180pt}
\times \left( 2\gamma_{\text{E}} + \log \left(\frac{\left(a + 2 \sqrt{a b} + b\right) \left( 10 \sqrt{a b} - 3 a - 3 b \right)^3 \left( 3 a \sqrt{a b} - 10 a b + 3 b \sqrt{a b} \right)^2}{g_{\text{s}}^2\, a b \left(\sqrt{a}-\sqrt{b}\right)^{12}} \right) \right) + \cdots. \nonumber
\end{align}

One \textit{a priori} clue that the results we just described were somehow bound to work, at least on what concerns the existence of negative-tension toric D-branes for the local curve, is that in the adequate double-scaling limit these D-branes become $c=0$ FZZT-branes \cite{m06}. On top of this, either matrix model \cite{msw07} or local curve \cite{m06, msw07} instanton actions are given by differences of holomorphic effective potentials, which is very much the same spirit in which ZZ-brane instantons are written as differences of FZZT branes \eqref{eq:ZZ-from-FZZTs} \cite{zz01, m03} (and this was further discussed from the point-of-view of toric D-branes in \cite{msw07}). It hence should come as no surprise that our earlier Liouville results in section~\ref{sec:ZZminimalstring} have uplifted to topological strings on a toric Calabi--Yau.

Let us make this explicit in the following (also serving as a double-check on our formulae above). The $c=0$ critical point is of course fully described by the Painlev\'e~I two-parameter resurgent transseries which has been greatly studied in the literature \cite{gikm10, asv11, as13, bssv22, mss22}. The aforementioned double-scaling of the local curve has been addressed at the level of the perturbative free energies and instanton action in \cite{cgmps06, m06} and we shall now do this at full (two parameter) transseries level. Our first observation is the (relative) simplicity of the above local-curve transseries-sectors. For instance, note how the inverse hyperbolic tangent function does not seem to play any relevant role at transseries level, whereas it is very much present in the moment function in \eqref{eq:local-curve-spectral-curve}. This will somehow conspire so as to make the upcomning double-scaling limit quite straightforward even at full nonperturbative level. In the conventions\footnote{Up to differences in conventions, this is almost the result of setting $k=2$ in our results of section~\ref{sec:NPminimalstring}; but see \cite{bssv22} for full details on the different conventions at play.} of \cite{m06, asv11}, the Painlev\'e~I equation is given by
\begin{equation}
u^2(z) - \frac{1}{6} u^{\prime\prime}(z) = z.
\end{equation}
\noindent
Let us now establish a strong consistency test on the local-curve results via double-scaling to the Painlev\'e~I transseries data (for example, see section~5 of \cite{asv11}). Following directly the setup in \cite{cgmps06, m06}, we write the local curve modulus $\xi$ in terms of the Painlev\'e~I variable as
\begin{equation}
\label{eq: XiDoubleScaleDefinition}
\xi=\xi_{\text{crit}}-\left(\frac{4 \left(1-\xi_{\text{crit}}\right)^3 g_{\text{s}}^2\, z^{5/2}}{\left(p-1\right)^8}\right)^{\frac{1}{5}},
\end{equation}
\noindent
with $\xi_{\text{crit}}=\frac{1}{\left(p-1\right)^2}$. As explained in \cite{m06}, the double-scaling limit of the local-curve at criticality is obtained for $p>2$, by taking $g_{\text{s}} \to 0$ and $\xi \to \xi_{\text{crit}}$ while keeping $z$ fixed. Choosing $p=3$, let us first make this calculation very explicit for our results of the $(\ell|0)$ sectors in \eqref{eq:3LocalCurvel0Sector}. Writing the endpoints of the cut $a,b$ in terms of $\xi$, and substituting in the definition \eqref{eq: XiDoubleScaleDefinition}, we find to lowest order in $g_{\text{s}}$ (note $\xi_{\text{crit}}=\frac{1}{4}$ here)
\begin{equation}
\frac{\mathcal{Z}_{X_3}^{(\ell|0)}(g_{\text{s}})}{\mathcal{Z}_{X_3}^{(0|0)}(g_{\text{s}})} \simeq \frac{G_{2}(\ell+1)}{\left(2\pi\right)^{\ell/2}} \left(-\frac{1}{96 \sqrt{3} z^{5/4}}\right)^{\frac{\ell^2}{2}} \rme^{-\ell \frac{8\sqrt{3}}{5} z^{\frac{5}{4}}} + \cdots.
\end{equation}
\noindent
This formula precisely matches what we would expect for the Painlev\'e~I multi-instantons. In fact, it is enough to show this matching for $\ell=1$ where the above becomes
\begin{equation}
\frac{\mathcal{Z}_{X_3}^{(1|0)}(g_{\text{s}})}{\mathcal{Z}_{X_3}^{(0|0)}(g_{\text{s}})} \simeq \frac{\rmi}{8 \cdot 3^{\frac{3}{4}}\sqrt{\pi}}\, \rme^{-\frac{8\sqrt{3}}{5} z^{\frac{5}{4}}} + \cdots.
\end{equation}
\noindent
Using the canonical Stokes coefficient straight from \cite{asv11, bssv22}, $\mathsf{S}_{(0,0)\to(1,0)}= -\frac{\rmi \sqrt[4]{3}}{2 \sqrt{\pi }}$ alongside the first non-zero free-energy coefficient in the $(1,0)$ sector $F_{0}^{(1,0)} = -\frac{1}{12}$, we find that their product precisely reproduces the above double-scaling limit. Moving on, we compute the double-scaling limit of the $(0|\bar{\ell})$ sectors (again for $p=3$). Without surprise, these are given by
\begin{equation}
\frac{\mathcal{Z}_{X_3}^{(0|\bar{\ell})}(g_{\text{s}})}{\mathcal{Z}_{X_3}^{(0|0)}(g_{\text{s}})} \simeq \frac{G_{2}(\bar{\ell}+1)}{\left(2\pi\right)^{\bar{\ell}/2}} \left(\frac{1}{96 \sqrt{3} z^{5/4}}\right)^{\frac{\bar{\ell}^2}{2}} \rme^{\bar{\ell} \frac{8\sqrt{3}}{5} z^{\frac{5}{4}}} + \cdots,
\end{equation}
\noindent
where again we find immediate exact matching by comparison against the data in \cite{asv11}. A more non-trivial check is to consider the double-scaling at $p=3$ of formula \eqref{eq:pLocalCurve11Sectorgenus1} for the $(1|1)$ transseries sector, at next-to-leading order in the genus\footnote{The calculation at lowest genus is effectively just the instanton-action, which to some extent had already been checked to double-scale correctly in \cite{cgmps06, m06}.}. In spite of the lengthy formula, it turns out we immediately find the rather simple
\begin{equation}
\frac{\mathcal{Z}_{X_3}^{(1|1)}(g_{\text{s}})}{\mathcal{Z}_{X_3}^{(0|0)}(g_{\text{s}})} \simeq  \frac{4 \rmi \sqrt{3}}{5\pi}\, z^{\frac{5}{4}} + \frac{17\rmi}{384\sqrt{3}\pi}\, \frac{1}{z^{5/4}} + \cdots,
\end{equation}
\noindent
again in perfect agreement with the expected Painlev\'e~I results. Finally, we compute the double-scaled results for the $(2|1)$ sector at $p=3$. This is
\begin{equation}
\frac{\mathcal{Z}_{X_3}^{(2|1)}(g_{\text{s}})}{\mathcal{Z}_{X_3}^{(0|0)}( g_{\text{s}})} - \frac{\mathcal{Z}_{X_3}^{(1|0)}(g_{\text{s}})}{\mathcal{Z}_{X_3}^{(0|0)}( g_{\text{s}})}\, \frac{\mathcal{Z}_{X_3}^{(1|1)}( g_{\text{s}})}{\mathcal{Z}_{X_3}^{(0|0)}( g_{\text{s}})} \simeq - \frac{4 \gamma_{\text{E}} + 5 \log z + 20 \log 2 + \log 729}{32 \cdot 3^{3/4} \pi^{3/2}\, z^{5/8}}\, \rme^{-\frac{8\sqrt{3}}{5} z^{\frac{5}{4}} } + \cdots,
\end{equation}
\noindent
once again in precise agreement with the expected results one obtains by substituting Painlev\'e~I data from \cite{asv11} into the double-scaled matrix models in \cite{mss22}. All these checks amount to rather strong motivation on the correctness of all our above topological-string expressions and their nonperturbative materialization for the local curve.

Having obtained the nonperturbative free energy for topological string theory on the local curve, we may ask about its enumerative-geometry content. On this regard, a relation between Stokes data and nonperturbative enumerative-invariants of toric Calabi--Yau geometries via topological strings was recently established in \cite{gm21}---and which we very briefly follow next. Consider the A-model genus-$g$ free energies written in the flat coordinate $T$, as $F_g \equiv F_g (T)$, and where $T$ is now chosen\footnote{Instead of the usual large-radius expansion yielding perturbative Gromov--Witten invariants as in \cite{cgmps06}.} so that the Painlev\'e~I $c=0$ critical point is located at $T=0$. Mapping the free energies back from the B-model via \eqref{eq:local-curve-mirror-map}, this then simply amounts to a shift of the standard K\"ahler parameter $t$ by \cite{cgmps06}
\begin{equation}
t_{\text{c}} = \log \left( \left( p \left(p-2\right) \right)^{p \left(2-p\right)}\left(p-1\right)^{2 \left(p-1\right)^2}\right).
\end{equation}
\noindent
In this frame, and much like in \cite{gm21}, the perturbative local-curve free-energies split as
\be
F_g (T) = F_g^{\text{sing}} (T) + F_g^{\text{reg}} (T).
\ee
\noindent
Herein, the singular piece---which was the familiar conifold contribution \cite{gv95} in \cite{gm21}---is now the ``A-model version of Painlev\'e~I''. It is an expansion in both integer and half-integer powers of $T=t-t_{\text{c}}$, and where the usual gap-condition is of course not present: instead all negative (half-integer) powers are present up to $-\frac{5}{2} \left(g-1\right)$. The general structure is
\begin{equation}
F_g^{\text{sing}}(T) = \sum_{j=0}^{5 \left(g-1\right)} \frac{F_{g,j}^{\text{sing}}(p)}{T^{j/2}}.
\end{equation} 
\noindent
For example, the leading singularity at $g=2$ is
\begin{align}
F_{g=2}^{\text{sing}}(T) = - \frac{7 \rmi}{5760\, \sqrt{2}}\, \frac{\left( p \left(p-2\right) \right)^{\frac{5}{2} p \left(p-2\right)+\frac{1}{2}}}{\left(p-1\right)^{5 p \left(p-2\right) + 4}}\, \frac{1}{T^{5/2}} + \cdots,
\end{align}
\noindent
and the full expansion at $g=2$ and $p=3$ is
\begin{align}
F_{g=2}^{\text{sing}} (T) &= \frac{5103\, \rmi}{335544320\, \sqrt{2}\, T^{5/2}} - \frac{675}{8388608\, T^2} - \frac{871\, \rmi}{10485760\, \sqrt{2}\, T^{3/2}} - \frac{3}{4096\, T} - \nonumber \\
&
- \frac{890827\, \rmi}{477757440 \sqrt{2}\, \sqrt{T}} - \frac{76849}{201553920}.
\end{align}
\noindent
As for the regular piece, it features the same inclusion of half-integer powers of $T$; and for example with $g=2$ and $p=3$ it reads
\begin{equation}
F_{g=2}^{\text{reg}}(T) = -\frac{65861\, \rmi\, \sqrt{T}}{725594112\, \sqrt{2}} - \frac{249923\, T}{573956280} - \frac{153426341\, \rmi\, T^{3/2}}{66119763456\, \sqrt{2}} + \frac{256655242\, T^2}{52301766015} + \cdots.
\end{equation}
\noindent
The interest of \cite{gm21} is in Stokes data which is purely associated to the \textit{regular} parts of these free energies (\textit{e.g.}, herein the singular-part Stokes data are the Painlev\'e~I Stokes data which were already obtained in \cite{bssv22}). These Stokes data (eventually relating to nonperturbative enumerative invariants of the local curve) should then follow by applying the same procedure as above to our diverse new transseries sectors. In this exact same fashion, we then find, for instance,
\begin{align}
F_{X_3}^{(\ell|0)} (T) &\simeq \frac{G_{2}(\ell+1)}{\left(2\pi\right)^{\ell/2}}\, \rme^{- \frac{\ell}{g_{\text{s}}} \left( 6\, \text{Li}_2 \left(-\frac{2}{3}\right) + \frac{7\pi^2}{2} - 15\rmi \pi \log 2 + \frac{3}{2} \log 3 \left\{ t_{\text{c}}^2 + 20 \log 3 - 42 \log 2 \right\} + \cdots \right)} \times \nonumber \\
&\times \left( - \frac{g_{\text{s}}}{48\, 2^{3/4} \sqrt[4]{3}\, T^{5/4}}\right)^{\frac{\ell^2}{2}} + \cdots, \\
F_{X_3}^{(0|\ell)} (T) &\simeq \frac{G_{2}(\ell+1)}{\left(2\pi\right)^{\ell/2}}\, \rme^{+ \frac{\ell}{g_{\text{s}}} \left( 6\, \text{Li}_2 \left(-\frac{2}{3}\right) + \frac{7\pi^2}{2} - 15\rmi \pi \log 2 + \frac{3}{2} \log 3 \left\{ t_{\text{c}}^2 + 20 \log 3 - 42 \log 2 \right\} + \cdots \right)} \times \nonumber \\
&\times \left( \frac{g_{\text{s}}}{48\, 2^{3/4} \sqrt[4]{3}\, T^{5/4}}\right)^{\frac{\ell^2}{2}} + \cdots, \\
F_{X_3}^{(1|1)} (T) &\simeq \frac{\rmi}{2\pi g_{\text{s}}} \left( 6\, \text{Li}_2 \left(-\frac{2}{3}\right) + \frac{7\pi^2}{2} - 15\rmi \pi \log 2 + \frac{3}{2} \log 3 \left\{ t_{\text{c}}^2 + 20 \log 3 - 42 \log 2 \right\} + \cdots \right) + \nonumber \\
&+ \frac{17\rmi g_{\text{s}}}{192\, 2^{3/4} \sqrt[4]{3} \pi\, T^{5/4}} + \cdots.
\end{align}
\noindent
It would be very interesting to address the complete resurgence of the local curve, and the relation of its Stokes data to enumerative geometry in future work.

One question we will have to address in future work is whether the above results for the local curve will uphold when considering more complicated toric geometries---to start-off with, the canonical topological-string examples of local $\BP^2$ and local $\BP^1 \times \BP^1$ (with a single and with two K\"ahler parameters, respectively). As for the local curve described above, these backgrounds also lead to spectral geometries of the type \eqref{eq:hyperelliptic-spectral-curve}, with non-polynomial moment functions, but geometries which are now intrinsically \textit{two-cut} configurations \cite{m06, bkmp07} (\textit{i.e.}, the mirror curve has genus one). Borel singularities of the perturbative series for local $\BP^2$ have actually been much addressed in the literature, with numerical plots to be found in, \textit{e.g.}, \cite{cesv14, cms16, gm21, gm22a, gm22b}. All these instances clearly illustrate the hallmark of resonance with the existence of \textit{symmetrically reflected} Borel singularities. There is less evidence for local $\BP^1 \times \BP^1$, but for the interesting analysis in \cite{dmp11} pertaining to its large $N$ dual, ABJM gauge theory on $\BS^3$ \cite{abjm08}---more precisely, the Chern--Simons matrix model on the lens space $\BS^3/\BZ_2$ \cite{dt09, mp09}. It was shown in \cite{dmp10, dmp11} via large-order analysis of the Borel singularities of the ABJM perturbative series that they always appear in \textit{symmetric pairs}. This was explicitly shown to be true at the relevant different points in moduli space (orbifold, conifold, large-radius), with instanton actions again always arising in symmetric pairs \cite{dmp11}. This is another distinct appearance of our by-now familiar hallmark of resonance. All evidence considered certainly begs for an extensive analysis of these models in future work.

Recall that the ABJM partition function on $\BS^3$ computes the type IIA string theory partition function on $\text{AdS}_4 \times \BC\BP^3$ \cite{abjm08}. Rather interestingly, Chern--Simons matrix-model instantons were identified in \cite{dmp11} as dual type IIA D2-branes wrapping an $\BR\BP^3$ inside the $\BC\BP^3$. In light of this, one is immediately led to ask if our resurgence negative-tension D-brane results may be in any way directly addressed for critical strings in an AdS background. This is what we shall turn to next, in the context of D-branes in $\text{AdS}_3$.


\subsection{On Critical Strings and Negative-Tension D-Branes in AdS}\label{subsec:ads-strings}

One string-theoretic example of a curved background where D-branes have been much studied is that of D-branes in $\text{AdS}_{3}$ \cite{s99a}. Perhaps the most studied D-branes in this background have been $\text{AdS}_{2}$ D-branes, but there are other types \cite{bp00, gks01, lop01, pst01, r05, gkv21}. Attempting a direct BCFT calculation in $\text{AdS}_{3}$ would of course take us far from the main line of this paper, but there is a remarkable correspondence between euclidean $\text{AdS}_{3}$ (usually denoted by $\BH_3^+$) and Liouville  (B)CFT's \cite{rt05, hr06, hs07} which allows for direct contact with our analysis in section~\ref{sec:ZZminimalstring}. This is usually known as the $\BH_3^+$--Liouville correspondence. In particular, we are interested in the results of \cite{r05}, mapping D-branes in $\text{AdS}_{3}$ to D-branes in Liouville theory. Roughly speaking, therein FZZT branes correspond to $\text{AdS}_{2}$ D-branes and ZZ branes correspond to ``discrete'' $\text{AdS}_{2}$ D-branes---which then allows us to translate the AdS discussion back to our results in section~\ref{sec:ZZminimalstring} and hence, focusing on ZZ-like or ``discrete'' $\text{AdS}_{2}$ D-branes, infer on the existence of negative-tension D-branes in $\text{AdS}_{3}$.

Consider (bosonic) string theory on (euclidean) $\text{AdS}_3$. The metric is taken in Poincar\'e-type coordinates
\be
\textsf{g} = R^2 \left( \rmd \phi \otimes \rmd \phi + \rme^{2\phi}\, \rmd \gamma \otimes \rmd \bar{\gamma} \right)
\ee
\noindent
(with $\phi \in \BR$ and $\gamma, \bar{\gamma} \in \BC$, and spacetime boundary the complex plane parametrized by $(\gamma,\bar{\gamma})$ at $\phi \to +\infty$), and the NS $B$-field is
\be
B = R^2\, \rme^{2\phi}\, \rmd\gamma \wedge \rmd\bar{\gamma}.
\ee
\noindent
The corresponding closed-string world-sheet CFT is \cite{gks98}
\be
\CS_{\text{AdS}_3} [\phi,\beta,\gamma] = \frac{1}{4\pi} \int \rmd^2 \sigma \left( \partial\phi \bar{\partial}\phi - \gamma \bar{\partial} \beta - \bar{\gamma} \partial \bar{\beta} - b^2\, \beta \bar{\beta}\, \rme^{2 b \phi} \right),
\ee
\noindent
alongside some compact manifold CFT\footnote{For example, in this bosonic-string setting we could consider the compact manifold to be a WZW three-sphere alongside an adequate torus; say: $\BH_{3,\mathsf{k}+2}^+ \oplus \text{SU} (2)_{\mathsf{k}-2} \oplus 20 \text{ free bosons} \oplus \mathfrak{b}\mathfrak{c}$ with vanishing total central charge.} and $\mathfrak{b}\mathfrak{c}$ reparametrization ghosts. Herein, $\phi$ is a free boson with background charge and the bosonic $\beta\gamma$-system alone has central charge $2$. At level $\mathsf{k}+2$ (relating to the $\text{AdS}_3$ radius as $R^2 = \mathsf{k}+2$) the $\text{AdS}_3$ CFT central charge is
\be
c_{\text{AdS}_3} = 3 + \frac{6}{\mathsf{k}}.
\ee
\noindent
Moving towards D-branes, in particular euclidean $\text{AdS}_2$ D-branes (see below), one finds the extra boundary term in the $\text{AdS}_3$ action \cite{fr07}
\be
\CS_{\text{AdS}_3\text{,B}} [\phi,\beta] = \upmu \oint \beta\, \rme^{b\phi}.
\ee

The $\text{AdS}_2$ D-branes we are interested in are most evident in the  ``AdS coordinates'' of \cite{bp00}, with metric description
\be
\label{eq:AdS3inAdScoord}
\textsf{g} = R^2 \left( \rmd\psi \otimes \rmd\psi + \cosh^2 \psi \left( \rmd \omega \otimes \rmd \omega - \cosh^2 \omega\, \rmd \tau \otimes \rmd \tau \right) \right).
\ee
\noindent
$\text{AdS}_{2}$ D-branes (defined via the twined conjugacy classes of $\text{SL}(2,\BR)$ \cite{bp00}) are the $\text{AdS}_{2}$ leaves of the constant $\psi=\uppsi$ foliation of $\text{AdS}_3$, \textit{i.e.}, they slice $\text{AdS}_3$ in terms of fixed radius $\text{AdS}_2$ leaves at fixed $\psi$. They are hence labeled by a single real parameter $\uppsi \in \BR$, and have varying radius $\rho_{\text{AdS}_2} = R_{\text{AdS}_3} \cosh \uppsi$. Denote them by
\be
\label{eq:ads2d-branes}
\ket{\uppsi}_{\text{AdS}_2}.
\ee
\noindent
As we discuss next, these D-branes are essentially the Liouville FZZT D-branes \cite{r05, hr06, fr07}.

Remarkably \cite{rt05, hs07} arbitrary correlation functions in the $\BH_3^+$ CFT may be simply written in terms of Liouville CFT correlation functions, on surfaces of arbitrary genus. This correspondence is extendable to correlation functions on the disk with $\text{AdS}_2$ D-brane boundary conditions, now rewritten in terms of Liouville BCFT correlation functions with FZZT D-brane boundary conditions \cite{hr06}. The bulk $\BH_3^+$--Liouville correspondence identifies the $\BH_3^+$ level $\mathsf{k}+2$ with the Liouville coupling $b$ as
\be
b^2 = \frac{1}{\mathsf{k}},
\ee
\noindent
and, at D-brane level \cite{r05, hr06}, it identifies (up to a choice of sign\footnote{The reader should not be misled to wonder if the negative sign choice has anything to do with negative-tension D-branes in Liouville theory---it does not. Recall from section~\ref{sec:ZZminimalstring} that standard and negative D-branes share the same value of $\mu_{\text{B}}$, as $x$ in \eqref{eq:uniformization-minimal-string-x} is the same for both. Because this is irrelevant for our argument, we choose to simply work with the positive sign everywhere below.}) the $\text{AdS}_2$ D-brane parameter $\uppsi$ with the Liouville boundary cosmological constant $\mu_{\text{B}}$ as\footnote{A word on conventions: in section~\ref{sec:ZZminimalstring} we followed the minimal string conventions of \cite{ss03, kopss04} for the Liouville bulk/boundary cosmological constants. Herein, to follow standard notation in the $\BH_3^+$--Liouville correspondence, it is convenient to undo such conventions and revert back to the original Liouville bulk/boundary cosmological constants present in \eqref{eq:Liouville-bulk-action}-\eqref{eq:Liouville-boundary-action}. Effectively this gives rise to the term $\sin \pi b^2$ in the following equation.}
\be
\label{eq:muBtoAdSuppsi}
\mu_{\text{B}} = \rmi\, \sqrt{\frac{\mu_{\text{L}}}{\sin \pi b^2}}\, \sinh \uppsi.
\ee
\noindent
This yields the relation between FZZT and $\text{AdS}_2$ D-branes as\footnote{Reverting Liouville bulk/boundary cosmological constant conventions as mentioned in the previous footnote changes \eqref{eq:uniformization-minimal-string-x} to $\frac{\mu_{\text{B}}}{\sqrt{\mu}}\, \sqrt{\sin \pi b^2} = T_p (\upzeta)$, in which case the ensuing equation immediately follows.}

\be
\label{eq:AdS2-brane->FZZT-brane}
\ket{\uppsi}_{\text{AdS}_2} \,\mapsto\, \ket{\upzeta=\cosh \frac{1}{p} \left( \uppsi + \rmi \frac{\pi}{2} \right)}_{\text{FZZT}}.
\ee

Now, as we already reviewed in \eqref{eq:ZZ-from-FZZTs}, Liouville ZZ-branes may be constructed as differences of FZZT-branes \cite{zz01, m03}. Having a map \eqref{eq:AdS2-brane->FZZT-brane} between FZZT and $\text{AdS}_2$ D-branes, naturally led \cite{r05} to construct \textit{discrete} ZZ-like $\text{AdS}_2$ D-branes from differences of standard \eqref{eq:ads2d-branes} $\text{AdS}_2$ D-branes---building on the $\BH_3^+$--Liouville correspondence and in complete analogy with ZZ-branes in \eqref{eq:ZZ-from-FZZTs}. These discrete D-instantons are defined and denoted by\footnote{At this stage one might recall how ZZ-branes are located at the pinches of the spectral curve as in \eqref{eq:xmn,ymn,upzetamn}. Herein, it is interesting to notice that, from an $\text{AdS}_3$ perspective, the discrete $\text{AdS}_2$ D-branes are associated with leaves of (symmetrically alternating) radii $\rho_{\text{AdS}_2} \left(m,n\right) = R_{\text{AdS}_3} \left(-1\right)^{m+1} \sin n \pi b^2$.}

\be
\label{eq:discrete-ads-branes}
\ket{m,n}_{\text{AdS}_2} = \ket{\rmi\pi \left(m - b^2 n - \frac{1}{2}\right)}_{\text{AdS}_2} - \ket{\rmi\pi \left(m + b^2 n - \frac{1}{2}\right)}_{\text{AdS}_2},
\ee
\noindent
with $n,m \in \BN$ \cite{r05}. Note that this discrete set of D-branes lives in the \textit{analytic continuation} of $\text{AdS}_3$ according to \eqref{eq:AdS3inAdScoord}. It is now immediate to further consider their negative-tension counterparts in full analogy with what was earlier done for Liouville D-branes in section~\ref{sec:ZZminimalstring}. First recall from section~\ref{sec:ZZminimalstring} how resonance arises in minimal string theory as a consequence of the switching of sheets in the FZZT moduli space, which, at the level of ZZ-branes, amounts to exchanging  $n$ with $-n$ as in formulae \eqref{eq:saddle-zeta-alpha}-\eqref{eq:saddle-zeta-beta}. In this spirit, we are led to conjecture negative-tension discrete $\text{AdS}_2$ D-branes as

\be
\label{eq:nt-discrete-ads-branes}
\overline{\ket{m,n}}_{\text{AdS}_2} \equiv \ket{m,-n}_{\text{AdS}_2} = \ket{\rmi\pi \left(m + b^2 n - \frac{1}{2}\right)}_{\text{AdS}_2} - \ket{\rmi\pi \left(m - b^2 n - \frac{1}{2}\right)}_{\text{AdS}_2}.
\ee
\noindent
A complete, explicit treatment of such nonperturbative contributions is quite complicated and outside the scope of this paper, but we can nevertheless perform some calculations in this direction which give clear support to our proposal and the resonant resurgence hypothesis. We start with the empty disk which was computed in \cite{r05} for standard D-branes, and which is then immediate to extend to negative-branes as
\begin{align}
\label{eq:ads3-forward-action}
\bra{0}\ket{m,n}_{\text{AdS}_2} &= + 2^{1/4} \left(-1\right)^{m+1} \sqrt{8 \pi b\, \Gamma\left(1-b^2\right) \Gamma\left(1+b^2\right)}\, \sin n \pi b^2, \\
\label{eq:ads3-backward-action}
\langle 0\overline{\ket{m,n}}_{\text{AdS}_2} &= - 2^{1/4} \left(-1\right)^{m+1} \sqrt{8 \pi b\,  \Gamma\left(1-b^2\right) \Gamma\left(1+b^2\right)}\, \sin n \pi b^2,
\end{align}
\noindent
where clearly a resonant structure is emerging by construction. Indeed, we do find an extra (at first seemingly harmless) minus-sign exactly as we did in section~\ref{sec:ZZminimalstring}, say in \eqref{eq:Adand-Ad} or \eqref{eq:22km1-negative-1-instanton-result}. Of course the main point of section~\ref{sec:ZZminimalstring} was to show how this innocent minus-sign can indeed lead to very non-trivial consequences, due to its appearance in the argument of the exponential terms sitting inside the integrals over D-brane moduli space---the reader may recall the comparison between \eqref{eq:20-contribution-minimal-strings} and \eqref{eq:11-contribution-minimal-strings}. As such, our goal in the present subsection is to show how a similar effect also occurs for the above (negative) discrete $\text{AdS}_2$ D-branes (albeit this time around there is no ``matrix model'' calculation to check the resulting transseries structure against).

The next step in this reasoning is to compute string-theoretic annulus amplitudes for (negative) discrete $\text{AdS}_2$ D-branes. This yields information on the nature of the interactions between, say, two discrete $\text{AdS}_2$ D-branes or a discrete $\text{AdS}_2$ D-brane with its negative-tension counterpart. Let us do this calculation following our strategy in section~\ref{sec:ZZminimalstring}; in particular by making manifest the \textit{different nature} of the interaction between two D-branes \textit{versus} the interaction between a D-brane and its negative-tension sibling, as in \eqref{eq:20-contribution-minimal-strings} versus \eqref{eq:11-contribution-minimal-strings}. First focus on the $(2,0)$ nonperturbative sector. Via the $\BH_3^+$--Liouville correspondence one can write the contribution of two discrete $\text{AdS}_{2}$ branes in terms of a double-integral over the moduli space of $\text{AdS}_{2}$ branes---as in the Liouville counterpart \eqref{eq:20-contribution-minimal-strings}. But in order to herein apply \eqref{eq:20-contribution-minimal-strings} a slight modification is however required. Recall how in section~\ref{sec:ZZminimalstring} swapping sheets was related to switching the sign of $m$ in $\ket{m,n}_{\text{ZZ}}$. In the present $\text{AdS}$ construction \cite{r05}, however, it is swapping the sign of $n$ in $\ket{m,n}_{\text{AdS}_2}$ which is employed to construct discrete $\text{AdS}_2$ branes. To be directly consistent with this choice let us now choose to foliate the Liouville surface with $q$ sheets instead of $p$. Specifically, in \eqref{eq:20-contribution-minimal-strings} we used $\alpha\beta$ to label the $p$ sheets of the minimal-string FZZT moduli-space, but we could equally have described the same surface using $q$ sheets---now given by the map $y=T_{q}(\upzeta)$ (recall \eqref{eq:uniformization-minimal-string-x}-\eqref{eq:uniformization-minimal-string-y}). As explained, this is what we will do from now on for the Liouville side, and hence use $\alpha,\beta \in \left\lbrace 0, \ldots, q-1 \right\rbrace$. Then, the $\BH_3^+$--Liouville correspondence replaces
\begin{align}
\mathsf{A}_{\text{D}}^{\text{L}}\left(\upzeta_{\alpha}\right) &\,\,\longrightarrow\,\, \mathsf{A}^{\text{AdS}_3}_{\text{D}}\left(\uppsi_{\alpha}\right),\\
\mathsf{A}_{\text{A}}^{\text{L}}\left(\upzeta_{\alpha}, \upzeta_{\beta}\right) &\,\,\longrightarrow\,\, \mathsf{A}^{\text{AdS}_3}_{\text{A}}\left(\uppsi_{\alpha}, \uppsi_{\beta}\right),
\end{align}
\noindent
where
\be
\label{eq:uppsi_alpha}
\uppsi_{\alpha} \equiv \uppsi + 2\pi\rmi b^2\,\alpha,
\ee
\noindent
and where the disk and annulus amplitudes are $\mathsf{A}_{\text{D}}^{\text{AdS}_3}(\uppsi) = \bra{0}\ket{\uppsi}$ and $\mathsf{A}_{\text{A}}^{\text{AdS}_3}(\uppsi, \tilde{\uppsi}) = \langle \uppsi|\tilde{\uppsi} \rangle$. Notice how the $q$ sheets of the FZZT moduli space (in the language of section~\ref{sec:ZZminimalstring}, where the Liouville surface was foliated with $p$ sheets, these were made explicit in \eqref{eq:upzeta_alpha}) are now translated to the $q$ values of the $\text{AdS}_2$ modulus in \eqref{eq:uppsi_alpha}, once we identify Liouville and $\text{AdS}$ data as in \eqref{eq:AdS2-brane->FZZT-brane}. Further notice how a shift of $\uppsi$ by $2\pi\rmi b^2\, \alpha$ corresponds to a switching of sheets in the Liouville analogue of section~\ref{sec:ZZminimalstring}, as follows from \eqref{eq:discrete-ads-branes}-\eqref{eq:nt-discrete-ads-branes}. Then, further including compact-manifold matter and $\mathfrak{b}\mathfrak{c}$ ghost contributions (dropping the $\text{AdS}_3$ CFT superscript, and keeping the $\text{AdS}_2$ BCFT label implicit as this is our only relevant boundary condition in full string theory) the string-theoretic $(2,0)$-sector becomes (compare with \eqref{eq:20-contribution-minimal-strings})
\begin{align}
\label{eq:20-contribution-ads-3}
\frac{\NCZ^{(2,0)}}{\NCZ_{\text{pert}}} &\simeq \frac{1}{2} \int_{\CC} \frac{\text{d}\uppsi}{2\pi} \int_{\CC} \frac{\text{d}\widetilde{\uppsi}}{2\pi}\, \exp \Bigg( \CA^{[-1]}_{\alpha\beta}(\uppsi) + \CA^{[-1]}_{\alpha\beta}(\widetilde{\uppsi}) + \CA^{[0]}_{\alpha\beta}(\uppsi) + \CA^{[0]}_{\alpha\beta}(\widetilde{\uppsi}) + \CA^{[0]}_{\alpha\beta,\alpha\beta}(\uppsi, \widetilde{\uppsi}) + o(g_{\text{s}}) \Bigg),
\end{align}
\noindent
where $\CC$ represents (eventually saddle-point) integration over the moduli space of an $\text{AdS}_2$ brane. In addition we have (compare with the corresponding Liouville definitions in \eqref{eq:differences-FZZT-disks}, \eqref{eq:FZZT-Annulus-Combination-1-inst} and \eqref{eq:FZZTAnnulusDifferences})
\begin{align}
\label{eq:Aalphabeta-1AdS}
\CA^{[-1]}_{\alpha\beta}(\uppsi) &=\mathsf{A}_{\text{D}} (\uppsi-2\pi\rmi b^2\, \alpha) - \mathsf{A}_{\text{D}} (\uppsi-2\pi\rmi b^2\, \beta), \\
\CA^{[0]}_{\alpha\beta}(\uppsi) &= \frac{1}{2} \Big( \mathsf{A}_{\text{A}}  (\uppsi_{\alpha}, \uppsi_{\alpha}) - \mathsf{A}_{\text{A}} (\uppsi_{\alpha}, \uppsi_{\beta}) - \mathsf{A}_{\text{A}} (\uppsi_{\beta}, \uppsi_{\alpha}) + \mathsf{A}_{\text{A}} (\uppsi_{\beta}, \uppsi_{\beta}) \Big), \\
\label{eq:ads3-annulus-general}
\CA^{[0]}_{\alpha\beta,\gamma\delta} (\uppsi,\widetilde{\uppsi}) &= \mathsf{A}_{\text{A}} (\uppsi_{\alpha}, \widetilde{\uppsi}_{\gamma}) - \mathsf{A}_{\text{A}} (\uppsi_{\alpha}, \widetilde{\uppsi}_{\delta}) - \mathsf{A}_{\text{A}} (\uppsi_{\beta}, \widetilde{\uppsi}_{\gamma}) + \mathsf{A}_{\text{A}} (\uppsi_{\beta}, \widetilde{\uppsi}_{\delta}).
\end{align}
\noindent
We can check that upon localization on the saddle-points of the integral, the leading contribution from \eqref{eq:20-contribution-ads-3} indeed reproduces the instanton action \eqref{eq:ads3-forward-action} (up to the expected factor of 2). Namely setting the derivative of \eqref{eq:Aalphabeta-1AdS} to zero we arrive at the condition\footnote{We are leaving some details under the rug. Indeed, there are two common ways of labeling operators in $\text{AdS}_3$, and the $\BH_3^+$--Liouville map is formulated in the Laplace-transformed basis often labeled by $\mu$ (unrelated to Liouville $\mu$). It is in this basis which the calculation is usually done. On the other hand, the disk amplitudes \eqref{eq:ads3-forward-action}-\eqref{eq:ads3-backward-action} were written in the physical $x$ basis for convenience. Indeed, for these empty disks all $x$ dependence drops out which is consistent with their correspondence to instanton actions.}
\begin{equation}
\sinh \left( \uppsi + 2\pi\rmi b^2\, \alpha - \frac{\rmi}{2} \right) = \sinh \left( \uppsi + 2\pi\rmi b^2\, \beta - \frac{\rmi}{2} \right).
\end{equation}
\noindent
The above condition is exactly solved by the moduli which give rise to the discrete ZZ-like branes in \eqref{eq:discrete-ads-branes}. However, in order to perform the saddle-point integration to higher orders an explicit and \textit{regularized} expression for the annulus amplitude is required (which at Liouville level is essentially our discussion in subsection~\ref{subsec:regularization-one-instanton}). Unfortunately, to the best of our knowledge, regularized $\text{AdS}$ annuli have not been addressed in the literature: from the standpoint of the $\BH_3^+$--Liouville correspondence, the map has only been properly established for disk amplitudes \cite{hr06}; and when directly addressing euclidean $\text{AdS}_2$ D-branes in $\text{AdS}_3$, the relevant annulus amplitude may be found in \cite{pst01}, but only the \textit{unregularized} version of this amplitude---which still prevents us from moving to next order. Nevertheless, for our purposes in the present subsection it turns out that leaving the result unintegrated, as in \eqref{eq:20-contribution-ads-3}, suffices.

Let us compare the $(2,0)$ contribution in \eqref{eq:20-contribution-ads-3} with the $(1,1)$ contribution which would result from the interaction of a discrete $\text{AdS}_2$ D-brane with its negative-tension counterpart. Following the exact same procedure as above we now find the integral (compare with \eqref{eq:11-contribution-minimal-strings})
\begin{align}
\label{eq:11-contribution-ads-3}
\frac{\NCZ^{(1,1)}}{\NCZ_{\text{pert}}} &\simeq \frac{1}{2} \int_{\CC} \frac{\text{d}\uppsi}{2\pi} \int_{\bar{\CC}} \frac{\text{d}\widetilde{\uppsi}}{2\pi}\, \exp \Bigg( \CA^{[-1]}_{\alpha\beta}(\uppsi) + \CA^{[-1]}_{\beta\alpha}(\widetilde{\uppsi}) + \CA^{[0]}_{\alpha\beta}(\uppsi) + \CA^{[0]}_{\beta\alpha}(\widetilde{\uppsi}) + \CA^{[0]}_{\alpha\beta,\beta\alpha}(\uppsi, \widetilde{\uppsi}) + o(g_{\text{s}}) \Bigg).
\end{align}
\noindent
Upon comparison with \eqref{eq:20-contribution-ads-3}, describing the interaction of two identical discrete $\text{AdS}_2$ D-branes, a trivial calculation using \eqref{eq:ads3-annulus-general} shows that, much like for Liouville, one has\footnote{This should come without surprise as it is a purely diagrammatic statement. It is likely generic.}
\begin{equation}
\CA^{[0]}_{\alpha\beta,\beta\alpha} (\uppsi, \widetilde{\uppsi}) = - \CA^{[0]}_{\alpha\beta,\alpha\beta} (\uppsi, \widetilde{\uppsi}).
\end{equation}
\noindent
This implies that, in complete analogy with the Liouville calculation, the present $(1,1)$ contribution differs from the $(2,0)$ calculation by \textit{two} minus signs. The \textit{first} one is almost harmless, and it can be found in front of the instanton action; whereas the \textit{second} one is non-trivial: it sits inside the exponential in the integrand and will produce \textit{inverse} integrand-terms upon comparison between $(2,0)$ and $(1,1)$ contributions---hence rather distinct final amplitudes (this is the same overall discussion as in section~\ref{sec:ZZminimalstring} and we need not repeat it). We hope to come back to this calculation in the near future, to make it as explicit as in the Liouville/minimal-string case.

The above reasoning supports the existence of negative-tension D-instantons in $\text{AdS}_3$, but it is by no means an absolute proof and further work is required. Going forward, one interesting point we have not discussed is the analogue extension of our Liouville BCFT analysis in section~\ref{sec:ZZminimalstring} to a \textit{direct} $\text{AdS}_3$ BCFT analysis following \cite{gks01, lop01, pst01}. It should be possible to compute both disk and annulus amplitudes involving negative-tension D-branes in $\text{AdS}_3$ and hence further build upon our discussion above---and this would also be very interesting future research. Such would-be negative-tension results could further have interesting applications, most notably due to the relation of the $\text{AdS}_3$ CFT to two- and three-dimensional black holes, as well as to the near-horizon geometry of NS5-branes. Another closely related future research venue would be to extend our results to the settings of \cite{rw05, gkv21} which are also in close Liouville proximity.

\acknowledgments
We would like to thank
\'Oscar~Dias,
Paolo~Gregori,
Amihay~Hanany,
Marcos~Mari\~no,
Sebastiano~Martinoli,
Sylvain~Ribault,
Jo\~ao~Rodrigues,
for useful discussions, comments and/or correspondence. In particular, we would very much like to thank Marcos~Mari\~no and Sylvain~Ribault for their valuable feedback on our first draft. The results in this paper were reported by RS at the ``ARA Summary Workshop'' held in December at the Isaac Newton Institute for Mathematical Sciences, University of Cambridge, to which the authors would like to thank for their hospitality. The authors would further like to thank the Isaac Newton Institute for Mathematical Sciences, Cambridge, for support and hospitality during the programme ``Applicable Resurgent Asymptotics: Towards a Universal Theory'' where work on this paper was undertaken. MS is supported by the LisMath Doctoral program and FCT-Portugal scholarship SFRH/PD/BD/135525/ 2018. NT is supported by a fellowship from ``la Caixa'' Foundation (ID 100010434), with code LCF/BQ/DI20/11780029. MS and NT were partially supported by grants from the Simons Foundation. This research was supported in part by CAMGSD/IST-ID and via the FCT-Portugal grants UIDB/04459/2020, UIDP/04459/2020, PTDC/MAT-OUT/28784/2017. This work was partially supported by EPSRC grant EP/R014604/1. This paper is partly a result of the ERC-SyG project, Recursive and Exact New Quantum Theory (ReNewQuantum) funded by the European Research Council (ERC) under the European Union's Horizon 2020 research and innovation programme, grant agreement 810573.

\newpage

\appendix

\section{On General Minimal-String String-Equations}\label{app:minimal-string-equation-setup}

This appendix includes some generalities alongside some new results concerning minimal-string string-equations, which are used for checks of our BCFT and matrix-model results in the main body to the text. We refer the reader to the discussion in \cite{gs21}  (which we will partially follow next) for further details and for a complete list of the relevant references.

Consider $(p,q)=(2,2k-1)$ minimal string theory. Its string (differential in $z$) equation, computing the string specific-heat $u(z)$, is given by
\be
\label{eq:minimalstrings}
\sum_{p=0}^{\left[\frac{k-1}{2}\right]} t_{k-2p}(k)\, \frac{k-2p}{\alpha_{k-2p,k-2p}}\, R_{k-2p} \left[ u(z) \right] = z,
\ee
\noindent
where the $\alpha$-coefficients are
\be
\alpha_{i j} = \left( -1 \right)^{j} \frac{\Gamma \left( 2 i \right) \Gamma \left( \left( i-j \right) + 1 \right)}{2^{2\ell}\, \Gamma \left( i \right) \Gamma \left( j \right) \Gamma \left( 2 \left( i-j \right) + 2 \right)},
\ee
\noindent
and where the KdV times $t_p (k)$ are given by
\be
t_p (k) = \frac{2^{p-\frac{5}{2}} \left(2k-1\right) \sqrt{\pi}}{\Gamma \left(p+1\right) \Gamma \left(\frac{3-k-p}{2}\right) \Gamma \left(\frac{2+k-p}{2}\right)}.
\ee
\noindent
The Gel'fand--Dikii KdV potentials $R_{k} \left[ u \right]$ are polynomials in the specific-heat $u(z)$ as well as its $z$-derivatives. Moreover, they are given by the recursion relation \cite{gd75}
\be
\label{eq:gfdpolys}
R_{k+1}' = \frac{1}{4}\, g_{\text{s}}^2\, R_{k}''' - u\, R_{k}' - \frac{1}{2} u'\, R_{k},
\ee
\noindent
with the starting coefficient $R_0 \left[u\right]=\frac{1}{2}$. The hierarchy of minimal-string string-equations are non-linear ordinary differential equations in $z$, which, at the end-of-the-day, must still be tuned to the conformal background \cite{mss91, ss03} so that the specific-heat is dependent only upon the string coupling, $u \equiv u (g_{\text{s}})$. Although we are mainly interested in this minimal-string hierarchy, it will prove useful to consider the multicritical hierarchy as a basis for what follows. This multicritical-string hierarchy of  equations is simpler, as \cite{gm90a, ds90, bk90, d90, gm90b}
\begin{equation}
\label{eq:multicrits}
\left(-1\right)^{k} \frac{2^{k+1}\, k!}{\left( 2k-1 \right)!!}\, R_k \left[ u(z) \right] = z.
\end{equation}

In the main body of the paper we are interested in understanding the generic $(2,2k-1)$ minimal string theory alongside its $k \to + \infty$ JT-gravity limit. In principle this might come about by writing down the generic string equation (minimal or multicritical alike) that follows from solving \eqref{eq:gfdpolys}. This is not so easy, since there is no known closed-form expression for an arbitrary Gel'fand--Dikii polynomial $R_k$. There is, however, some work constructing \textit{partial} closed-forms for these polynomials \cite{gz90b, gz91, gs21}. It turns out that even with only partial knowledge of these $R_k$, some calculations may be made yielding certain quantities in the transseries solutions to the generic $k$th string equation. For example, it was obtained in \cite{gs21} a closed-form expression for the first non-zero coefficient of the one-instanton sector of the $k$th minimal string (MS), as
\begin{equation}
\label{eq:MS10Sectoru}
u^{(1,0)}_{0,{\text{MS-}}k}(z) = \sigma \left( \sum_{n=0}^{k-1} \sum_{\ell=k-2\left[\frac{k-1}{2}\right]}^{k} \ell\, n\, t_{\ell} (k)\, \frac{\alpha_{\ell,\ell-n}}{\alpha_{\ell\ell}}\, u_0^{\ell-n-1}(z)\, A'(z)^{2n-1} \right)^{-\frac{1}{2}},
\end{equation}
\noindent
for some arbitrary normalization constant $\sigma$, where $u_0(z)$ is the perturbative solution to the string equation at lowest genus, and where $A(z)$ is the instanton action for  this instanton sector.

\paragraph{Gel'fand--Dikii Coefficients:}

The analysis in \cite{gs21}, from which the above-mentioned arbitrary-$k$ statement for $u^{(1,0)}_{0,\text{MS-}k}$ followed, only made use of closed-form coefficients for terms in the Gel'fand--Dikii polynomials with the following schematic structure
\begin{equation}
R_k \equiv R_k \left\{ u^k, u^{\bullet}\, \frac{\rmd^{\bullet}u}{\rmd z^{\bullet}}, u^{\bullet}\, \frac{\rmd^{\bullet}u}{\rmd z^{\bullet}}\, \frac{\rmd u}{\rmd z} \right\}.
\end{equation}
\noindent
Herein $\bullet$ is an element of some index set that is summed over (see equation (3.50) of \cite{gs21}). As it turns out, however, knowledge of closed-form Gel'fand--Dikii coefficients must now be extended if we are to check the $(1,1)$ nonperturbative sector at arbitrary $k$ in this paper using string equations. Via similar methods to those in \cite{gs21}, and moreover by using the relation between the Gel'fand--Dikii polynomials and the Miura--Gardner--Kruskal recurrence \cite{g15}, we were able to obtain closed-forms of all coefficients of the enlarged schematic structure
\begin{equation}
R_k \equiv R_k \left\{ u^k, u^{\bullet}\, \frac{\rmd^{\bullet}u}{\rmd z^{\bullet}}, u^{\bullet} \left( \frac{\rmd^{\bullet}u}{\rmd z^{\bullet}} \right)^2, u^{\bullet} \left( \frac{\rmd^{\bullet}u}{\rmd z^{\bullet}} \right)^3 \right\}.
\end{equation}
\noindent
As expected, these coefficients become more and more complicated and it is not enlightening to show the full results here. Below we will present formulae only up to terms like $u^{\bullet} \left( \frac{\rmd^{\bullet}u}{\rmd z^{\bullet}} \right)^2$, for which we find
\begin{align}
R_k &= \upalpha_{k}\, u^k + \sum_{p=1}^{k-1} \upalpha_{k,p}\, u^{k-p-1}\, \frac{\rmd^{2p}u}{\rmd z^{2p}}\, g_{\text{s}}^{2p} + \\
& + \sum_{p_1=4}^{k+1} \sum_{p_2=0}^{p_1-4} \upalpha_{k,p_1,p_2}\, u^{k+1-p_1}\, \frac{\rmd^{2p_1-p_2-7}u}{\rmd z^{2p_1-p_2-7}}\, \frac{\rmd^{p_2+1}u}{\rmd z^{p_2+1}}\, g_{\text{s}}^{2p_1-6} + \cdots, \nonumber
\end{align}
\noindent
where the coefficients above are given respectively by 
\bea
\upalpha_k &=& \left(-4\right)^{-k}\, \frac{\Gamma \left(2 k \right)}{k\, \Gamma \left(k\right)^2}, \\
\upalpha_{k,p} &=& \left(-1\right)^{k+p}\, 2^{-2(p+1)}\, \frac{\Gamma \left(k+\frac{1}{2}\right)}{\Gamma \left(p+\frac{3}{2}\right) \Gamma \left(k-p\right)}, \\
\upalpha_{k,p_1,p_2} &=& (-1)^{k+p_1+p_2+1}\, 4^{1-p_1} \left(\delta_{4,p_1-p_2}-2\right) \frac{\Gamma \left(k+\frac{1}{2}\right)}{\Gamma \left(p_1-\frac{1}{2}\right) \Gamma \left(k-p_1+2\right)} \times \\
&&
\times \left( 1 - \frac{\Gamma \left(p_2-2p_1+6\right)}{\Gamma \left(4-2 p_1\right) \Gamma \left(p_2+3\right)} \right). \nonumber
\eea
\noindent
Note the relation to the previously defined coefficients, $\upalpha_k = \frac{\alpha_{k,k}}{k}$ and $\upalpha_{k,p} =\alpha_{k,k-p}$.

Using the above formulae, we were able to extend the current predictions for the specific-heat leading-coefficients of order-$k$ multicritical or minimal strings; which we present below. In the main text, we will use these results to obtain a non-trivial check of matrix model and BCFT results for the free energies. But before moving on to the new results, we recall from \cite{gs21} that the derivatives of the minimal-string instanton-actions obey certain algebraic equations. Since these results will be used in the main text as well, we include them explicitly herein as 
\begin{equation}
\label{eq:instActionODEMS}
\sum_{n=0}^{k-1} \left( \sum_{\ell=k-2\left[\frac{k-1}{2}\right]}^{k} \ell\, t_{\ell} (k)\, \frac{\alpha_{\ell,\ell-n}}{\alpha_{\ell\ell}}\, u_0^{\ell-n-1}(z) \right) A'(z)^{2n} = 0.
\end{equation}

\paragraph{Transseries Coefficients:}

For simplicity we begin with the multicritical theory, whose order-$k$ specific-heat satisfies the string equation \eqref{eq:multicrits}. Using a two-parameter\footnote{See \cite{gs21} for a more generic set-up. However, a one-parameter transseries \textit{ansatz} was mainly used as there were not enough closed-form Gel'fand--Dikii coefficients available to obtain data in the ``bulk'' of the transseries.} transseries \textit{ansatz}, alongside the newly-found closed-form Gel'fand--Dikii coefficients, this allows us to write the first non-zero contribution to the $(2,0)$ sector of the $k$th multicritical string (MC) as 
\begin{equation}
\label{eq:MC20Sector}
u_{0,{\text{MC-}}k}^{(2,0)}(z) = \frac{\sum\limits_{i=0}^{k-2} (-1)^{i+k+1} \frac{ \left(\frac{1}{4}-4^{-2-i}\right)\, \Gamma \left(k+\frac{1}{2}\right)}{\Gamma \left(i+\frac{5}{2}\right) \Gamma \left(k-1-i\right)}\, u_0^{k-i-2}(z)\, A'(z)^{2i}}{\sum\limits_{i=0}^{k-1} 2^{2i}\, \upalpha_{k,i}\, u_0^{k-i-1}(z)\, A'(z)^{2i}} \left( u_0^{(1,0)}(z) \right)^2.
\end{equation}
\noindent
Note that on the right-hand side $u_0^{(1,0)}(z)$ would be better denoted by $u^{(1,0)}_{0,{\text{MC-}}k}(z)$. However this will not be convenient in the following where we will use this very same equation but rather with $u^{(1,0)}_{0,{\text{MS-}}k}(z)$, as in \eqref{eq:MS10Sectoru}, in its right-hand side. Hence we leave it herein unspecified. In the same fashion, the first non-zero contribution to the $(1,1)$ sector of the $k$th multicritical string is given by
\begin{equation}
\label{eq:MC11Sector}
u_{0,{\text{MC-}}k}^{(1,1)}(z) = \sum_{i=0}^{k-2} (-1)^{i+k+1} \frac{4^{-1-i}\, \Gamma \left(k+\frac{1}{2}\right)}{\alpha_{k,k}\, \Gamma \left(i+\frac{3}{2}\right) \Gamma \left(k-1-i\right)}\, u_0^{-i-1}(z)\, A'(z)^{2i} \left(u_0^{(1,0)}(z)\, u_0^{(0,1)}(z) \right).
\end{equation}
\noindent
Using the above formulae it is then straightforward to immediately obtain the results we want for the specific heat of the $(2,2k-1)$ minimal string, now satisfying the string equation \eqref{eq:minimalstrings}. Writing these explicitly, we find that the first non-zero contribution to the minimal-string $(2,0)$ sector is given by
\begin{equation}
\label{eq:MS20SectorSpecificHeat}
u_{0,{\text{MS-}}k}^{(2,0)}(z) = \frac{\sum\limits_{p=0}^{\left[\frac{k-1}{2}\right]} \frac{t_{k-2p}(k)}{\alpha_{k-2p,k-2p}} \left( \sum\limits_{i=0}^{k-2p-1} \frac{(-1)^{k-2p-i}\, \Gamma \left(k-2p+\frac{1}{2}\right)\, u_0^{k-2p-i-1} (z)\, A^{\prime}(z)^{2i}}{4\, \Gamma \left(i+\frac{3}{2}\right)\, \Gamma \left(k-2p-i\right)} \right)\, u_{0,{\text{MC-}}(k-2p)}^{(2,0)} (z)}{\sum\limits_{p=0}^{\left[\frac{k-1}{2}\right]} \frac{t_{k-2p}(k)}{\alpha_{k-2p,k-2p}} \left( \sum\limits_{i=0}^{k-2p-1} \frac{(-1)^{k-2p-i}\, \Gamma \left(k-2p+\frac{1}{2}\right)\, u_0^{k-2p-i-1} (z)\, A^{\prime}(z)^{2i}}{4\, \Gamma \left(i+\frac{3}{2}\right)\, \Gamma \left(k-2p-i\right)} \right)},
\end{equation}
\noindent
while the first non-zero contribution to the minimal-string $(1,1)$ sector simplifies and is given by
\begin{equation}
\label{eq:MS11Sectoru}
u_{0,{\text{MS-}}k}^{(1,1)}(z) = \frac{\sum\limits_{p=0}^{[\frac{k-1}{2}]} t_{k-2p}(k)\, \left(k-2p\right)\, u_0^{k-2p-1}(z)\, u_{0,{\text{MC-}}(k-2p)}^{(1,1)}(z)}{\sum\limits_{p=0}^{[\frac{k-1}{2}]} t_{k-2p}(k)\, \left(k-2p\right)\, u_0^{k-2p-1}(z)}.
\end{equation}
\noindent
Keep in mind that in \eqref{eq:MS20SectorSpecificHeat} and \eqref{eq:MS11Sectoru} we are using \eqref{eq:MC20Sector} and \eqref{eq:MC11Sector} but whose latter respective right-hand sides use lower-instanton \textit{minimal string} data (and not multicritical data).

\section{On the $(2,5)$ Minimal-String String-Equation}\label{app:minimal-string-25}

Having discussed generics on minimal-string string-equations in the previous appendix~\ref{app:minimal-string-equation-setup}, let us now focus on a concrete, non-trivial example. We will address the string equation for $(2,5)$ minimal string theory, following upon the analysis in \cite{gs21}. The string equation \eqref{eq:minimalstrings} becomes 
\begin{equation}
\frac{5}{4\sqrt{2}} \left( - u + u^3 - g_{\text{s}}^2\, u\, u^{\prime\prime} - \frac{1}{2} g_{\text{s}}^2 \left( u^{\prime} \right)^2 + \frac{1}{10} g_{\text{s}}^4\, u^{\prime\prime\prime\prime} \right) = z,
\end{equation}
\noindent
where, at the end-of-the-day, we will need to fix $z=0$ in order to reach the conformal background.

The full four-parameter transseries solution to this equation will be of the form \eqref{eq:generic-4parameter-TS}, which we write for the specific heat as
\begin{equation}
u \left(z, g_{\text{s}}, \boldsymbol{\sigma} \right) = \sum_{\boldsymbol{n} \in \BN_{0}^{4}} \boldsymbol{\sigma}^{\boldsymbol{n}}\, \rme^{-(n_1-m_1)\frac{A_1 (z)}{g_{\text{s}}}-(n_2-m_2)\frac{A_2 (z)}{g_{\text{s}}}}\, u^{(\boldsymbol{n})} (z, g_{\text{s}}),
\end{equation}
\noindent
where $\boldsymbol{n} = (n_1,m_1,n_2,m_2)$ and where the transseries sectors are given by asymptotic series with starting genus $\beta_{n_1 m_1 n_2 m_2}$ as
\begin{equation}
u^{(\boldsymbol{n})} (z, g_{\text{s}}) \simeq \sum_{g=0}^{+\infty} g_{\text{s}}^{g+\beta_{n_1 m_1 n_2 m_2}}\, u^{\boldsymbol{n}}_g (z).
\end{equation}
\noindent
Details of the calculation have been outlined in \cite{gs21}, together with the perturbative and the one-instanton sectors. Let us therefore only list a few nonperturbative sectors herein. Recall that for this, and at either perturbative or nonperturbative levels, it makes sense to introduce additional variables \cite{biz80, iz92, gs21} (implicitly encoding $z$-dependence in a user-friendly way for the specific calculations; see \cite{gs21}) as
\begin{align}
\label{eq:class-gen-0-s-e}
& \frac{5}{4\sqrt{2}} \left( u_0(z)^{3}-u_0(z) \right) = z, & \\
& U(z) = \frac{\sqrt{5}}{u_0(z)}\, \sqrt{2-u_0^2(z)}, & \mathcal{U}(z) = \sqrt{\frac{5-U(z)}{5+U(z)}},
\end{align}
\noindent
where \eqref{eq:class-gen-0-s-e} is the classical, genus-zero string equation. Then we have the sectors
\begin{align}
u^{(1,1,0,0)}_0 (z) &= 0, & 
u^{(1,1,0,0)}_1 (z) &= \sqrt{10}\, \frac{\left(U-1\right) \left(U^2+5\right)^{5/4}}{U\, \sqrt{U+5} \left(U^2-25\right)}, \\
u^{(0,1,1,0)}_0 (z) &= 0, & 
u^{(0,1,1,0)}_1 (z) &= \frac{2^{1/4}}{5^{7/4}}\, \frac{\left(3\, \mathcal{U}^4 - 4\, \mathcal{U}^2 + 3\right)^{5/4}}{\left(-\mathcal{U}\right)^{3/2} \left(\mathcal{U}^2-1\right)},\\
u_0^{(2,1,0,0)}(z) &= \frac{\left(\frac{5}{2}\right)^{3/4} \left(U^2+5\right)^{3/8}}{8\,  \sqrt{U}\, \sqrt[4]{5-U}}\, \log \frac{U^4 \left(5-U\right)^{10}}{\left(U^2+5\right)^7}, \span
\end{align}
\noindent
and furthermore the one that we are interested-in for our comparison is the $(1,1,1,0)$ sector. For that we interestingly find the non-rational term
\begin{align}
\label{eq:compare-string-equation-1110-guy}
u^{(1,1,1,0)}_0(z) &= 10^{3/8}\, \frac{\left(3\, \mathcal{U}^4 - 4\, \mathcal{U}^2 + 3\right)^{3/8}}{\left(1-\mathcal{U}^2\right)^{1/2}}\, \tanh^{-1} \mathcal{U}.
\end{align}
\noindent
All we have left to do is to translate these results to the free energy. As this involves the term $A^{\prime}(z)$ it is convenient to revert back to the variable $U(z)$. With the procedure outlined in \cite{gs21} we readily obtain 
\begin{align}
\label{eq:compare25fe21}
F^{(2,1,0,0)}_0(z) &= -\frac{\left(\frac{5}{2}\right)^{1/4} \left(U^2+5\right)^{7/8}}{32\,  \sqrt{U} \left(5-U\right)^{5/4}}\, \log \frac{U^4 \left(5-U\right)^{10}}{\left(U^2+5\right)^7}, \\
\label{eq:compare25fe1110}
F^{(1,1,1,0)}_0(z) &= \frac{1}{2} \left( \frac{5}{2} \right)^{1/4} \frac{\left(U^2+5\right)^{7/8}}{U^{1/2} \left(5+U\right)^{5/4}}\, \tanh^{-1} \sqrt{\frac{10}{5+U}-1}.
\end{align}
\noindent
In the $(2,5)$ minimal-string conformal-background where $z \to 0$, $u_0 \to 1$, $U \to \sqrt{5}$, we find 
\begin{align}
F^{(1,1,1,0)}_0 = \frac{1}{2}\, \frac{2^{5/8} \cdot 5^{7/8}}{\left( 5+\sqrt{5} \right)^{5/4}}\, \tanh^{-1} \sqrt{\frac{1}{2} \left( 3-\sqrt{5} \right)}.
\end{align}

\section{Useful Formula for Matrix-Model Computations}\label{app:matrix-formula}

For completeness and the convenience of the reader who wishes to reproduce our calculations in full detail, let us include one useful formula in this appendix. The following relation is very useful as it captures the derivative method used at length in the calculations in \cite{mss22},
\begin{align}
\CI &= \int_{\CC^{\star}} \text{d}x \int_{\bar{\CC}^{\star}} \text{d}\bar{x}\, \frac{1}{(x-\bar{x})^2}\, \rme^{-\frac{1}{g_{\text{s}}} \left(V(x) - V(\bar{x})\right)}\, \Big\{ \CF_0 (x, \bar{x}) + g_{\text{s}}\, \CF_1 (x, \bar{x}) + \cdots \Big\}\, = \nonumber \\
&= - g_{\text{s}}\, \frac{\rmi\pi}{12}\, \frac{1}{V^{\prime\prime}\left(x^{\star}\right)^3}\, \Bigg\{ \CF_0 (x^{\star}, x^{\star}) \left( \left(V^{(3)}(x^{\star})\right)^2 - V^{\prime\prime}(x^{\star})\, V^{(4)}(x^{\star})\right) - \nonumber \\
&- V^{\prime\prime}(x^{\star})\, V^{(3)}(x^{\star}) \left( \CF_0^{(1,0)} (x^{\star}, x^{\star}) + \CF_0^{(0,1)} (x^{\star}, x^{\star}) \right) + \nonumber \\
&+ 3 \left(V^{\prime\prime}(x^{\star})\right)^2 \left( \CF_0^{(2,0)} (x^{\star}, x^{\star}) - 2 \CF_0^{(1,1)} (x^{\star}, x^{\star}) + \CF_0^{(0,2)} (x^{\star}, x^{\star}) \right) \Bigg\} + o(g_{\text{s}}^2),
\end{align}
\noindent
where $\CC^{\star}$ and $\bar{\CC}^{\star}$ are steepest-descent contours through a saddle $x^{\star}$ of $V(x)$, and where $\CF_i (x,\bar{x})$ is any function (with its derivatives also featured). For example, when addressing the calculation of the $(1|1)$ configuration, we need to compute
\begin{equation}
\frac{\mathcal{Z}^{(1|1)}(t, g_{\text{s}})}{\mathcal{Z}^{(0|0)}(t, g_{\text{s}})} = \frac{1}{(2\pi)^2}\, \CI,
\end{equation}
\noindent
where we pick 
\begin{equation}
\CF_0(x,\bar{x}) = \exp \Big( 2 A_{0;2} (x, x) + 2 A_{0;2} (\bar{x}, \bar{x}) - 4 A_{0;2} (x, \bar{x}) \Big)
\end{equation}
\noindent
and
\begin{align}
\CF_1(x,\bar{x}) &= \CF_0(x,\bar{x})\, \Big( 2 A_{1;1}(x) - 2 A_{1;1}(\bar{x}) + \frac{4}{3} A_{0;3}(x,x,x) - \frac{4}{3} A_{0;3}(\bar{x}, \bar{x}, \bar{x}) + \nonumber \\
&+ 4 A_{0;3}(x,x, \bar{x}) - 4 A_{0;3}(x, \bar{x}, \bar{x}) \Big).
\end{align}
\noindent
Herein the $A_{g;h}$'s denote adequate integrated versions of the multi-resolvent $W$ correlators, which depend on the choice of Bergman kernel. See \cite{mss22} for details.

\newpage

\bibliographystyle{plain}

\end{document}